\documentclass[final,3p,times]{elsarticle}

\usepackage{color}

\usepackage{epsfig}
\usepackage{amsmath,mathtools,blkarray,bm}
\usepackage{dcolumn}

\allowdisplaybreaks

\newcolumntype{.}{D{.}{.}{10}}
\newcolumntype{L}[1]{>{\raggedright\let\newline\\\arraybackslash\hspace{0pt}}m{#1}}
\newcolumntype{C}[1]{>{\centering\let\newline\\\arraybackslash\hspace{0pt}}m{#1}}

\newcommand\f[2]{\frac{#1}{#2}}

\def\be{\begin{equation}}
\def\ee{\end{equation}}
\def\bea{\begin{eqnarray}}
\def\eea{\end{eqnarray}}

\def\nn{\nonumber}
\def\hs{\hspace{.1mm}}

\def\as{{\alpha_s}}
\def\msbar{$\overline{\mathrm{MS}}$ }

\def\mR{\mu^2_R}
\def\mF{\mu^2_F}

\def\ep{\epsilon}
\def\cm{\mathcal{M}}
\def\cf{\mathcal{F}}
\def\cT{\bm{\mathrm{T}}}
\def\sP{\hat{\bm{\mathrm{P}}}}
\def\Ph{\hat{P}}
\def\kper{k_{\perp}}
\def\ktil{\tilde{k}}

\def\R{\mathrm{Re}}

\def\la{\langle}
\def\ra{\rangle}
\def\lra{\leftrightarrow}

\def\pipj{(p_I\cdot p_J)}
\def\pipjo{(p_I\cdot p_{j_0})}
\def\piq{(p_I\cdot q)}
\def\pjq{(p_J\cdot q)}
\def\pjoq{(p_{j_0}\cdot q)}
\def\ai{\alpha_I}
\def\aj{\alpha_J}
\def\Li{\mathrm{Li}}

\def\Nep#1{\Biggl( \f{\mu_R^2 e^{\gamma_{\rm{E}}}}{4\pi} \Biggr)^{#1\ep}}

\def\mReg{\mu_R^2 e^{\gamma_{\rm{E}}}}

\journal{Nuclear Physics B}

\begin{document}

\begin{frontmatter}

\title{Four-dimensional formulation of the
  sector-improved residue subtraction scheme}

    \author{M. Czakon and D. Heymes}

    \address{
      Institute for Theoretical Particle Physics and Cosmology,
      RWTH Aachen University, D-52056 Aachen, Germany
    }

\cortext[thanks]{Preprint number: TTK-14-16}

\begin{abstract}

  \noindent
  Four years ago, one of us introduced a novel subtraction scheme
  \cite{Czakon:2010td} for the evaluation of double-real radiation
  contributions to cross sections at next-to-next-to-leading order
  (NNLO) in QCD. This approach, named SecToR Improved Phase sPacE for
  Real radiation (STRIPPER), has already found several non-trivial
  applications. In particular, it has allowed for the determination of
  NNLO corrections to hadronic top-quark pair production, fully
  differential top-quark decays, inclusive semileptonic charmless
  b-quark decays, associated Higgs boson and jet production in gluon
  fusion, muon decay spin asymmetry, and t-channel single-top
  production. Common to these calculations was the use of conventional
  dimensional regularization (CDR). In this publication, we present a
  complete formulation of the subtraction scheme for arbitrary
  processes with any number of colored partons in the final state, and
  up to two partons in the initial state. Furthermore, we modify the
  integrated subtraction terms of the double-real radiation to enable
  the introduction of the 't Hooft-Veltman version of dimensional
  regularization (HV), in which resolved states are
  four-dimensional. We demonstrate the correctness of our approach on
  the example of top-quark pair production in the gluon fusion
  channel.

\end{abstract}

\end{frontmatter}

%%%%%%%%%%%%%%%%%%%%%%%%%%%%%%%%%%%%%%%%%%%%%%%%%%%%%%%%%%%%%%%%%%%%%%%%%%%%%%%%

\section{Introduction}

\noindent
Perturbative calculations beyond the next-to-leading order of Quantum
Chromodynamics are notorious for their complexity. Already the
next-to-next-to-leading order presents tremendous obstacles. One of
them is the evaluation of infrared and ultraviolet divergent two-loop
virtual amplitudes. This is an unsolved issue in the general case
despite the existence of a few analytic results for low multiplicity
processes. Another obstacle is the evaluation of double-real and mixed
real-virtual contributions containing infrared singular
phase space integrals. In principle, these contributions can be
obtained by means of Monte Carlo techniques, once suitable
subtractions have been introduced to generate numerically integrable
functions. The form of the subtractions defines a {\it subtraction
  scheme}. Unfortunately, it turns out that subtraction schemes are
extremely complex. At present, there are several on-going multi-year
efforts at the construction of general solutions. Antenna subtraction
\cite{GehrmannDeRidder:2005cm}, and $q_T$ subtraction
\cite{Catani:2007vq}, are amongst the most advanced initiatives, which
have already found several non-trivial applications. Another scheme
under construction has been introduced in
Ref.~\cite{Somogyi:2005xz}. Here, we will be concerned with the subtraction
scheme STRIPPER introduced by one of us in
Ref.~\cite{Czakon:2010td}. Inspired in some aspects by ideas of
Frixione, Kunszt and Signer \cite{Frixione:1995ms}, and in other
aspects by ideas of Binoth and Heinrich \cite{Binoth:2000ps}, the
scheme has proven its worth in many applications: it has allowed for
the determination of NNLO corrections to hadronic top-quark pair
production \cite{Czakon:2011ve, Baernreuther:2012ws, Czakon:2012zr,
  Czakon:2012pz, Czakon:2013goa}, fully differential top-quark decays
\cite{Brucherseifer:2013iv}, inclusive semileptonic charmless b-quark
decays \cite{Brucherseifer:2013cu}, associated Higgs boson and jet
production in gluon fusion \cite{Boughezal:2013uia}, muon decay spin
asymmetry \cite{Caola:2014daa}, and t-channel single-top production
\cite{Brucherseifer:2014ama}. The listed advanced STRIPPER
applications performed independently of the inventor have been
preceded by the much simpler case of QED corrections to $Z$-boson
decay into a pair of massless leptons \cite{Boughezal:2011jf}. The
purpose of the present work is to complete the construction of the
scheme in order to allow for the evaluation of cross sections for
arbitrary processes.

The original idea of Ref.~\cite{Czakon:2010td} was to concentrate on
the numerical calculation of the coefficients of the Laurent expansion
in $\ep$ (dimensional regularization parameter, with space-time
dimension $d = 4-2\ep$) of the double-real cross  section
contribution. The latter requires a phase space integral over the
momenta not only of the partons present in the Born approximation, but
also of two  additional massless partons. It seemed obvious that other
cross section contributions are much easier to obtain, since their
kinematics is, in the worst case, the same as that of next-to-leading
order real-radiation contributions. Furthermore, the concept was to
refrain from (almost) any analytic  integration. By inspection of
other efforts, it was clear that it is the insistence on analytic
integration that makes the subtraction schemes difficult to develop.
Finally, in order to avoid as many complications as possible, the
construction was performed uniformly in $d$ dimensions. This
corresponds to conventional dimensional regularization, where both
momenta and spin degrees-of-freedom of external particles are $d$
dimensional. This implies that even Born matrix elements will have a
non-trivial expansion in $\ep$. While the other basic ideas of the
scheme stood the test of time, the use of CDR is now an important
drawback. Indeed, software implementations of tree-level matrix
elements only provide them at $\ep = 0$, i.e.\ in four
dimensions. This makes it necessary to recalculate the matrix elements
for each project from scratch. Furthermore, the need to parameterize
an increasing number of dimensions depending on the multiplicity of
the process seems not only a major annoyance, but also a source of
inefficiency. In this publication, we will solve the problem by
introducing a number of corrections to the integrated subtraction
terms in the double-real radiation contribution. As a result, we will
formulate the scheme in 't Hooft-Veltman regularization. In
particular, we will only need four-dimensional external momenta and
polarizations in the evaluation of actual matrix elements. There will
still be a trace of higher dimensionality in the integration over
unresolved momenta. However, we will only have to consider six
dimensional unresolved momenta in the worst case. On the example of
top-quark pair production, we will also demonstrate that the
introduced improvements of STRIPPER fulfill their purpose. We stress
that we only present the algorithm to obtain the expressions needed
for the implementation of the subtraction scheme. This algorithm
requires the knowledge of soft and collinear limits of QCD amplitudes
and provides the expressions for the subtraction and integrated
subtraction terms by simple substitutions. Due to the
number and size of the resulting final formulae, we do not reproduce
them here. Instead, we plan to provide a software package in the
future. Nevertheless, existing calculations can be converted to 't
Hooft-Veltman regularization with moderate effort following this
publication.

The paper is organized as follows. In the next section, we present an
outline of the subtraction scheme together with a complete list of
cross section contributions to evaluate in a general
next-to-next-to-leading order calculation. Subsequently, we proceed
with the construction of the scheme by considering its major elements:
phase space decomposition (Section~\ref{sec:decomposition}), phase
space parameterization (Section~\ref{sec:parameterization}), and
derivation of the subtraction and integrated subtraction terms
(Section~\ref{sec:subtraction}). This part is common to the CDR and HV
formulations. Afterwards, we discuss the modifications necessary to
introduce HV regularization: average over the azimuthal angles
(Section~\ref{sec:AzimuthalAverage}), separation of finite
contributions (Section~\ref{sec:FiniteContributions}), and finally the
regularization itself (Section~\ref{sec:HVregularization}). This part
is followed by the example of top-quark pair production in the gluon
fusion channel (Section~\ref{sec:example}). The main text is closed
with conclusions. Appendices form an important part of the
publication. They are self-contained, and the information  provided
therein is sufficient to construct the complete subtraction scheme in
the most general case. Our notation is summarized in
\ref{sec:Notation}. Spherical angles in $d$ dimensions are discussed
in \ref{sec:spherical}. Divergences of virtual amplitudes are
considered in \ref{sec:VirtualIR}, whereas limits of amplitudes are
contained in \ref{sec:treelimits} and \ref{sec:1Llimits}.
\ref{sec:AltarelliParisi} contains the well-known Altarelli-Parisi
splitting kernels necessary for initial-state collinear
renormalization of partonic cross sections.

%%%%%%%%%%%%%%%%%%%%%%%%%%%%%%%%%%%%%%%%%%%%%%%%%%%%%%%%%%%%%%%%%%%%%%%%%%%%%%%%

\section{Outline of the subtraction scheme}
\label{sec:outline}

\noindent
We are interested in the calculation of next-to-next-to-leading order
QCD corrections to scattering cross sections. The subtraction scheme
we will describe can also be used for decay processes, since the phase
space integrals present there share the same structure. The formulae
below would only have to be trivially modified.  Furthermore, it is
possible to include QED corrections by a modification of the color
algebra (see \ref{sec:Notation}). We will not discuss this issue in
any more detail. We now need to specify the possible initial and final
states. At present, the most relevant case is hadron-hadron
scattering. Therefore, we will give expressions with this class of
processes in mind. Nevertheless, we will separate the hadron scale
physics from the parton scale physics. Indeed, we will only ever
manipulate partonic cross sections. This makes it possible to include
lepton-hadron and lepton-lepton scattering processes by simply
removing factorization contributions. As far as the final state is
concerned, we only require that there be at least two particles at the
Born level. The reason is that in the special case of a single
particle, the partonic cross section becomes a distribution, and we
assume that we can work with ordinary functions only. In the
following, we will concentrate on external partons carrying a color
charge. Any additional particles, which are not charged under the
color gauge group play no role whatsoever in the construction of the
scheme, and can be trivially included in the phase space integrals.

The hadronic cross section is known to factorize into a convolution
of parton distribution functions and the renormalized partonic cross
section
\be
\sigma_{h_1h_2}(P_1, P_2) = \sum_{ab} \iint_0^1 \mathrm{d}x_1
\mathrm{d}x_2 \, f_{a/h_1}(x_1, \mF) \,
f_{b/h_2}(x_2, \mF) \, \hat{\sigma}_{ab}(x_1P_1, x_2P_2;
\, \as(\mR), \, \mR, \, \mF) \; ,
\ee
where $P_{1,2}$ are the momenta of the hadrons $h_{1,2}$, and
$f_{a/h}(x,\mF)$ is the parton distribution function (PDF) of parton
$a$ within the hadron $h$, at the factorization scale $\mu_F$. In the
following, $p_{1,2} = x_{1,2} P_{1,2}$ will denote the parton
momenta. If we are able to numerically evaluate the partonic cross
section, $\hat{\sigma}_{ab}$, for an arbitrary initial-state energy,
then we can also obtain the hadronic cross section by an additional
integration over the parton momentum fractions. This is the reason,
why the construction of the subtraction scheme will never make
reference to the initial-state hadrons, but only to the initial-state
partons. The partonic cross section can be expanded in a series in
$\as(\mR)$, where $\mu_R$ is the renormalization scale. Up to
next-to-next-to-leading order (NNLO) it reads
\be
\hat{\sigma}_{ab} = \hat{\sigma}^{(0)}_{ab} + \hat{\sigma}^{(1)}_{ab}
+ \hat{\sigma}^{(2)}_{ab} \; .
\ee
Each term of the expansion can be further decomposed according to the
multiplicity of the final state. At leading order
\be
\hat{\sigma}^{(0)}_{ab} = \hat{\sigma}^{\mathrm{B}}_{ab} =
\f{1}{2\hat{s}} \f{1}{N_{ab}} \int \mathrm{d} \bm{\Phi}_n \, \la
\cm_n^{(0)} | \cm_n^{(0)} \ra \, \mathrm{F}_n
\; ,
\ee
where $\hat{s} = (p_1+p_2)^2$ is the square of the partonic
center-of-mass energy, while $N_{ab}$ is the spin and color average
factor, defined as the product of the number of spin and color degrees
of freedom of the partons $a$ and $b$. The subscript $n$ points to the
number of final states in this contribution. The notation and
normalization of phase spaces and matrix elements is specified in
\ref{sec:Notation}. $\mathrm{F}_n$ is the measurement function
defining the observable. It is a function of the final state momenta
and flavors. Up to the next-to-next-to-leading order, we need three
such functions, $F_n$, $F_{n+1}$ and $F_{n+2}$. They must fulfill the
requirements of infrared safety, i.e.\ if the energy of a final state
parton vanishes, or if the parton becomes collinear to another one,
then $F_{n+m} \to F_{n+m-1}$ for $m > 0$ and $F_n \to 0$.

At next-to-leading order there is
\be
\hat{\sigma}^{(1)}_{ab} = \hat{\sigma}^{\mathrm{R}}_{ab} + 
\hat{\sigma}^{\mathrm{V}}_{ab} + \hat{\sigma}^{\mathrm{C}}_{ab} \; ,
\ee
with
\be
\begin{gathered}
\hat{\sigma}^{\mathrm{R}}_{ab} =
\f{1}{2\hat{s}} \f{1}{N_{ab}} \int \mathrm{d} \bm{\Phi}_{n+1} \, \la
\cm_{n+1}^{(0)} | \cm_{n+1}^{(0)} \ra \, \mathrm{F}_{n+1} \; , \quad
\hat{\sigma}^{\mathrm{V}}_{ab} = \f{1}{2\hat{s}} \f{1}{N_{ab}} \int
\mathrm{d} \bm{\Phi}_n \, 2 \R \, \la \cm_n^{(0)} | \cm_n^{(1)} \ra \,
\mathrm{F}_n \; , \\[0.2cm] 
\hat{\sigma}^{\mathrm{C}}_{ab}(p_1,p_2) = \f{\as}{2\pi} \f{1}{\ep}
\left( \f{\mR}{\mF} \right)^{\ep} \sum_c \int_0^1 \mathrm{d}z
\left[P^{(0)}_{ca}(z) \, \hat{\sigma}^{\mathrm{B}}_{cb}(zp_1,p_2) +
  P^{(0)}_{cb}(z) \, \hat{\sigma}^{\mathrm{B}}_{ac}(p_1,zp_2) \right]
\; ,
\end{gathered}
\ee
where the splitting functions required are reproduced in
\ref{sec:AltarelliParisi}. Finally, at next-to-next-to-leading order,
we have
\be
\hat{\sigma}^{(2)}_{ab} = \hat{\sigma}^{\mathrm{RR}}_{ab} +
\hat{\sigma}^{\mathrm{RV}}_{ab} + \hat{\sigma}^{\mathrm{VV}}_{ab} +
\hat{\sigma}^{\mathrm{C1}}_{ab} + \hat{\sigma}^{\mathrm{C2}}_{ab} \; ,
\ee
with
\be
\begin{gathered}
\hat{\sigma}^{\mathrm{RR}}_{ab} =
\f{1}{2\hat{s}} \f{1}{N_{ab}} \int \mathrm{d} \bm{\Phi}_{n+2}
\, \la \cm_{n+2}^{(0)} | \cm_{n+2}^{(0)} \ra \, \mathrm{F}_{n+2} \; ,
\quad \hat{\sigma}^{\mathrm{RV}}_{ab} =
\f{1}{2\hat{s}} \f{1}{N_{ab}} \int \mathrm{d} \bm{\Phi}_{n+1} \, 2 \R
\, \la \cm_{n+1}^{(0)} | \cm_{n+1}^{(1)} \ra \, \mathrm{F}_{n+1} \; ,
\\[0.2cm] \hat{\sigma}^{\mathrm{VV}}_{ab} =
\f{1}{2\hat{s}} \f{1}{N_{ab}} \int \mathrm{d} \bm{\Phi}_n \, \Big( 2
\R \, \la \cm_n^{(0)} | \cm_n^{(2)} \ra + \la \cm_n^{(1)} |
\cm_n^{(1)} \ra \Big) \, \mathrm{F}_n \; ,
\end{gathered}
\ee
and
\be
\begin{split}
\label{eq:C1andC2}
\hat{\sigma}^{\mathrm{C1}}_{ab}(p_1,p_2) &= \f{\as}{2\pi}
\f{1}{\ep} \left( \f{\mR}{\mF} \right)^{\ep} \sum_c \int_0^1\mathrm{d}z
\left[ P^{(0)}_{ca}(z) \, \hat{\sigma}^{\mathrm{R}}_{cb}(zp_1,p_2) +
  P^{(0)}_{cb}(z)\, \hat{\sigma}^{\mathrm{R}}_{ac}(p_1,zp_2) \right] \; ,
\\[0.2cm]
\hat{\sigma}^{\mathrm{C2}}_{ab}(p_1,p_2) &=
\f{\as}{2\pi}\f{1}{\ep}
\left(\f{\mR}{\mF}\right)^{\ep} \sum_c \int_0^1\mathrm{d}z
\left[P^{(0)}_{ca}(z) \, \hat{\sigma}^{\mathrm{V}}_{cb}(zp_1,p_2) +
  P^{(0)}_{cb}(z)\, \hat{\sigma}^{\mathrm{V}}_{ac}(p_1,zp_2)
  \right] \\
&+\left(\f{\as}{2\pi}\right)^2\f{1}{2\ep}\left(\f{\mR}{\mF}\right)^{2\ep}
\sum_c \int_0^1\mathrm{d}z \left[P^{(1)}_{ca}(z)\,
  \hat{\sigma}^{\mathrm{B}}_{cb}(zp_1,p_2) + P^{(1)}_{cb}(z)\,
  \hat{\sigma}^{\mathrm{B}}_{ac}(p_1,zp_2)
  \right] \\
&+\left(\f{\as}{2\pi}\right)^2\f{\beta_0}{4\ep^2}
\left[\left(\f{\mR}{\mF}\right)^{2\ep}-2\left(\f{\mR}{\mF}\right)^{\ep}\right]
\sum_c \int_0^1\mathrm{d}z \left[P^{(0)}_{ca}(z)\,
  \hat{\sigma}^{\mathrm{B}}_{cb}(zp_1,p_2) + P^{(0)}_{cb}(z)
  \,\hat{\sigma}^{\mathrm{B}}_{ac}(p_1,zp_2)
  \right] \\
&+\left(\f{\as}{2\pi}\right)^2\f{1}{2\ep^2}\left(\f{\mR}{\mF}\right)^{2\ep}
\sum_{cd}\int_0^1\mathrm{d}z \left[\Big(P^{(0)}_{cd}\otimes P^{(0)}_{da}\Big)(z)\,
  \hat{\sigma}^{\mathrm{B}}_{cb}(zp_1,p_2)  + \Big(P^{(0)}_{cd}\otimes
  P^{(0)}_{db}\Big)(z)\, \hat{\sigma}^{\mathrm{B}}_{ac}(p_1,zp_2)
  \right] \\
&+\left(\f{\as}{2\pi}\right)^2\f{1}{\ep^2}\left(\f{\mR}{\mF}\right)^{2\ep}
\sum_{cd}\iint_0^1\mathrm{d}z\,\mathrm{d}\bar{z}
\left[P^{(0)}_{ca}(z)\,P^{(0)}_{db}({\bar z})\,
  \hat{\sigma}^{\mathrm{B}}_{cd}(zp_1,\bar{z}p_2) 
    \right] \; ,
\end{split}
\ee
where
\be
\left(f \otimes g\right)(x)=
\iint_0^1\mathrm{d}y\,\mathrm{d}z\, f(y)g(z)\,\delta(x-yz) \; .
\ee

In general, it is only possible to evaluate the cross section
contributions listed above using numerical Monte Carlo integration
methods. Unfortunately, phase spaces with $n+1$ and $n+2$ final states
contain infrared singular configurations, which are regulated with
dimensional regularization. The latter, if applied naively, spoils the
stability of the numerics, since a limit has to be taken once all
contributions are combined. This problem is resolved with a
subtraction scheme, which extracts the explicit singularities and
provides integrable functions, which do not depend on the
parameter of dimensional regularization.

The idea behind the construction of the subtraction scheme STRIPPER
is to derive Laurent expansions in $\epsilon$ for each of the cross
section contributions independently. The basic algorithm has three
stages:
\begin{description}

\item[phase space decomposition (Section~\ref{sec:decomposition}):]
  phase spaces with $n+1$ and $n+2$ final-state particles are
  decomposed into sectors, in which only certain types of
  singularities may occur;

\item[phase space parameterization
  (Section~\ref{sec:parameterization}):] in each sector, a special
  parameterization is introduced using spherical coordinates in $d$
  dimensions (\ref{sec:spherical}), in which singularities are only
  parameterized with 2 variables for $n+1$-final-state-particles phase
  spaces, and with 4 variables for $n+2$-final-state-particles phase
  spaces;

\item[generation of subtraction and integrated subtraction terms
  (Section~\ref{sec:subtraction}):] in each parameterization,
  subtraction terms in the relevant variables are introduced, which
  make it possible to obtain an expansion in $\epsilon$, the
  coefficients of which are integrable. The subtraction terms only
  require the knowledge of the singular limits of QCD amplitudes
  (\ref{sec:treelimits} and \ref{sec:1Llimits}), and are process
  independent in the sense that the process dependence is confined to
  the matrix elements. Furthermore, pointwise convergence of phase
  space integration is guaranteed.

\end{description}
After application of this algorithm, cross sections may in principle
be evaluated numerically. Nevertheless, since dimensional
regularization involves infinite dimensional vectors, the effective
dimension of the vectors, which actually occur in the calculation
increases with multiplicity. In fact, any new vector requires an
increase of the effective dimension by one. For two-to-two processes
at leading order, one already needs five dimensions at
next-to-next-to-leading order. Furthermore, matrix elements must
be provided as expansions in $\ep$. In order to simplify the
calculation, we introduce the 't Hooft-Veltman version of dimensional
regularization, in which resolved particle momenta and spin
degrees-of-freedom are four-dimensional. In this case, we also only
need four-dimensional matrix elements. The construction proceeds in
additional three stages:
\begin{description}

\item[average over azimuthal angles
  (Section~\ref{sec:AzimuthalAverage}):] integrated subtraction terms,
  which have been derived in relation to a collinear limit, are
  averaged over the unphysical transverse direction. For most cases,
  this is equivalent to the use of averaged splitting functions, but
  there are important exceptions. This step is important in order not
  to have contractions of four-dimensional matrix elements with
  $d$-dimensional transverse vectors;

\item[separation of finite contributions
  (Section~\ref{sec:FiniteContributions}):] the different contributions
  listed in this section are further decomposed into classes with
  different kinematics and loop order. The sum of the terms in each
  class is finite. This requires a modification of the integrated
  subtraction terms for the double-real radiation. In practice,
  counter\-terms are introduced, which are added to one class of
  contributions and subtracted from another;

\item['t Hooft-Veltman regularization of separately finite
  contributions (Section~\ref{sec:HVregularization}):] the measurement
  function is modified to contain delta-functions restricting the
  momenta of resolved particles to be four-dimensional. For most
  classes of finite contributions, this is already sufficient to
  fulfill the requirements of 't Hooft-Veltman regularization, and the
  matrix elements can be evaluated in four dimensions. Nevertheless,
  one class, the single-unresolved contributions to double-real
  radiation, requires a further modification of the integrated
  subtraction terms.

\end{description}
After these steps, the subtraction scheme does not require higher
orders of the $\ep$-expansion of the matrix elements, and all resolved
momenta are four-dimensional. The calculation still involves
unresolved momenta, which may need up to two additional dimensions,
but only occur in soft and splitting functions. In general,
two-to-two processes at leading order require five-dimensional
unresolved momenta at next-to-next-to-leading order. For higher
multiplicity, six-dimensional momenta must be introduced.

%%%%%%%%%%%%%%%%%%%%%%%%%%%%%%%%%%%%%%%%%%%%%%%%%%%%%%%%%%%%%%%%%%%%%%%%%%%%%%%%

\section{Phase space decomposition}
\label{sec:decomposition}

\noindent
The first step in the construction of the subtraction scheme is the
decomposition of the phase space into sectors. In each of the sectors,
the momenta of selected partons are parameterized with a few relevant
variables (see Section~\ref{sec:parameterization}), which allows
for easy extraction of explicit poles in the dimensional
regularization parameter $\ep$, and generation of numerically
integrable coefficients of the Laurent series (see
Section~\ref{sec:subtraction}). The decomposition is performed with
the help of {\it selector functions}. The only essential property
required of the selector functions, besides that they form a
decomposition of unity, is that they vanish in case of soft and/or
collinear limits not covered by the parameterization of a given
sector.

Cross sections are defined with a measurement function, which
specifies an infrared safe observable. This measurement function
reduces the number of possible singular configurations. For instance,
at next-to-leading order, the measurement function $\mathrm{F}_{n+1}$
only allows a parton to become soft (vanishing energy), two massless
partons to become collinear to each other (vanishing relative angle),
or a parton to become both soft and collinear to another massless
parton. This is achieved by the vanishing of $\mathrm{F}_{n+1}$ for
any other singular configuration. $\mathrm{F}_{n+1}$ also enters the
definition of cross sections at next-to-next-to-leading order, if they
involve $n+1$ partons in the final state. The selector functions for
$n+1$-parton phase space integrals are used to regulate the
singularities allowed by $\mathrm{F}_{n+1}$. Since at most a pair of
partons may be involved, the selector functions for this case,
$\mathcal{S}_{i,k}$, have two indices. The first index, $i$, specifies
the {\it unresolved} parton, which is massless and in the final
state, while the second index, $k \neq i$, specifies
the {\it reference} parton, which is massless as well, but may be
either in the initial or in the final state. $\mathcal{S}_{i,k}$ does
not vanish if the unresolved and reference partons become collinear to
each other, or if the unresolved parton becomes soft. However, it
vanishes if any gluon besides $i$ becomes soft, or if any two partons
besides $i$ and $k$ become collinear and this kinematical
configuration is singular. In consequence, if the reference parton is
in the initial state, we only have to consider the following flavor
pairs, $(f_i,f_k)$, of the flavor $f_i$ of parton $i$ and flavor $f_k$
of parton $k$
\be
(g,g), \, (g,q), \, (g,\bar{q}), \, (q,g), \, (\bar{q},g), \, (q,q),
\, (\bar{q},\bar{q}) \; .
\ee
If the reference parton, on the other hand, is in the final state,
then the number of possible pairs is reduced, because we may use the
symmetry between $i$ and $k$. Nevertheless, if the
unresolved-reference-parton pair contains a gluon, the latter has to
be the unresolved parton. The list is thus
\be
(g,g), \, (g,q), \, (g,\bar{q}), \, (q,\bar{q}) \; .
\ee
While the choice of $\mathcal{S}_{i,k}$ is by no means unique, we have
to provide a valid set. We choose a form very similar to the one
described in \cite{Frixione:2007vw}. To this end, we introduce
\be
\begin{aligned}
d_{i,k} &= \bigg( \f{E_i}{\sqrt{\hat{s}}} \bigg)^\alpha
(1-\cos\theta_{ik})^\beta \; ,
\end{aligned}
\ee
where $\alpha = \gamma$ for a gluon and $\alpha = 0$ for a quark,
while $\beta,\gamma > 0$. $E_i$ is the energy of parton $i$, and
$\theta_{ik}$ is the relative angle between partons $i$ and $k$. With
the help of $d_{i,k}$, we define
\be
\mathcal{S}_{i,k} = \f{1}{D_1 \, d_{i,k}} \; ,
\quad D_1 = \sum_{ik} \f{1}{d_{i,k}} \; .
\ee
Clearly, $\mathcal{S}_{i,k}$ form a decomposition of unity
\be
\sum_{ik} \mathcal{S}_{i,k} = 1 \; .
\ee
We note that the contributions of two sectors, which have the same
flavors of the unresolved and reference partons, are equal. This
implies that there is no need to calculate the contributions of such
sectors independently. In consequence, even for high-multiplicity
processes, the number of independent sectors is moderate. For example,
the most computationally intensive pure gluon amplitudes in multi-jet
production will only require three sectors independently of the
multiplicity.

The next-to-next-to-leading order measurement function occurring in
double-real radiation, $\mathrm{F}_{n+2}$, allows for
more singular cases than $\mathrm{F}_{n+1}$. In particular, it does
not vanish if three partons, or up to two pairs of partons, become
collinear. Similarly, it does not vanish if up to two partons become
soft. In other words, it allows for two unresolved partons. In order
to accommodate the possible {\it triple-collinear} (three partons
becoming collinear) and {\it double-collinear} (two pairs of partons
becoming collinear) limits, we must introduce two types of selector
functions: $\mathcal{S}_{ij,k}$ for unresolved partons $i$ and $j$
with reference parton $k$; and $\mathcal{S}_{i,k;j,l}$ for unresolved
parton $i$ with reference parton $k$, and unresolved parton $j$ with
reference parton $l$. Of course, all indices must be different in both
cases, and the ordering of $i$ and $j$ in $\mathcal{S}_{ij,k}$, as
well as the ordering of the two pairs, $(i,k)$ and $(j,l)$ in
$\mathcal{S}_{i,k;j,l}$ is irrelevant. In the triple-collinear sector,
the flavor sets allowing for singular kinematical configurations
are
\be
\label{eq:flav}
\{g,g,g\}, \, \{g,g,q\}, \, \{g,g,\bar{q}\} \, \{g,q,\bar{q}\},
\{q,\bar{q},q'\} \; ,
\ee
where $q'$ can be any quark or anti-quark including $q$ and $\bar{q}$,
and we have used crossing to define the flavors of all partons as if
they were out-going. If $k$ is in the initial state, the possible
flavor triples, $(f_i,f_j,f_k)$, we have to consider, are obtained by
selecting any of the different flavors in each set of the list
\eqref{eq:flav} and crossing it to the initial state. This procedure
results in the following list
\be
(g,g,g), \, (g,g,q), \, (g,g,\bar{q}), \, (g,q,g), \, (g,\bar{q},g), \,
(g,q,q), \, (g,\bar{q},\bar{q}), \, (q,\bar{q},g), \, (q,\bar{q},q'),
\, (q',q,q), \, (q',\bar{q},\bar{q}) \; .
\ee
If $k$ is in the final state, on the other hand, we have additional
freedom to reorder the partons within the triple. Minding the
necessity to regulate final state soft gluons and double-soft
quark-anti-quark pairs, we obtain the following list
\be
(g,g,g), \, (g,g,q), \, (g,g,\bar{q}), \, (g,q,\bar{q}), \,
(q,\bar{q},g), \, (q,\bar{q},q') \; .
\ee
The flavor assignments we have to consider in the
double-collinear sector are simply a composition of two
next-to-leading order cases for the two involved pairs. Let us now
introduce
\be
\begin{gathered}
d_{ij,k} = \bigg( \f{E_i}{\sqrt{\hat{s}}} \bigg)^{\alpha_i}
\bigg( \f{E_j}{\sqrt{\hat{s}}} \bigg)^{\alpha_j}
\left[ (1-\cos\theta_{ij})(1-\cos\theta_{ik})(1-\cos\theta_{jk})
  \right]^\beta \; ,
\end{gathered}
\ee
where $\alpha_{i,j} = \gamma$ for gluons, $\alpha_i = \alpha_j
= \gamma$ for a quark-anti-quark pair, and $\alpha_{i,j} = 0$
otherwise. We define
\be
\mathcal{S}_{ij,k} = \f{1}{D_2 \, d_{ij,k}} \; , \quad
\mathcal{S}_{i,k;j,l} = \f{1}{D_2 \, d_{i,k} d_{j,l}} \; , \quad
D_2 = \sum_{ij} \Big[ \sum_k \f{1}{d_{ij,k}} + \sum_{kl}
\f{1}{d_{i,k} d_{j,l}} \Big] \; .
\ee
Note that in the case of $\mathcal{S}_{i,k;j,l}$, if $(i,j)$ is a
quark-anti-quark pair, we set $\alpha = \gamma$ in both $d_{i,k}$ and
$d_{j,l}$, contrary to the next-to-leading order case. The
decomposition of unity is
\be
\sum_{ij} \Big[ \sum_k \mathcal{S}_{ij,k} + \sum_{kl}
  \mathcal{S}_{i,k;j,l} \Big] = 1 \; .
\ee
The number of sectors to actually evaluate is again reduced by the
fact that sectors with identical flavors give equal
contributions. Returning to the example of pure gluon amplitudes, we
would always only have to calculate double-real radiation
contributions of seven sectors: three triple-collinear and four
double-collinear.

We must finally point out that the sectors we have introduced are
sufficient for any problem with at least three massless partons. In
the case of only two massless partons (say heavy-quark pair-production
from colorless initial states) in the final state, it is necessary to
introduce fictitious reference momenta.

%%%%%%%%%%%%%%%%%%%%%%%%%%%%%%%%%%%%%%%%%%%%%%%%%%%%%%%%%%%%%%%%%%%%%%%%%%%%%%%%

\section{Phase space parameterization}
\label{sec:parameterization}

\noindent
After having decomposed the phase space into sectors using selector
functions of Section~\ref{sec:decomposition}, we now need to introduce
appropriate parameterizations for the single-collinear sector (one
unresolved momentum), and the triple- and double-collinear sectors
(two unresolved momenta). The phase space integrals always assume the
form
\be
\int \mathrm{d}\bm{\Phi}_{n+n_u} = \int
\mathrm{d}\bm{\Phi}_{\substack{\text{reference} \\ \text{unresolved}}}
\int \mathrm{d}\bm{\Phi}_{n-n_{fr}}(Q)
\; , \quad n \geq 2 \; , \quad n_u \in \{1,2\} \; , \quad 0 \leq
n_{fr} \leq n_u \; ,
\ee
where $n$ is the number of final state momenta in the Born
approximation, while $n_u$ is the number of unresolved momenta, and
$n_{fr}$ the number of final state reference momenta. $Q$ is the total
momentum of the remaining final-state particles. In particular, the
phase space $\int \mathrm{d}\bm{\Phi}_{n-n_{fr}}(Q)$ describes the
decay process of a state with momentum $Q$ into $n-n_{fr}$ particles
with momenta $q_i$
\be
Q \to q_1 + \dots + q_{n-n_{fr}} \; .
\ee
For future reference, we define the notation for the minimal invariant
mass of the final state
\be
Q_{\mathrm{min}} = \sum_{i=1}^n m_i \; ,
\ee
where $m_i$ is mass of the final state particle $i$ in the Born
approximation. Furthermore, we introduce the related maximal energy of
a massless final state parton
\be
E_{\mathrm{max}} = \f{\sqrt{\hat{s}}}{2} \left( 1 -
\f{Q_{\mathrm{min}}^2}{\hat{s}} \right) \; ,
\ee
where $\hat{s}$ is the square of the partonic center-of-mass energy.

The parameterizations are constructed with several rules in mind. In
particular, the phase space is defined in the partonic center-of-mass
frame in all sectors but the single-collinear sector for use in the
calculation of the factorization contribution
$\hat{\sigma}^{\mathrm{C1}}$. The initial state momenta, $p_1$ and
$p_2$ are always in the center-of-mass system. The unresolved momenta
are parameterized by rescaled energy and angle variables, to allow for
a straightforward identification of the singular limits of
amplitudes. The unresolved phase space integrals contain
\be
\iint_0^1 \mathrm{d}\eta \, \mathrm{d}\xi \, \eta^{a_1 - b_1 \ep} \,
\xi^{a_2 - b_2\ep} \; ,
\ee
for a single unresolved momentum, and
\be
\iiiint_0^1 \mathrm{d}\eta_1 \mathrm{d}\eta_2 \mathrm{d}\xi_1
\mathrm{d}\xi_2  \, \eta_1^{a_1 - b_1 \ep} \, \eta_2^{a_2 - b_2 \ep} \,
\xi_1^{a_3 - b_3\ep} \, \xi_2^{a_4 - b_4\ep} \; ,
\ee
for two unresolved momenta. In general $\eta$ are related to angles, and
$\xi$ to energies. The factors $x^{a-b\ep}$, $x \in \{\eta,\xi\}$ or $x
\in \{\eta_1,\eta_2,\xi_1,\xi_2\}$, regulate singularities at $x =
0$. This requires $a_i \geq 0$ and $b_i > 0$. The latter restriction
follows from the fact that infrared divergences are regulated by $\ep
< 0$. If the product of a matrix element and the measurement function
is integrated with the given phase space parameterization together
with the appropriate selector function, there are no other
singularities than those located at $x = 0$, and the singular
behavior in the limit is described by the factor $x^{-1-b\ep}$,
i.e.\ the coefficient of this factor is finite at the limit for any
value of $\ep$. The method of proof of this fact has been discussed in
\cite{Czakon:2010td}, and the reader is referred to that publication
for details.

The parameterizations make ample use of the rotation invariance of the
single-particle measure. They are valid in CDR with $d$-dimensional
momenta of both reference and unresolved momenta. Nevertheless, if
reference momenta and additional final state particle momenta are
restricted to four dimensions, then one unresolved momentum is at most
five dimensional, while a second at most six dimensional. This
statement should be interpreted in the sense that the additional
dimensions can be trivially integrated out, since nothing depends on
them.

A final comment is necessary on the order of integration. The phase
spaces are written in such a way that every next integral inherits the
parameters of the previous one. In other words, the integration range
of a given integral is restricted depending on the particular values of
the parameters of all the previous integrals. In order to achieve a
pointwise cancellation of singularities during Monte Carlo
integration, as discussed in the next section, it is necessary to
parameterize $\int \mathrm{d}\bm{\Phi}_{n-n_{fr}}(Q)$ in a
specific way. This phase space depends on $3(n-n_{fr})-4$ parameters
in four dimensions. These parameters must be rescaled to a fixed
finite (usually unit) range, as natural for Monte Carlo
integration. The rescaling itself depends on $Q$. During the
evaluation, the rescaled parameters are kept fixed for both the
unsubtracted matrix element with phase space weight, and for all the
subtraction terms with their phase space weights. The same comment
also applies to the reference parton energy.

Let us now specify the parameterizations using the notation of
\ref{sec:spherical} for vectors and integrals in spherical coordinates
in $d$-dimensions.

%%%%%%%%%%%%%%%%%%%%%%%%%%%%%%%%%%%%%%%%%%%%%%%%%%%%%%%%%%%%%%%%%%%%%%%%%%%%%%%%

\subsection{Single-collinear sector parameterization: one reference
  momentum, one unresolved momentum}
\label{sec:single-collinear}

\noindent
The resolved and unresolved parton momenta read
\be
r^\mu = r^0 \, \hat{r}^\mu = r^0 \begin{pmatrix} 1
  \\ \bm{\hat{r}} \end{pmatrix} \; , \quad 
u^\mu = u^0 \, \hat{u}^\mu = u^0 \begin{pmatrix} 1
  \\ \bm{\hat{u}} \end{pmatrix} \; ,
\ee
with the angular parameterization
\begin{align}
\bm{\hat{r}} &= \bm{\hat{n}}^{(3-2\ep)}(\alpha_1, \alpha_2, \dots)
\; ,
\nn \\[0.2cm]
\bm{\hat{u}} &= \bm{R}^{(3-2\ep)}_1(\alpha_1, \alpha_2, \dots)
\bm{\hat{n}}^{(3-2\ep)}(\theta, \phi, \rho_1, \rho_2, \dots) \; .
\end{align}
For reasons explained in Section~\ref{sec:caseIV}, we will allow for a
boosted frame, where the initial state momentum $p_1$ is rescaled with
$z$. Of course, the symmetric case, where the rescaling is applied to
$p_2$ can be treated by relabeling. In most applications, we will set
$z = 1$. In the case of an initial state reference momentum, the phase
space is
\be
\label{eq:unresolved0}
\int \mathrm{d}\bm{\Phi}_{n+1} =
\int \mathrm{d}\bm{\Phi}_{\text{unresolved}}
\int \mathrm{d}\bm{\Phi}_{n}( z p_1+p_2-u ) \; ,
\ee
while for a final state reference momentum
\be
\int \mathrm{d}\bm{\Phi}_{n+1} = \Nep{} \int_{\mathcal{S}^{2-2\ep}_1}
\mathrm{d}\bm{\Omega}(\alpha_1, \alpha_2, \dots)
\int \mathrm{d}\bm{\Phi}_{\text{unresolved}}
\int_0^{r_{\mathrm{max}}^0} \f{\mathrm{d}r^0 \,
  (r^0)^{1-2\ep}}{2(2\pi)^{3-2\ep}} \int
\mathrm{d}\bm{\Phi}_{n-1}( zp_1+p_2-r-u ) \; ,
\ee
where
\be
r_\mathrm{max}^0 = \f{2\sqrt{\hat{s}} \, \big(
  E_{\mathrm{max}} - u^0 \big) - \big( \hat{s} - 2 p_1 \cdot u )
  (1-z)}{2 \, \big[ \sqrt{\hat{s}} - \hat{r} \cdot \big( u + p_1 (1-z)
    \big) \big]} \; ,
\ee
and in the particular case $n = 2$
\be
\int_0^{r_\mathrm{max}^0} \f{\mathrm{d}r^0 \,
  (r^0)^{1-2\ep}}{2(2\pi)^{3-2\ep}} \int \mathrm{d}\bm{\Phi}_1(zp_1 +
p_2 - r -u) = \f{(r_\mathrm{max}^0)^{1-2\ep}}{4(2\pi)^{2-2\ep}}
\f{1}{\sqrt{\hat{s}} - \hat{r} \cdot \big( u + p_1 (1-z) \big)} \; .
\ee
The unresolved phase space reads
\begin{multline}
\label{eq:unresolved1}
\int \mathrm{d}\bm{\Phi}_{\text{unresolved}} = \Nep{}
\int_{\mathcal{S}^{2-2\ep}_1} \mathrm{d}\bm{\Omega}(\theta, \phi,
\rho_1, \dots) \int_0^{u^0_{\mathrm{max}}} \f{\mathrm{d}u^0 \,
  (u^0)^{1-2\ep}}{2(2\pi)^{3-2\ep}} = \\
\f{E_{\mathrm{max}}^2}{(2\pi)^3}
\bigg( \f{\pi\mReg}{4E_{\mathrm{max}}^2} \bigg)^\ep
\int_{\mathcal{S}^{1-2\ep}_1} \mathrm{d}\bm{\Omega}(\phi, \rho_1,
\dots) \iint_0^1 \mathrm{d}\eta \, \mathrm{d}\xi \, \eta^{-\ep} \,
\xi^{1-2\ep} \big(1-\eta\big)^{-\ep} \xi_{\mathrm{max}}^{2-2\ep} \; ,
\end{multline}
where
\be
\label{eq:single-collinear-u0}
u^0 = E_{\mathrm{max}} \, \xi \, \xi_{\mathrm{max}} \; , \quad
\cos\theta = 1-2\eta \; , \quad \xi_{\mathrm{max}} = \f{1 -
  \f{\sqrt{\hat{s}}}{2E_{\mathrm{max}}}(1-z)}{1 - \f{1}{\sqrt{\hat{s}}}
      ( p_1 \cdot \hat{u} ) (1-z) } \; .
\ee
The behavior of amplitudes in the collinear limit is characterized by
the transverse vector
\be
\label{eq:uperp}
u_\perp^\mu = \begin{pmatrix} 0 \\ \bm{\hat{u}_\perp} \end{pmatrix} \;
, \quad \bm{\hat{u}_{\perp}} = \lim_{\theta \to 0}
\f{\bm{\hat{u}}-\bm{\hat{r}}}
{\lVert \bm{\hat{u}}-\bm{\hat{r}} \rVert} = \left. 
\f{\partial \bm{\hat{u}}}{\partial \theta} \right|_{\theta = 0}
= \bm{\hat{R}}^{(3-2\ep)}_1(\alpha_1, \alpha_2, \dots)
\bm{\hat{n}}^{(3-2\ep)}\left(\f{\pi}{2}, \phi, \rho_1, \rho_2,
\dots\right)
\; .
\ee
%

%%%%%%%%%%%%%%%%%%%%%%%%%%%%%%%%%%%%%%%%%%%%%%%%%%%%%%%%%%%%%%%%%%%%%%%%%%%%%%%%

\subsection{Triple-collinear sector parameterization: one reference
  momentum, two unresolved momenta}
\label{sec:triple-collinear}

\begin{figure}[t]
  \begin{center}
    \includegraphics[scale=.75]{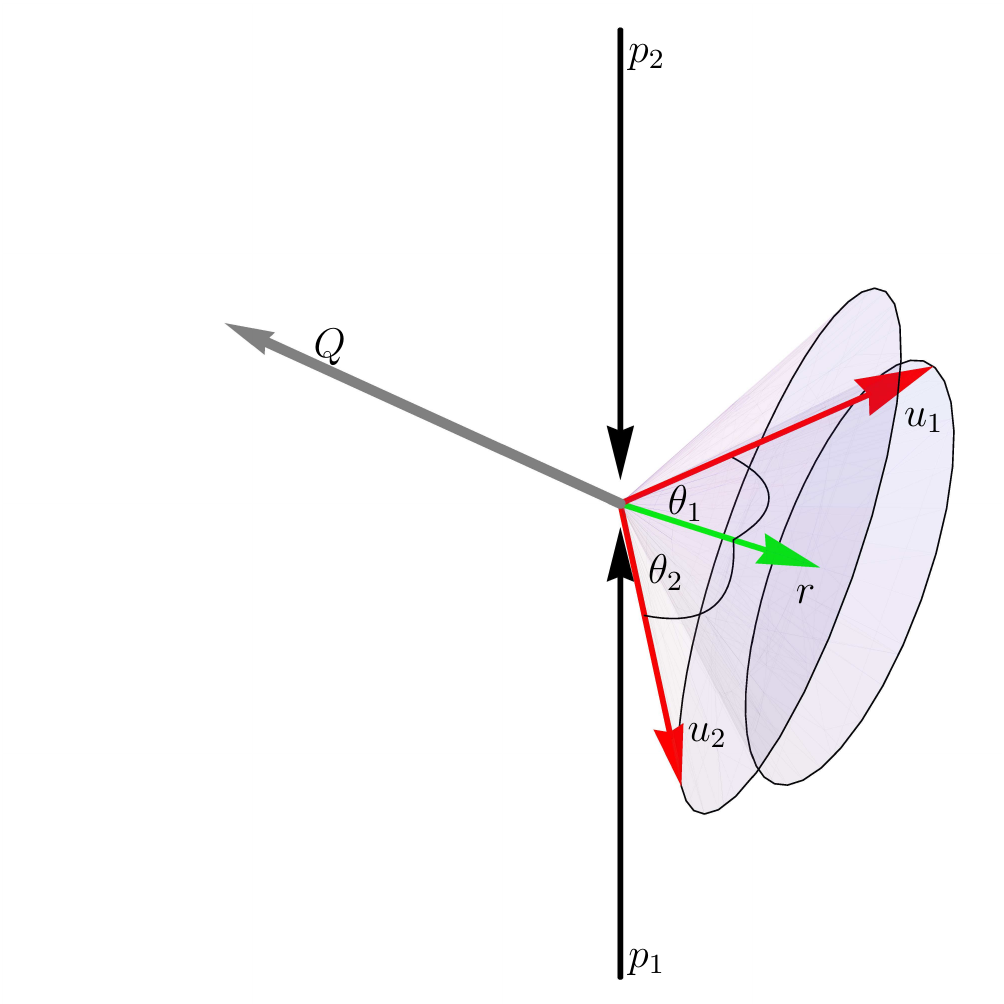}
    \includegraphics[scale=.75]{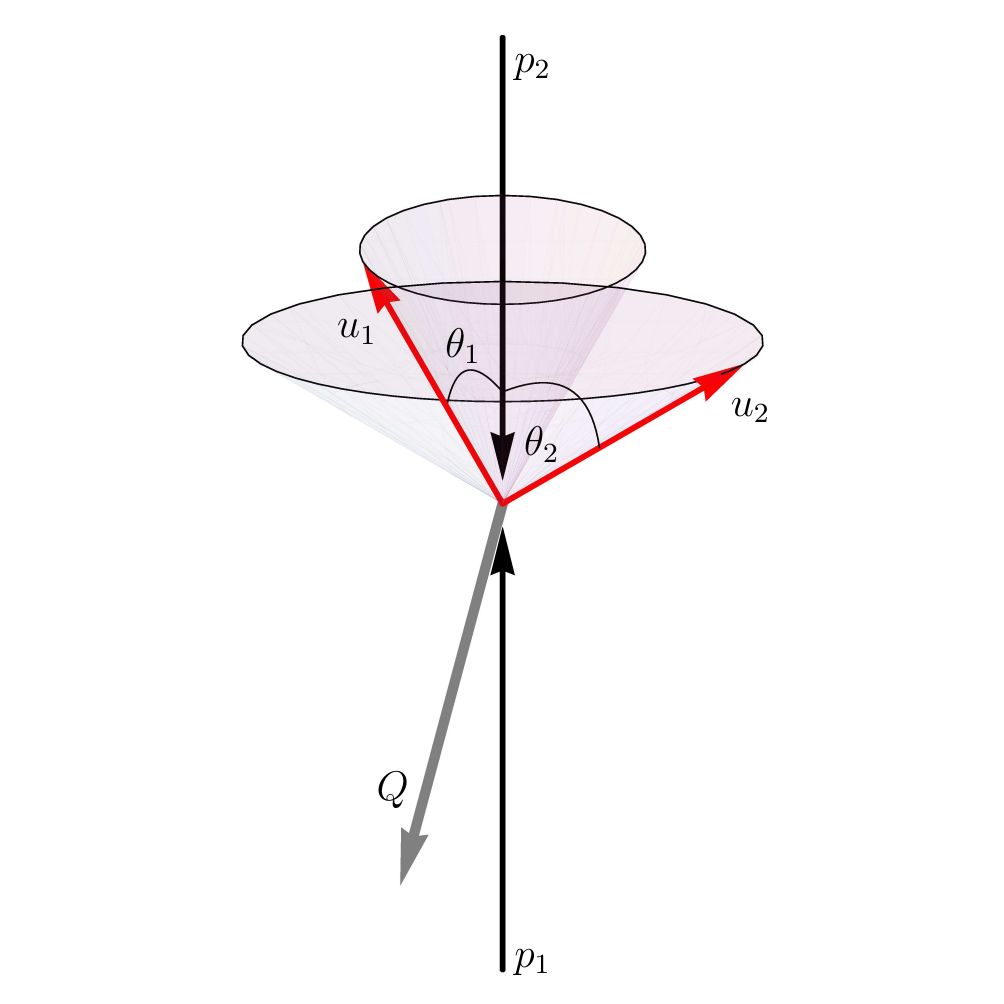}
  \end{center}
  \caption{\label{fig:triple} Example momentum parameterizations in
    the case of two unresolved momenta, $u_1$ and $u_2$, and one
    reference momentum, $r$, which may be either in the final state
    (left), or in the initial state (right, with $r = p_1$). The
    angles $\theta_1$ and $\theta_2$ allow for a straightforward
    parameterization of the collinear limits with respect to the
    reference momentum. $Q$ is the total momentum of the remaining
    final-state particles, while $p_1$ and $p_2$ are the initial-state
    momenta.}
\end{figure}

\noindent
The resolved and unresolved parton momenta read
\be
r^\mu = r^0 \, \hat{r}^\mu = r^0 \begin{pmatrix} 1
  \\ \bm{\hat{r}} \end{pmatrix} \; , \quad 
u_1^\mu = u_1^0 \, \hat{u}_1^\mu = u_1^0 \begin{pmatrix} 1
  \\ \bm{\hat{u}_1} \end{pmatrix} \; , \quad
u_2^\mu = u_2^0 \, \hat{u}_2^\mu = u_2^0 \begin{pmatrix} 1
  \\ \bm{\hat{u}_2} \end{pmatrix} \; ,
\ee
with the angular parameterization (see Fig.~\ref{fig:triple})
\begin{align}
\label{eq:tripleparameterization}
\bm{\hat{r}} &= \bm{\hat{n}}^{(3-2\ep)}(\alpha_1, \alpha_2, \dots)
\; ,
\nn \\[0.2cm]
\bm{\hat{u}_1} &= \bm{R}^{(3-2\ep)}_1(\alpha_1, \alpha_2, \dots)
\bm{\hat{n}}^{(3-2\ep)}(\theta_1, \phi_1, \rho_1, \rho_2, \dots) \; ,
\nn \\[0.2cm]
\bm{\hat{u}_2} &= \bm{R}^{(3-2\ep)}_1(\alpha_1, \alpha_2, \dots)
\bm{R}^{(3-2\ep)}_2(\phi_1, \rho_1, \rho_2, \dots)
\bm{\hat{n}}^{(3-2\ep)}(\theta_2, \phi_2, \sigma_1, \sigma_2, \dots)
\; .
\end{align}
In the case of an initial state reference momentum, the phase space is
\be
\int \mathrm{d}\bm{\Phi}_{n+2} =
\int \mathrm{d}\bm{\Phi}_{\text{unresolved}}
\int \mathrm{d}\bm{\Phi}_{n}( p_1+p_2-u_1-u_2 ) \; ,
\ee
while for a final state reference momentum
\be
\int \mathrm{d}\bm{\Phi}_{n+2} = \Nep{} \int_{\mathcal{S}^{2-2\ep}_1}
\mathrm{d}\bm{\Omega}(\alpha_1, \alpha_2, \dots)
\int \mathrm{d}\bm{\Phi}_{\text{unresolved}}
\int_0^{r_{\mathrm{max}}^0} \f{\mathrm{d}r^0 \,
  (r^0)^{1-2\ep}}{2(2\pi)^{3-2\ep}} \int
\mathrm{d}\bm{\Phi}_{n-1}( p_1+p_2-r-u_1-u_2 ) \; ,
\ee
where
\be
\label{eq:r0max2}
r_\mathrm{max}^0 = \f{\sqrt{\hat{s}} \left( E_{\mathrm{max}} - u_1^0 -
  u_2^0 \right) + u_1 \cdot u_2}{\sqrt{\hat{s}} - \hat{r} \cdot \left(
  u_1 + u_2 \right)} \; ,
\ee
and in the particular case $n = 2$
\be
\label{eq:n=2case}
\int_0^{r_{\mathrm{max}}^0} \f{\mathrm{d}r^0 \,
  (r^0)^{1-2\ep}}{2(2\pi)^{3-2\ep}} \int \mathrm{d}\bm{\Phi}_1(p_1 +
p_2 - r - u_1 - u_2) =
\f{(r_\mathrm{max}^0)^{1-2\ep}}{4(2\pi)^{2-2\ep}} \f{1}{\sqrt{\hat{s}}
  - \hat{r} \cdot \left( u_1 + u_2 \right)} \; .
\ee
\begin{figure}[t]
  \begin{center}
    \includegraphics[scale=.6]{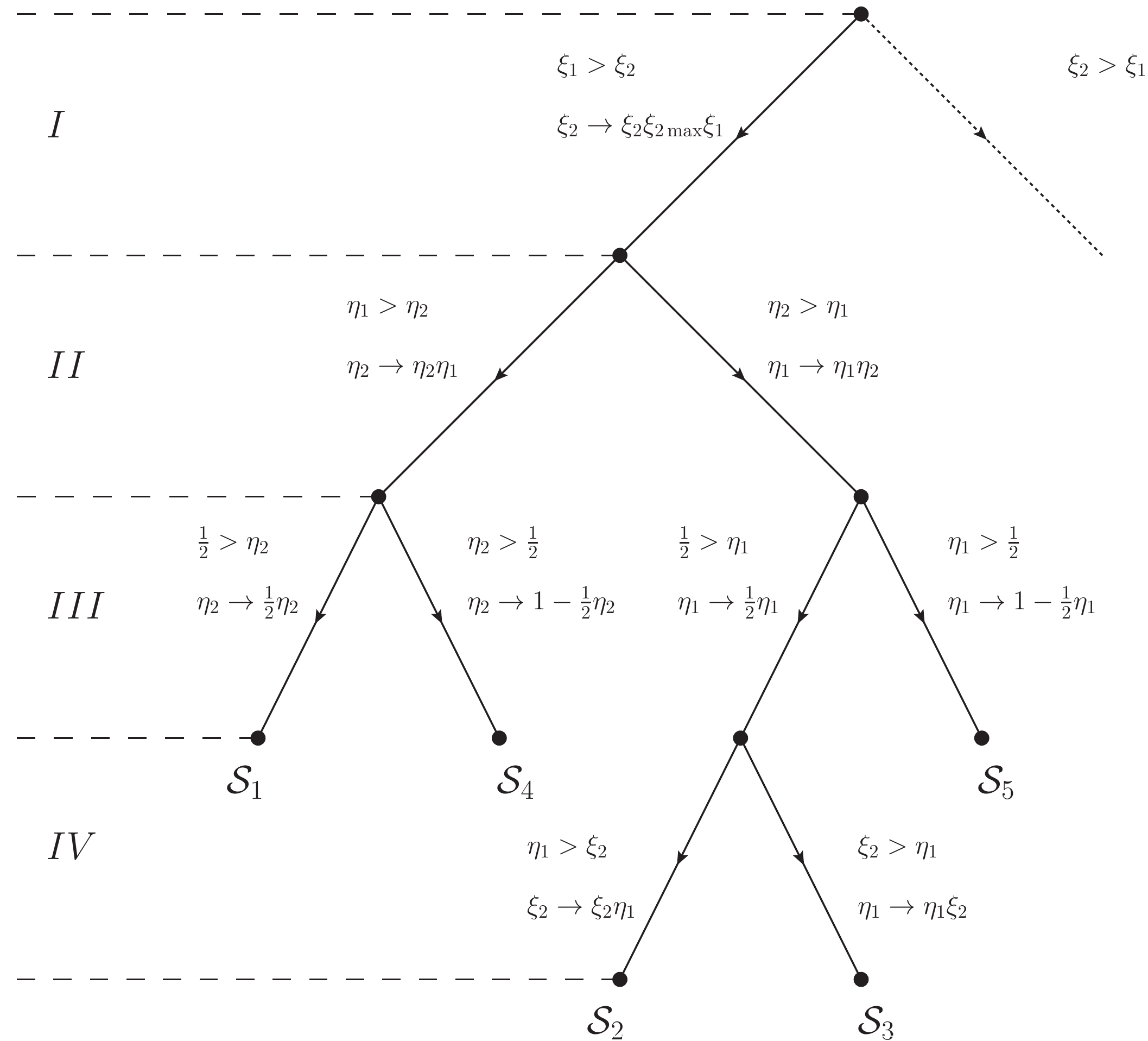}
  \end{center}
  \caption{\label{fig:decomposition} Decomposition tree of the
    triple-collinear sector unresolved phase space. Starting at the
    root with $\eta_i = \hat{\eta}_i$, $\xi_i = \hat{\xi}_i$,
    substitutions are performed at four levels corresponding to the
    factorization of the soft (I), collinear (II and III) and
    soft-collinear (IV) limits. The omitted right branch of the tree
    corresponds to a different ordering of the energies of the
    unresolved partons, and can be obtained by renaming the indices of
    the variables, $1 \leftrightarrow 2$. The function
    $\xi_{2\,\mathrm{max}}$ is defined in Eq.~(\ref{eq:ximax}).}
\end{figure}
The parameters $u_{1,2}^0,\theta_{1,2}$, and $\phi_2$ of the
unresolved momenta are replaced by $\hat{\xi}_{1,2},
\hat{\eta}_{1,2}$, and $\zeta$
\be
\label{eq:parameters3}
\begin{gathered}
u_1^0 = E_{\mathrm{max}} \, \hat{\xi}_1 \; , \quad u_2^0 =
E_{\mathrm{max}} \, \hat{\xi}_2 \; , \\[0.2cm]
\cos\theta_1 = 1-2\hat{\eta}_1 \; , \quad 
\cos\theta_2 = 1-2\hat{\eta}_2 \; , \quad 
\cos\phi_2 =  \f{1-2\eta_3-(1-2 \hat{\eta}_1) (1-2 \hat{\eta}_2)}{4
  \sqrt{(1-\hat{\eta}_1) \hat{\eta}_1 (1-\hat{\eta}_2) \hat{\eta}_2}}
\; , \\[0.2cm]
\eta_3 = \f{\hat{u}_1 \cdot \hat{u}_2}{2} = \f{1 - \cos\theta_{12}}{2} =
\f{(\hat{\eta}_1-\hat{\eta}_2)^2}{\hat{\eta}_1+\hat{\eta}_2-2\hat{\eta}_1
  \hat{\eta}_2-2(1-2\zeta)
  \sqrt{\hat{\eta}_1(1-\hat{\eta}_1)\hat{\eta}_2(1-\hat{\eta}_2)}} \; .
\end{gathered}
\ee
The unresolved phase space is split according to the ordering of the
energies of the unresolved partons
\be
\int \mathrm{d}\bm{\Phi}_{\text{unresolved}}
= \int \mathrm{d}\bm{\Phi}_{\text{unresolved}} \,
\Big( \theta(u_1^0 - u_2^0) + \theta(u_2^0 - u_1^0 ) \Big) \; .
\ee
Each of the two resulting contributions is further decomposed
according to Fig.~\ref{fig:decomposition} into five sectors
\begin{multline}
\label{eq:unresolved3}
\!\!\!\!\!\! \int \mathrm{d}\bm{\Phi}_{\text{unresolved}} \,
\theta(u_1^0 - u_2^0) = \\ \Nep2 \int_{\mathcal{S}^{2-2\ep}_1}
\mathrm{d}\bm{\Omega}(\theta_1, \phi_1, \rho_1, \dots)
\int_{\mathcal{S}^{2-2\ep}_1} \mathrm{d}\bm{\Omega}(\theta_2, \phi_2,
\sigma_1, \sigma_2, \dots) \int_0^{u^0_{\mathrm{max}}}
\f{\mathrm{d}u_1^0 \, (u_1^0)^{1-2\ep}}{2(2\pi)^{3-2\ep}}
\int_0^{u^0_{2\, \mathrm{max}}} \f{\mathrm{d}u_2^0 \,
  (u_2^0)^{1-2\ep}}{2(2\pi)^{3-2\ep}} \, \theta\big( u_1^0 - u_2^0
\big) \, = \\ \f{E_{\mathrm{max}}^4}{(2\pi)^6}
\bigg( \f{\pi\mReg}{8E_{\mathrm{max}}^2} \bigg)^{2\ep}
\int_{\mathcal{S}^{1-2\ep}_1} \mathrm{d}\bm{\Omega}(\phi_1, \rho_1,
\dots) \int_{\mathcal{S}^{-2\ep}_1} \mathrm{d}\bm{\Omega}(\sigma_1,
\sigma_2, \dots) \int_0^1 \mathrm{d} \zeta \, \Big( \zeta \big(
1-\zeta \big) \Big)^{-\f{1}{2}-\ep} \iiiint_0^1 \mathrm{d}\eta_1
\mathrm{d}\eta_2 \mathrm{d}\xi_1 \mathrm{d}\xi_2
\sum_{i=1}^5 \mu_{\mathcal{S}_i} \; ,
\end{multline}
where $\eta_{1,2}, \xi_{1,2}$ parameterize $\hat{\eta}_{1,2}, \hat{\xi}_{1,2}$
as in Tab.~\ref{tab:kinmap}, while $\mu_{\mathcal{S}_i}$ can be found
in Tab.~\ref{tab:measure}, with
\be
\label{eq:ximax}
\begin{gathered}
\xi_{2 \, \mathrm{max}} = \min \left[ 1, \, \f{1}{\hat{\xi}_1}
  \f{1-\hat{\xi}_1}{1-\f{2E_{\mathrm{max}}}{\sqrt{\hat{s}}} \,
    \hat{\xi}_1 \, \eta_3} \right] \; , \\[0.2cm]
\eta_{31}(\eta_1,\eta_2) = \left. \f{\eta_3}{\eta_1}
\right|_{\begin{subarray}{l} \hat{\eta}_1 = \eta_1 \\ \hat{\eta}_2 =
    \eta_1 \eta_2/2 \end{subarray}} = \f{(2-\eta_2)^2}{2
  \left(2+\eta_2(1-2\eta_1)-2 (1-2 \zeta)
  \sqrt{\eta_2(1-\eta_1)(2-\eta_1 \eta_2)}\right)} \; ,
\\[0.2cm] \eta_{32}(\eta_1,\eta_2) =
\left. \f{\eta_3}{\eta_1\eta_2^2} \right|_{\begin{subarray}{l} \hat{\eta}_1 =
    \eta_1 \\ \hat{\eta}_2 = \eta_1(2-\eta_2)/2 \end{subarray}} = \f{1}{2\left(
  2+(1-2\eta_1)(2-\eta_2)-2(1-2\zeta)\sqrt{(1-\eta_1)(2-\eta_2)(2-\eta_1(2
    -\eta_2))} \right)} \; .
\end{gathered}
\ee
The essential property of these functions is that they do not vanish
in any of the singular limits of amplitudes indicated by the vanishing
of any of $\eta_{1,2}, \xi_{1,2}$.
\begin{table}[t]
  \begin{center}
    \renewcommand{\arraystretch}{1.5}
    \begin{tabular}{cC{2cm}C{2cm}C{2cm}C{2cm}C{2cm}}
      \hline
      & ${\cal S}_1$ & ${\cal S}_2$ & ${\cal S}_3$ &
      ${\cal S}_4$ & ${\cal S}_5$ \\ \hline $\hat{\eta}_1$ &
      $\eta_1$ & $\f{1}{2}\eta_1\eta_2$ & $\f{1}{2}
      \eta_1\eta_2\xi_2$ & $\eta_1$ & $\f{1}{2}(2-\eta_1)\eta_2$
      \\ $\hat{\eta}_2$ & $\f{1}{2}\eta_1\eta_2$ & $\eta_2$
      & $\eta_2$ & $\f{1}{2}\eta_1(2-\eta_2)$ & $\eta_2$
      \\ $\hat{\xi}_1$ & $\xi_1$ & $\xi_1$ & $\xi_1$ & $\xi_1$
      & $\xi_1$
      \\ $\hat{\xi}_2$ & $\xi_1\xi_2\xi_{2\,\mathrm{max}}$ &
      $\eta_1\xi_1\xi_2\xi_{2\,\mathrm{max}}$ &
      $\xi_1\xi_2\xi_{2\,\mathrm{max}}$ &
      $\xi_1\xi_2\xi_{2\,\mathrm{max}}$ &
      $\xi_1\xi_2\xi_{2\,\mathrm{max}}$ \\ \hline
    \end{tabular}
  \end{center}
  \caption{\label{tab:kinmap} Original kinematic variables of the
    triple-collinear sector parameterization, $\hat{\eta}_1,
    \hat{\eta}_2, \hat{\xi}_1, \hat{\xi}_2$, expressed through the
    sector variables, $\eta_1, \eta_2, \xi_1, \xi_2$, of the five
    sectors, $\mathcal{S}_1, \dots, \mathcal{S}_5$, defined in
    Fig.~\ref{fig:decomposition}. The function $\xi_{2\,\mathrm{max}}$
    is defined in Eq.~(\ref{eq:ximax}).}
\end{table}
\begin{table}[t]
  \begin{center}
    \renewcommand{\arraystretch}{1.5}
    \begin{tabular}{L{1.5cm}l}
      \hline
      & \multicolumn{1}{c}{$\mu_{\mathcal{S}_i}$}
      \\ \hline \\[-.6cm] $\mathcal{S}_1$ &
      $\displaystyle
      \eta_1^{1-2\epsilon}\eta_2^{-\epsilon}\xi_1^{3-4\epsilon}\xi_2^{1-2\epsilon}
      \left((1-\eta_1)(2-\eta_1\eta_2)\right)^{-\epsilon}
      \left( \f{\eta_{31}(\eta_1,\eta_2)}{2-\eta_2}
      \right)^{1-2\epsilon} \xi_{2 \, \mathrm{max}}^{\; 2-2\epsilon}$
      \\[0.2cm] $\mathcal{S}_2$ &
      $\displaystyle
      \eta_1^{2-3\epsilon}\eta_2^{1-2\epsilon}\xi_1^{3-4\epsilon}\xi_2^{1-2\epsilon}
      \left((1-\eta_2)(2-\eta_1\eta_2)\right)^{-\epsilon}
      \left( \f{\eta_{31}(\eta_2,\eta_1)}{2-\eta_1}
      \right)^{1-2\epsilon} \xi_{2 \, \mathrm{max}}^{\; 2-2\epsilon}$
      \\[0.2cm] $\mathcal{S}_3$ &
      $\displaystyle
      \eta_1^{-\epsilon}\eta_2^{1-2\epsilon}\xi_1^{3-4\epsilon}\xi_2^{2-3\epsilon}
      \left((1-\eta_2)(2-\eta_1\eta_2\xi_2)\right)^{-\epsilon}
      \left( \f{\eta_{31}(\eta_2,\eta_1\xi_2)}{2-\eta_1\xi_2}
      \right)^{1-2\epsilon} \xi_{2 \, \mathrm{max}}^{\; 2-2\epsilon}$
      \\[0.3cm] $\mathcal{S}_4$ &
      $\eta_1^{1-2\epsilon}\eta_2^{1-2\epsilon}\xi_1^{3-4\epsilon}\xi_2^{1-2\epsilon}
      \left((1-\eta_1)(2-\eta_2)(2-\eta_1(2-\eta_2))\right)^{-\epsilon}
      \eta_{32}^{1-2\epsilon}(\eta_1,\eta_2) \, \xi_{2 \, \mathrm{max}}^{\; 2-2\epsilon}$
      \\[0.3cm] $\mathcal{S}_5$ &
      $\eta_1^{1-2\epsilon}\eta_2^{1-2\epsilon}\xi_1^{3-4\epsilon}\xi_2^{1-2\epsilon}
      \left((1-\eta_2)(2-\eta_1)(2-\eta_2(2-\eta_1))\right)^{-\epsilon}
      \eta_{32}^{1-2\epsilon}(\eta_2,\eta_1) \, \xi_{2 \,
        \mathrm{max}}^{\; 2-2\epsilon}$ \\ \hline
    \end{tabular}
  \end{center}
  \caption{\label{tab:measure} Integration measures,
    $\mu_{\mathcal{S}_i}$, of the five sectors $\mathcal{S}_1, \dots,
    \mathcal{S}_5$, of the triple-collinear sector
    parameterization. The functions $\eta_{31}, \eta_{32}$ and $\xi_{2
      \, \mathrm{max}}$ are defined in Eq.~(\ref{eq:ximax}).}
\end{table}

The behavior of amplitudes in collinear limits is characterized by
the transverse vectors
\be
u_{i\perp}^\mu = \begin{pmatrix} 0
  \\ \bm{\hat{u}_{i\perp}} \end{pmatrix} \; , \quad i = 1,2,3 \; ,
\ee
with
\begin{align}
\bm{\hat{u}_{1\perp}} = \lim_{\theta_1 \to 0}
\f{\bm{\hat{u}_1}-\bm{\hat{r}}}
{\lVert \bm{\hat{u}_1}-\bm{\hat{r}} \rVert} &= \left. 
\f{\partial \bm{\hat{u}_1}}{\partial \theta_1} \right|_{\theta_1 = 0}
= \bm{\hat{R}}^{(3-2\ep)}_1(\alpha_1, \alpha_2, \dots)
\bm{\hat{n}}^{(3-2\ep)}\left(\f{\pi}{2}, \phi_1, \rho_1, \rho_2,
\dots\right)
\; ,
\\[0.2cm]
\bm{\hat{u}_{2\perp}} = \lim_{\theta_2 \to 0}
\f{\bm{\hat{u}_2}-\bm{\hat{r}}}
{\lVert \bm{\hat{u}_2}-\bm{\hat{r}} \rVert} &= \left. 
\f{\partial \bm{\hat{u}_2}}{\partial \theta_2} \right|_{\theta_2 = 0}
= \bm{\hat{R}}^{(3-2\ep)}_1(\alpha_1, \alpha_2, \dots)
\bm{\hat{R}}^{(3-2\ep)}_2(\phi_1, \rho_1, \rho_2, \dots)
\bm{\hat{n}}^{(3-2\ep)}\left(\f{\pi}{2}, \phi_2, \sigma_1, \sigma_2,
\dots\right)
\; .
\end{align}
In order to determine the third transverse vector, let us consider a
general parameterization
\be
\label{eq:azimuthalparameterization}
\begin{gathered}
\phi_2 = \phi_2(\theta_1, \theta_2, \zeta) = \phi_2(\theta_2,
\theta_1, \zeta) \; , \quad \phi_2(\theta_1, \theta_1, \zeta) = 0 \; ,
\\[0.3cm] \phi_2(\theta_1, \theta_2, \zeta) =
\partial^+_{\theta_2}\phi_2(\theta_1, \zeta) \, \big|
\theta_2 - \theta_1 \big| + \mathcal{O}\big( (\theta_2 - \theta_1)^2
\big) \; , \\[0.2cm] \partial^+_{\theta_2}\phi_2(\theta_1, \zeta) =
\lim_{\theta_2 \to \theta_1^+} \f{\phi_2(\theta_1, \theta_2, \zeta) -
  \phi_2(\theta_1, \theta_1, \zeta)}{\theta_2 - \theta_1} \; .
\end{gathered}
\ee
We then have
\begin{multline}
\bm{\hat{u}_{3\perp}}^\pm = \lim_{\theta_2 \to \theta_1^\pm}
\f{\bm{\hat{u}_2}-\bm{\hat{u}_1}}
{\lVert \bm{\hat{u}_2}-\bm{\hat{u}_1} \rVert} = \pm
\mathcal{N}_{3\perp}(\theta_1, \zeta)
\lim_{\theta_2 \to \theta_1^\pm}
\f{\bm{\hat{u}_2}-\bm{\hat{u}_1}}{\theta_2 - \theta_1} =
\pm \mathcal{N}_{3\perp}(\theta_1, \zeta) \, 
\bm{\hat{R}}^{(3-2\ep)}_1(\alpha_1, \alpha_2, \dots)
\bm{\hat{R}}^{(3-2\ep)}_1(\theta_1, \phi_1, \rho_1, \rho_2, \dots)
\\ \times \left( \bm{\hat{n}}^{(3-2\ep)}\left(\f{\pi}{2}, 0, 0,
\dots\right) \pm \sin\theta_1 \, \partial^+_{\theta_2}\phi_2(\theta_1,
\zeta) \, \bm{\hat{n}}^{(3-2\ep)}\left(\f{\pi}{2}, \f{\pi}{2},
\sigma_1, \sigma_2, \dots\right) \right) \; ,
\end{multline}
where $\mathcal{N}_{3\perp}(\theta_1, \zeta)$ is the positive
normalization factor
\be
\mathcal{N}_{3\perp}(\theta_1, \zeta) = \left[ 1 + \left(
  \sin\theta_1 \, \partial^+_{\theta_2}\phi_2(\theta_1, \zeta)
  \right)^2 \right]^{-\f{1}{2}} \; .
\ee
The transverse vector can be reexpressed as
\be
\label{eq:transverse3}
\bm{\hat{u}_{3\perp}}^\pm = \bm{\hat{R}}^{(3-2\ep)}_1(\alpha_1,
\alpha_2, \dots) \bm{\hat{R}}^{(3-2\ep)}_1(\theta_1, \phi_1, \rho_1,
\rho_2, \dots) \bm{\hat{n}}^{(3-2\ep)}\left(\f{\pi}{2},
\tilde{\phi}_2^\pm, \sigma_1, \sigma_2, \dots\right)
\; ,
\ee
with
\be
\label{eq:azimuthalangle}
\tan \tilde{\phi}^\pm_2(\theta_1, \zeta) = \pm \sin\theta_1 \,
\partial^+_{\theta_2}\phi_2(\theta_1, \zeta) \; , \quad
\tilde{\phi}^+_2 \in \left[0, \f{\pi}{2} \right[ \; , \quad
\tilde{\phi}^-_2 \in \left[\f{\pi}{2}, \pi \right[ \; .
\ee
For our particular choice of the dependence of $\phi_2$ on $\zeta$,
Eq.~(\ref{eq:parameters3}), there is
\be
\cos\tilde{\phi}^\pm_2 = \pm \sqrt{\zeta} \; , \quad
\sin\tilde{\phi}^\pm_2 = \sqrt{1-\zeta} \; .
\ee
%

%%%%%%%%%%%%%%%%%%%%%%%%%%%%%%%%%%%%%%%%%%%%%%%%%%%%%%%%%%%%%%%%%%%%%%%%%%%%%%%%

\subsection{Double-collinear sector parameterization: two reference
  momenta, two unresolved momenta}
\label{sec:double-collinear}

%%%%%%%%%%%%%%%%%%%%%%%%%%%%%%%%%%%%%%%%%%%%%%%%%%%%%%%%%%%%%%%%%%%%%%%%%%%%%%%%

\subsubsection{General case with $n > n_{fr}$}
\label{sec:double-collinear-general}

\begin{figure}[t]
  \begin{center}
    \includegraphics[scale=.75]{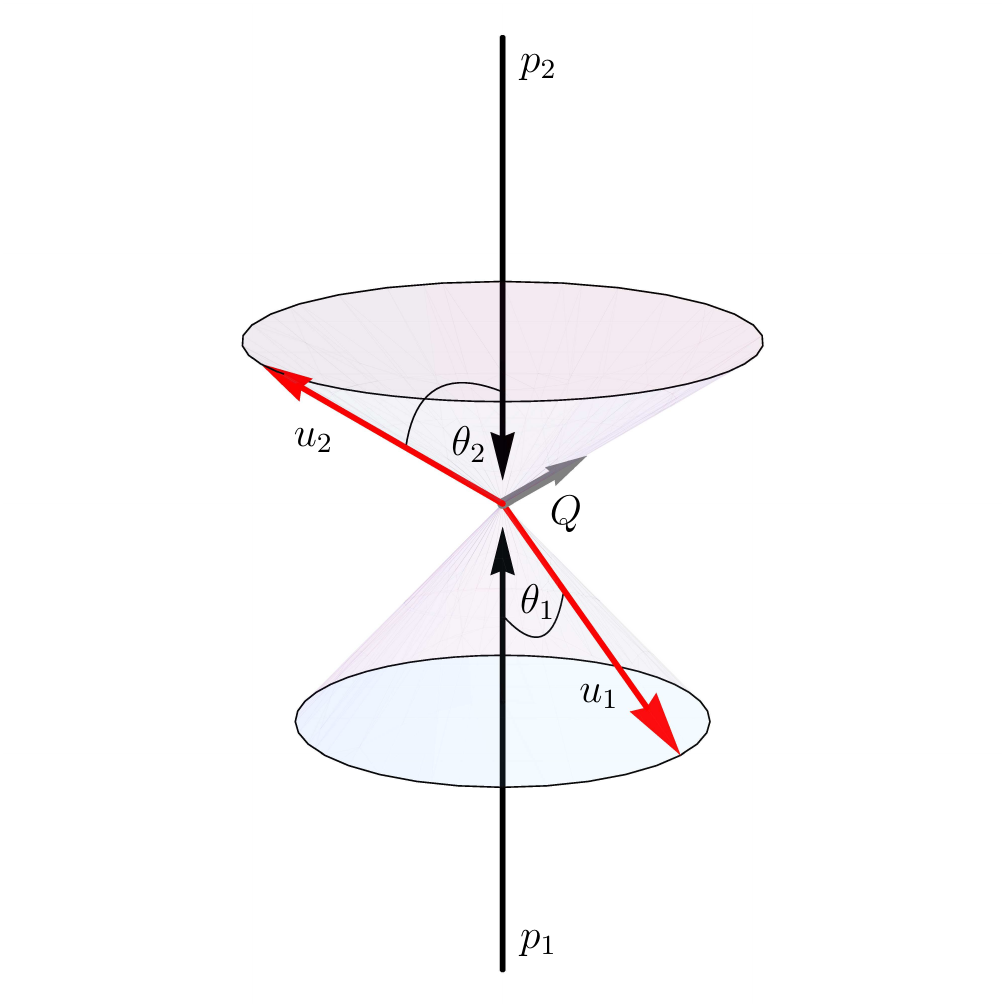}
    \includegraphics[scale=.75]{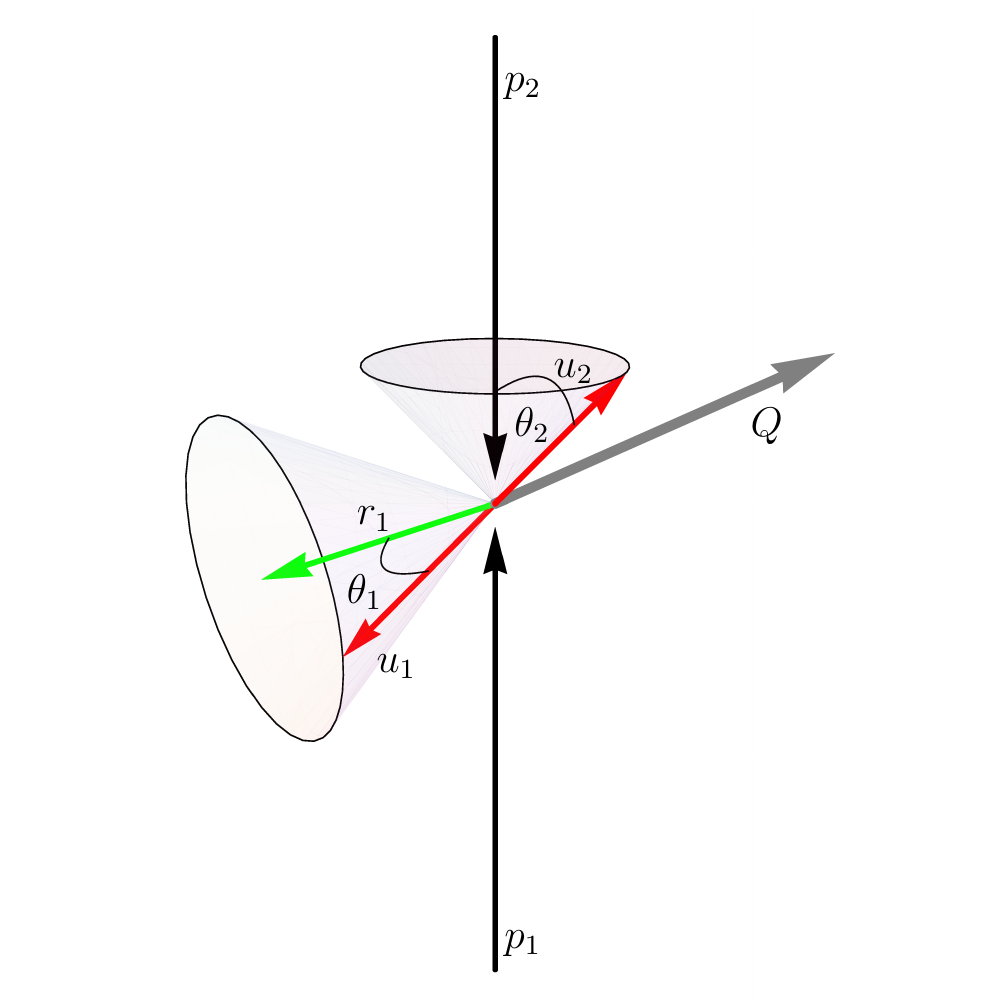}
  \end{center}
  \caption{\label{fig:double} Example momentum parameterizations in
    the case of two unresolved momenta, $u_1$ and $u_2$, and two
    reference momenta, $r_1$ and $r_2$. The left picture shows a case
    with initial state reference momenta with $r_1 = p_2$ and $r_2 =
    p_1$, while the right picture shows a mixed case with one final
    state reference momentum, $r_1$, and one initial state reference
    momentum, $r_2 = p_1$. The angles $\theta_1$ and $\theta_2$ allow
    for a straightforward parameterization of the collinear limits
    with respect to the reference momenta. $Q$ is the total momentum
    of the remaining final-state particles, while $p_1$ and $p_2$ are
    the initial-state momenta.}
\end{figure}

\noindent
The resolved and unresolved parton momenta read
\be
r_1^\mu = r_1^0 \, \hat{r}_1^\mu = r_1^0 \begin{pmatrix} 1
  \\ \bm{\hat{r}_1} \end{pmatrix} \; , \quad 
r_2^\mu = r_2^0 \, \hat{r}_2^\mu = r_2^0 \begin{pmatrix} 1
  \\ \bm{\hat{r}_2} \end{pmatrix} \; , \quad 
u_1^\mu = u_1^0 \, \hat{u}_1^\mu = u_1^0 \begin{pmatrix} 1
  \\ \bm{\hat{u}_1} \end{pmatrix} \; , \quad
u_2^\mu = u_2^0 \, \hat{u}_2^\mu = u_2^0 \begin{pmatrix} 1
  \\ \bm{\hat{u}_2} \end{pmatrix} \; ,
\ee
with the angular parameterization (see Fig.~\ref{fig:double})
\begin{align}
\bm{\hat{r}_1} &= \bm{\hat{n}}^{(3-2\ep)}(\alpha_1, \alpha_2, \dots)
\; ,
\nn \\[0.2cm]
\bm{\hat{r}_2} &= \bm{\hat{n}}^{(3-2\ep)}(\beta_1, \beta_2, \dots)
\; ,
\nn \\[0.2cm]
\bm{\hat{u}_1} &= \bm{R}^{(3-2\ep)}_1(\alpha_1, \alpha_2, \dots)
\bm{\hat{n}}^{(3-2\ep)}(\theta_1, \phi_1, \rho_1, \rho_2, \dots) \; ,
\nn \\[0.2cm]
\bm{\hat{u}_2} &= \bm{R}^{(3-2\ep)}_1(\beta_1, \beta_2, \dots)
\bm{R}^{(3-2\ep)}_4(\rho_2, \rho_3, \dots)
\bm{\hat{n}}^{(3-2\ep)}(\theta_2, \phi_2, \sigma_1, \sigma_2, \dots)
\; .
\end{align}
In the case of two initial state reference momenta, the phase space is
\be
\int \mathrm{d}\bm{\Phi}_{n+2} =
\int \mathrm{d}\bm{\Phi}_{\text{unresolved}}
\int \mathrm{d}\bm{\Phi}_{n}( p_1+p_2-u_1-u_2 ) \; ,
\ee
while for one final state ($r_1$) and one initial state ($r_2$)
reference momentum
\be
\int \mathrm{d}\bm{\Phi}_{n+2} = \Nep{} \int_{\mathcal{S}^{2-2\ep}_1}
\mathrm{d}\bm{\Omega}(\alpha_1, \alpha_2, \dots)
\int \mathrm{d}\bm{\Phi}_{\text{unresolved}}
\int_0^{r_{\mathrm{max}}^0} \f{\mathrm{d}r_1^0 \,
  (r_1^0)^{1-2\ep}}{2(2\pi)^{3-2\ep}} \int \mathrm{d}\bm{\Phi}_{n-1}(
p_1+p_2-r_1-u_1-u_2 ) \; ,
\ee
and similarly if the r\^oles of $r_1$ and $r_2$ are reversed.
For two final state reference momenta, there is
\begin{multline}
\int \mathrm{d}\bm{\Phi}_{n+2} = \Nep2 \int_{\mathcal{S}^{2-2\ep}_1}
\mathrm{d}\bm{\Omega}(\alpha_1, \alpha_2, \dots)
\int_{\mathcal{S}^{2-2\ep}_1} \mathrm{d}\bm{\Omega}(\beta_1, \beta_2,
\dots) \int \mathrm{d}\bm{\Phi}_{\text{unresolved}} \\ \times
\int_0^{r_{\mathrm{max}}^0} \f{\mathrm{d}r_1^0 \,
  (r_1^0)^{1-2\ep}}{2(2\pi)^{3-2\ep}} \int_0^{r_{2\,\mathrm{max}}^0}
\f{\mathrm{d}r_2^0 \, (r_2^0)^{1-2\ep}}{2(2\pi)^{3-2\ep}} \int
\mathrm{d}\bm{\Phi}_{n-2}( p_1+p_2-r_1-r_2-u_1-u_2 ) \; .
\end{multline}
$r_{\mathrm{max}}^0$ is defined in Eq.~(\ref{eq:r0max2}) with
$\hat{r} = \hat{r}_1$. If the second reference momentum is in the
initial state and $n = 2$, then Eq.~(\ref{eq:n=2case})
applies. Furthermore, in the case of two final state reference momenta
\be
r_{2\,\mathrm{max}}^0 =
\f{\sqrt{\hat{s}} \left( E_{\mathrm{max}} - u_1^0 - u_2^0 - r_1^0
  \right) + r_1 \cdot ( u_1 + u_2 ) + u_1 \cdot u_2}{\sqrt{\hat{s}} -
  \hat{r}_2 \cdot \left( u_1 + u_2 + r_1 \right)} \; ,
\ee
and in the particular case $n = 3$
\be
\int_0^{r_{2\,\mathrm{max}}^0} \f{\mathrm{d}r_2^0 \,
  (r_2^0)^{1-2\ep}}{2(2\pi)^{3-2\ep}} \int \mathrm{d}\bm{\Phi}_1( p_1
+ p_2 - u_1 - u_2 ) = \f{(r_{2\,\mathrm{max}}^0)^{1-2\ep}}{4(2\pi)^{2-2\ep}}
\f{1}{\sqrt{\hat{s}} - \hat{r}_2 \cdot \left( u_1 + u_2 + r_1 \right)}
\; .
\ee
The unresolved phase space is split according to the ordering of the
energies of the unresolved partons
\be
\int \mathrm{d}\bm{\Phi}_{\text{unresolved}}
= \int \mathrm{d}\bm{\Phi}_{\text{unresolved}} \,
\Big( \theta(u_1^0 - u_2^0) + \theta(u_2^0 - u_1^0 ) \Big) \; ,
\ee
with
\begin{multline}
\label{eq:unresolved2}
\!\!\!\!\!\! \int \mathrm{d}\bm{\Phi}_{\text{unresolved}} \,
\theta(u_1^0 - u_2^0) = \\ \Nep2 \int_{\mathcal{S}^{2-2\ep}_1}
\mathrm{d}\bm{\Omega}(\theta_1, \phi_1, \rho_1, \dots)
\int_{\mathcal{S}^{2-2\ep}_1} \mathrm{d}\bm{\Omega}(\theta_2, \phi_2,
\sigma_1, \sigma_2, \dots) \int_0^{u^0_{\mathrm{max}}}
\f{\mathrm{d}u_1^0 \, (u_1^0)^{1-2\ep}}{2(2\pi)^{3-2\ep}}
\int_0^{u^0_{2\, \mathrm{max}}} \f{\mathrm{d}u_2^0 \,
  (u_2^0)^{1-2\ep}}{2(2\pi)^{3-2\ep}} \, \theta\big( u_1^0 - u_2^0
\big) \, = \\ \f{E_{\mathrm{max}}^4}{(2\pi)^6}
\bigg( \f{\pi\mReg}{4E_{\mathrm{max}}^2} \bigg)^{2\ep}
\int_{\mathcal{S}^{1-2\ep}_1} \mathrm{d}\bm{\Omega}(\phi_1, \rho_1,
\dots) \int_{\mathcal{S}^{1-2\ep}_1} \mathrm{d}\bm{\Omega}(\phi_2,
\sigma_1, \sigma_2, \dots) \\ \times \iiiint_0^1 \mathrm{d}\eta_1
\mathrm{d}\eta_2 \mathrm{d}\xi_1 \mathrm{d}\xi_2 \, \eta_1^{-\ep}
\eta_2^{-\ep} \xi_1^{3-4\ep} \xi_2^{1-2\ep} \Big( \big(1-\eta_1\big)
\big(1-\eta_2\big) \Big)^{-\ep} \xi_{2\,\mathrm{max}}^{2-2\ep} \; ,
\end{multline}
where
\be
\begin{gathered}
\label{eq:parameters2}
u_1^0 = E_{\mathrm{max}} \, \xi_1 \; , \quad u_2^0 = E_{\mathrm{max}} 
\, \xi_1 \xi_2 \xi_{2\,\mathrm{max}} \; , \quad \xi_{2 \,
  \mathrm{max}} = \min \left[ 1, \, \f{1}{\xi_1}
  \f{1-\xi_1}{1-\f{E_{\mathrm{max}}}{\sqrt{\hat{s}}} \, \xi_1 \, \hat{u}_1
    \cdot \hat{u}_2} \right] \; , \\[0.2cm] \cos\theta_1 = 1-2\eta_1
\; , \quad \cos\theta_2 = 1-2\eta_2 \; .
\end{gathered}
\ee
Notice that if $\alpha_i = \beta_i = 0$ for $i > 2$, then $\hat{u}_1
\cdot \hat{u}_2$ is only a function of $\alpha_{1,2}, \beta_{1,2},
\theta_{1,2}, \phi_{1,2}, \rho_1$, and $\sigma_{1,2}$. This fact will
be important for the four-dimensional formulation of the subtraction
scheme.

The behavior of amplitudes in collinear limits is characterized by
the transverse vectors
\be
u_{i\perp}^\mu = \begin{pmatrix} 0
  \\ \bm{\hat{u}_{i\perp}} \end{pmatrix} \; , \quad i = 1,2 \; ,
\ee
with
\begin{align}
\bm{\hat{u}_{1\perp}} &= \lim_{\theta_1 \to 0}
\f{\bm{\hat{u}_1}-\bm{\hat{r}_1}}
{\lVert \bm{\hat{u}_1}-\bm{\hat{r}_1} \rVert} = \left. 
\f{\partial \bm{\hat{u}_1}}{\partial \theta_1} \right|_{\theta_1 = 0}
= \bm{\hat{R}}^{(3-2\ep)}_1(\alpha_1, \alpha_2, \dots)
\bm{\hat{n}}^{(3-2\ep)}\left(\f{\pi}{2}, \phi_1, \rho_1, \rho_2,
\dots\right)
\; , \\[0.2cm]
\bm{\hat{u}_{2\perp}} &= \lim_{\theta_2 \to 0}
\f{\bm{\hat{u}_2}-\bm{\hat{r}_2}}
{\lVert \bm{\hat{u}_2}-\bm{\hat{r}_2} \rVert} = \left. 
\f{\partial \bm{\hat{u}_2}}{\partial \theta_2} \right|_{\theta_2 = 0}
= \bm{\hat{R}}^{(3-2\ep)}_1(\beta_1, \beta_2, \dots)
\bm{\hat{R}}^{(3-2\ep)}_4(\rho_2, \rho_3, \dots)
\bm{\hat{n}}^{(3-2\ep)}\left(\f{\pi}{2}, \phi_2, \sigma_1, \sigma_2,
\dots\right)
\; .
\end{align}
%

%%%%%%%%%%%%%%%%%%%%%%%%%%%%%%%%%%%%%%%%%%%%%%%%%%%%%%%%%%%%%%%%%%%%%%%%%%%%%%%%

\subsubsection{Special case with $n = n_u = n_{fr} = 2$}
\label{sec:double-collinear-special}

\begin{figure}[t]
  \begin{center}
    \includegraphics[scale=.75]{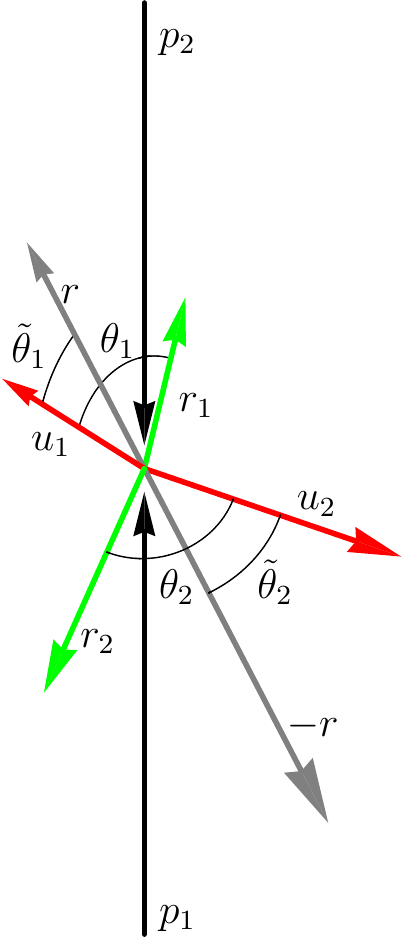}
    \hspace{2cm}
    \includegraphics[scale=.65]{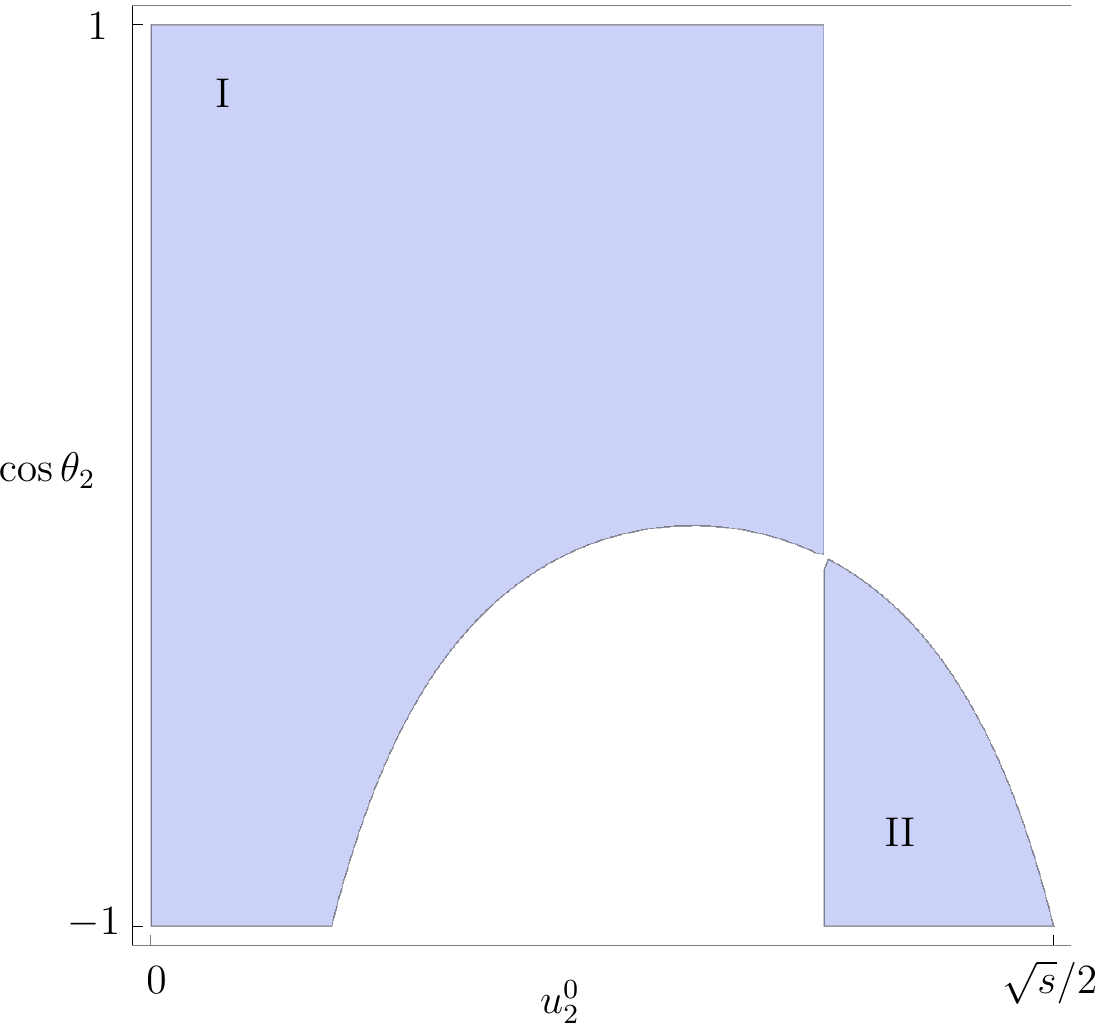}
  \end{center}
  \caption{\label{fig:nr0} Parameterization of a fully massless
    four-particle phase space (left), where two of the final states
    are used as reference vectors, $r_1$ and $r_2$, while the other
    two, $u_1$ and $u_2$, are considered to be unresolved. $p_1$ and
    $p_2$ are initial state momenta, while $r$ is a reference vector
    equal to the sum of one reference and one unresolved momentum, $r
    = r_1 + u_1$. If the parameters of one of the unresolved momenta
    are specified, then the available range of parameters of the other
    one is split into two disjunct regions (right). There are neither
    soft nor collinear singularities in region II.}
\end{figure}

\noindent
This is a parameterization of a fully massless four-particle
phase space. Unlike previous cases, it uses an auxiliary reference
vector, $\bm{r}$. The configuration is depicted in Fig.~\ref{fig:nr0},
and defines the following reference and unresolved momenta
\be
r_1^\mu = r_1^0 \, \hat{r}_1^\mu = r_1^0 \begin{pmatrix} 1
  \\ \bm{\hat{r}_1} \end{pmatrix} \; , \quad 
r_2^\mu = r_2^0 \, \hat{r}_2^\mu = r_2^0 \begin{pmatrix} 1
  \\ \bm{\hat{r}_2} \end{pmatrix} \; , \quad 
u_1^\mu = u_1^0 \, \hat{u}_1^\mu = u_1^0 \begin{pmatrix} 1
  \\ \bm{\hat{u}_1} \end{pmatrix} \; , \quad
u_2^\mu = u_2^0 \, \hat{u}_2^\mu = u_2^0 \begin{pmatrix} 1
  \\ \bm{\hat{u}_2} \end{pmatrix} \; ,
\ee
where
\be
\bm{r} = r_1^0 \, \bm{\hat{r}_1} + u_1^0
\, \bm{\hat{u}_1} = - r_2^0 \, \bm{\hat{r}_2} - u_2^0 \,
\bm{\hat{u}_2} \; , \quad r = \lVert \bm{r} \rVert \; , \quad
\bm{\hat{r}} = \f{\bm{r}}{r}
\; ,
\ee
with the angular parameterization
\begin{align}
\bm{\hat{r}} &= \bm{\hat{n}}^{(3-2\ep)}(\alpha_1, \alpha_2, \dots)
\; ,
\nn \\[0.2cm]
\bm{\hat{u}_1} &= \bm{R}^{(3-2\ep)}_1(\alpha_1, \alpha_2, \dots)
\bm{\hat{n}}^{(3-2\ep)}(\tilde{\theta}_1, \phi_1, \rho_1, \rho_2, \dots)
\; ,
\nn \\[0.2cm]
\bm{\hat{u}_2} &= \bm{R}^{(3-2\ep)}_1(\alpha_1, \alpha_2, \dots)
\bm{R}^{(3-2\ep)}_4(\rho_2, \rho_3, \dots)
\bm{\hat{n}}^{(3-2\ep)}(\tilde{\theta}_2, \phi_2, \sigma_1, \sigma_2,
\dots)
\; .
\end{align}
We start with a general discussion of the phase space. A
parameterization suitable for the derivation of the subtraction and
integrated subtraction terms will be provided near the end of the
section. We write
\be
\begin{split}
\int \mathrm{d}\bm{\Phi}_{4} &= \Nep3 \int_{\mathcal{S}^{2-2\ep}_1}
\mathrm{d}\bm{\Omega}(\alpha_1, \alpha_2, \dots) \\
&\times \int_{\mathcal{S}^{1-2\ep}_1} \mathrm{d}\bm{\Omega}(\phi_1, \rho_1,
\dots) \int_{\mathcal{S}^{1-2\ep}_1} \mathrm{d}\bm{\Omega}(\phi_2,
\sigma_1,\sigma_2, \dots) \\
&\times \int_0^\infty\f{\mathrm{d}u_1^0 \,
  (u_1^0)^{1-2\ep}}{2(2\pi)^{3-2\ep}}
\int_{-1}^{1}\mathrm{d}\cos\theta_1
\,\left(1-\cos^2\theta_1\right)^{-\ep}
\int_0^\infty\f{\mathrm{d}u_2^0
  (u_2^0)^{1-2\ep}}{2(2\pi)^{3-2\ep}} \int_{-1}^{1}
\mathrm{d}\cos\theta_2 \,\left(1-\cos^2\theta_2\right)^{-\ep} \\
&\times\left(\f{1-\cos^2\tilde{\theta}_1}{1-\cos^2\theta_1}
 \right)^{-\ep}\left(\f{1-\cos^2\tilde{\theta}_2}{1-\cos^2\theta_2}
 \right)^{-\ep} \f{r_1^0\,r_2^0}{4\left(2\pi\right)^{2-2\ep}\,
   r^{1+2\ep} \, \left|r_1^0 + r_2^0 + u_1^0\cos\theta_1 +
   u_2^0\cos\theta_2\right|} \; ,
\end{split}
\ee
where 
\be
 \cos\tilde{\theta}_i=\f{r_i^0\cos\theta_i+u_i^0}{r} \; ,
\ee
\begin{align}
 r_1^0&=\f{\hat{s}-2\sqrt{\hat{s}}\left(u_1^0+u_2^0\right) +
   2u_2^0\left(u_1^0+u_2^0\right)+2\left(\sqrt{\hat{s}}-u_1^0-u_2^0\right)
   u_2^0\cos\theta_2}{2\left(\sqrt{\hat{s}}-u_1^0\left(1-\cos\theta_1\right)
   - u_2^0\left(1-\cos\theta_2\right)\right)} \; ,
 \nn \\    
 r_2^0&=\f{\hat{s}-2\sqrt{\hat{s}}\left(u_1^0+u_2^0\right)+2u_1^0 \left(u_1^0
   + u_2^0\right)+2\left(\sqrt{\hat{s}}-u_1^0-u_2^0\right)
   u_1^0\cos\theta_1}{2\left(\sqrt{\hat{s}}-u_1^0\left(1-\cos\theta_1\right)
   - u_2^0\left(1-\cos\theta_2\right)\right)} \; ,
 \nn \\
 r&=\sqrt{\Bigl(u_1^0\Bigr)^2+\Bigl(r_1^0\Bigr)^2+2u_1^0r_1^0\cos\theta_1} =
 \sqrt{\Bigl(u_2^0\Bigr)^2+\Bigl(r_2^0\Bigr)^2+2u_2^0r_2^0\cos\theta_2}
 \; .
\end{align}
Notice that in the collinear limit $\theta_i \to 0$, also $\tilde{\theta}_i
\to 0$. This implies that the integration measure factors
\be
\left(\f{1-\cos^2\tilde{\theta}_i}{1-\cos^2\theta_i}\right)^{-\ep} \; ,
\ee
do not influence the scaling of the integrand in the limit.

The integration region splits into two, region I and region II. We
choose the energy and angle of the first parton to provide
restrictions on the energy and angle of the second parton, see
Fig.~\ref{fig:nr0}, thus
\be
0\leq u_1^0 \leq \f{\sqrt{\hat{s}}}{2} \; , \quad-1\leq \cos\theta_1\leq1
\; .
\ee
Region I reads
\begin{gather}
0 \, \leq  \, u_2^0 \, < \, \f{\hat{s} - 2u_1^0 \left( \sqrt{\hat{s}} -
  u_1^0 \right) \left( 1 - \cos\theta_1 \right)}{2 \left( \sqrt{\hat{s}} -
  u_1^0 \left( 1 - \cos\theta_1 \right) \right)} \; , \nn \\[0.2cm]
- \, \mathrm{min} \left[ 1 , \, \f{\hat{s} - 2 \sqrt{\hat{s}} \left(
    u_1^0 + u_2^0 \right) + 2 u_2^0 \left( u_1^0 + u_2^0 \right)}{2
    u_2^0 \left( \sqrt{\hat{s}} - u_1^0 - u_2^0 \right)} \right] \,
\leq \, \cos\theta_2 \, \leq \, 1 \; ,
\end{gather}
with
\be
r_1^0 + r_2^0 + u_1^0\cos\theta_1 + u_2^0\cos\theta_2 \, > \,  0 \; .
\ee
Region II reads
\begin{gather}
\f{\hat{s} - 2u_1^0 \left( \sqrt{\hat{s}} - u_1^0 \right) \left( 1 -
  \cos\theta_1 \right)}{2 \left( \sqrt{\hat{s}} - u_1^0 \left( 1 -
  \cos\theta_1 \right) \right)} \, \leq  \, u_2^0 \, \leq \,
\f{\sqrt{\hat{s}}}{2} \; , \nn \\[0.2cm]
 -1 \, \leq \, \cos\theta_2 \, \leq \, \, - \f{\hat{s} - 2
   \sqrt{\hat{s}} \left( u_1^0 + u_2^0 \right) + 2 u_2^0 \left( u_1^0
   + u_2^0 \right)}{2 u_2^0 \left( \sqrt{\hat{s}} - u_1^0 - u_2^0
   \right)} \; ,
\end{gather}
with
\be
r_1^0 + r_2^0 + u_1^0\cos\theta_1 + u_2^0\cos\theta_2 \, \leq \, 0 \; .
\ee
Region II does not contain soft or collinear singularities. Indeed, as
long as $u_1^0 > 0$ and $\cos\theta_1 < 1$, the endpoints $u_2^0 = 0$
and $\cos\theta_2 = 1$ are not reachable. On the other hand, if either
$u_1^0 = 0$ or $\cos\theta_1 = 1$, region II has zero volume.

We are now ready to present the final parameterization of the phase
space. We have
\begin{multline}
\int \mathrm{d}\bm{\Phi}_4
= \int \mathrm{d}\bm{\Phi}_4 \, \Big[
\Big( \theta(u_1^0 - u_2^0) + \theta(u_2^0 - u_1^0 ) \Big) \,
\theta\big( r_1^0 + r_2^0 + u_1^0\cos\theta_1 + u_2^0\cos\theta_2
\big) \\ + \theta\big( -r_1^0 - r_2^0 - u_1^0\cos\theta_1 -
u_2^0\cos\theta_2 \big) \Big] \; .
\end{multline}
The contribution on the second line represents region II. Since there
are no singularities there, we will keep the formulae we have
given before without any further modification. The integral can be
performed in four dimensions, i.e.\ $\ep = 0$. On the other hand, we
will write
\begin{multline}
\label{eq:dijet}
\int \mathrm{d}\bm{\Phi}_4 \, \theta(u_1^0 - u_2^0)
\, \theta\big( r_1^0 + r_2^0 + u_1^0\cos\theta_1 + u_2^0\cos\theta_2
\big) = \int_{\mathcal{S}^{2-2\ep}_1} \mathrm{d}\bm{\Omega}(\alpha_1,
\alpha_2, \dots) \\ \times
\f{E_{\mathrm{max}}^4}{4(2\pi)^8}
\bigg( \f{\pi\mReg}{E_{\mathrm{max}}^2} \bigg)^{3\ep}
\int_{\mathcal{S}^{1-2\ep}_1} \mathrm{d}\bm{\Omega}(\phi_1, \rho_1,
\dots) \int_{\mathcal{S}^{1-2\ep}_1} \mathrm{d}\bm{\Omega}(\phi_2,
\sigma_1,\sigma_2, \dots) \iiiint_0^1 \mathrm{d}\eta_1
\mathrm{d}\eta_2 \mathrm{d}\xi_1 \mathrm{d}\xi_2 \, \eta_1^{-\ep}
\eta_2^{-\ep} \xi_1^{3-4\ep} \xi_2^{1-2\ep}
\\ \times \eta_{2 \, \mathrm{max}} \, \xi_{2 \, \mathrm{max}}^{2-2\ep}
\left(\f{1}{\eta_1} \left( 1-\cos^2\tilde{\theta}_1 \right) \right)^{-\ep}
\left(\f{1}{\eta_2} \left( 1-\cos^2\tilde{\theta}_2 \right) \right)^{-\ep}
\bigg( \f{E_{\mathrm{max}}}{r} \bigg)^{2\ep} \f{r_1^0\,r_2^0}{r \,
  \big( r_1^0 + r_2^0 + u_1^0\cos\theta_1 + u_2^0\cos\theta_2 \big)}
\; ,
\end{multline}
where
\be
\begin{gathered}
\label{eq:parameters20}
u_1^0 = E_{\mathrm{max}} \, \xi_1 \; , \quad u_2^0 = E_{\mathrm{max}}
\, \xi_1 \xi_2 \xi_{2 \, \mathrm{max}} \; , \quad
\xi_{2 \, \mathrm{max}} = \min \left[ 1 , \f{1}{\xi_1} \f{1 - \eta_1
    \xi_1 (2-\xi_1)}{1 - \eta_1 \xi_1} \right] \; , \\[0.2cm]
\cos\theta_1 = 1-2\eta_1 \; , \quad \cos\theta_2 =
1-2\eta_2\eta_{2\,\mathrm{max}} \; , \quad \eta_{2 \, \mathrm{max}} =
\min \left[ 1 , \f{1}{\xi_1} \f{1-\xi_1}{\xi_2 \xi_{2 \, \mathrm{max}}
    \big( 2 - \xi_1 ( 1 + \xi_2 \xi_{2 \, \mathrm{max}} ) \big) }
  \right] \; .
\end{gathered}
\ee
The factor in the last line of Eq.~(\ref{eq:dijet}) is regular in all
limits.

The behavior of amplitudes in collinear limits is characterized by
the transverse vectors
\be
u_{i\perp}^\mu = \begin{pmatrix} 0
  \\ \bm{\hat{u}_{i\perp}} \end{pmatrix} \; , \quad i = 1,2 \; ,
\ee
with
\begin{align}
\bm{\hat{u}_{1\perp}} = \lim_{\theta_1 \to 0}
\f{\bm{\hat{u}_1}-\bm{\hat{r}_1}}
{\lVert \bm{\hat{u}_1}-\bm{\hat{r}_1} \rVert} &= \left. 
\f{\partial \bm{\hat{u}_1}}{\partial \tilde{\theta}_1}
\right|_{\tilde{\theta}_1 = 0}
= \bm{\hat{R}}^{(3-2\ep)}_1(\alpha_1, \alpha_2, \dots)
\bm{\hat{n}}^{(3-2\ep)}\left(\f{\pi}{2}, \phi_1, \rho_1, \rho_2,
\dots\right)
\; ,
\\[0.2cm]
\bm{\hat{u}_{2\perp}} = \lim_{\theta_2 \to 0}
\f{\bm{\hat{u}_2}-\bm{\hat{r}_2}}
{\lVert \bm{\hat{u}_2}-\bm{\hat{r}_2} \rVert} &= \left. 
\f{\partial \bm{\hat{u}_2}}{\partial \tilde{\theta}_2}
\right|_{\tilde{\theta}_2 = 0}
= \bm{\hat{R}}^{(3-2\ep)}_1(\alpha_1, \alpha_2, \dots)
\bm{\hat{R}}^{(3-2\ep)}_4(\rho_2, \rho_3, \dots)
\bm{\hat{n}}^{(3-2\ep)}\left(\f{\pi}{2}, \phi_2, \sigma_1, \sigma_2,
\dots\right)
\; .
\end{align}
%

%%%%%%%%%%%%%%%%%%%%%%%%%%%%%%%%%%%%%%%%%%%%%%%%%%%%%%%%%%%%%%%%%%%%%%%%%%%%%%%%

\subsection{Angular integrations beyond four dimensions}

\noindent
The parameterizations we have presented involve $d$-dimensional
angular integrations. In practice, the number of relevant dimensions
depends on the number of vectors present in the problem. However, we
will later define the scheme in 't Hooft-Veltman regularization, in
which the resolved momenta (the final state momenta $q_i$, the
reference momenta $r_i$, and possibly up to two of the unresolved
momenta), are four-dimensional. Inspecting the explicit
parameterizations of the unresolved momenta from the previous
subsections, we notice that the scalar products amongst themselves
and with four-dimensional vectors involve parameters from at most two
additional dimensions. The angular integrations over the parameters,
which do not occur in the integrand can be performed explicitly. If
the integrand depends on four-dimensional parameters only, then we use
\be
\int_{\mathcal{S}_1^{-2\ep}} \mathrm{d}\bm{\Omega} \, 1 = 2
\f{(4\pi)^{-\ep}\Gamma(1-\ep)}{\Gamma(1-2\ep)} \; .
\ee
If the integrand depends on the parameters of a single unresolved
momentum, then
\begin{multline}
\int_{\mathcal{S}_1^{-2\ep}} \mathrm{d}\bm{\Omega}(\rho_1, \dots) =
\f{(4\pi)^{-\ep}\Gamma(1-\ep)}{\Gamma(1-2\ep)} \\
\times \int_{-1}^{+1} \mathrm{d}\cos\rho_1 \left( \delta(1-\cos\rho_1)
+ \delta(1 + \cos\rho_1) - 2 \ep \, \f{4^\ep
  \Gamma(1-2\ep)}{\Gamma^2(1-\ep)} \left[
  \f{1}{(1-\cos^2\rho_1)^{1+\ep}} \right]_+\right) \; .
\end{multline}
Finally, if the integrand depends on the parameters of two unresolved
momenta, then the integration over the parameters of the first is done
with the previous formula, while the integration over the parameters
of the second requires
\begin{multline}
\int_{\mathcal{S}_1^{-2\ep}} \mathrm{d}\bm{\Omega}(\sigma_1, \sigma_2,
\dots) = \f{(4\pi)^{-\ep}\Gamma(1-\ep)}{2\Gamma(1-2\ep)} \\
\begin{split}
\times
\int_{-1}^{+1} \mathrm{d}\cos\sigma_1\int_{-1}^{+1}
\mathrm{d}\cos\sigma_2 \, \Biggl(& \big( \delta(1-\cos\sigma_1) +
\delta(1 + \cos\sigma_1) \big) \big( \delta(1-\cos\sigma_2) +
\delta(1 + \cos\sigma_2) \big) \\ &
- 2 \ep \, \f{4^\ep\Gamma(1-2\ep)}{\Gamma^2(1-\ep)} \left[
  \f{1}{(1-\cos^2\sigma_1)^{1+\ep}} \right]_+ \big(
\delta(1-\cos\sigma_2) + \delta(1 + \cos\sigma_2) \big) \\ &
-\f{2+4\ep}{\pi}\left[
  \f{1}{(1-\cos^2\sigma_1)^{1+\ep}} \right]_+\left[
  \f{1}{(1-\cos^2\sigma_2)^{\f{3}{2}+\ep}} \right]_+
\Biggr) \; .
\end{split}
\end{multline}
The integrands contain the distribution
\be
\int_{-1}^{+1} \mathrm{d}\cos\rho \left[
  \f{1}{(1-\cos^2\rho)^\alpha} \right]_+ f(\cos\rho) =
\int_{-1}^{0} \mathrm{d}\cos\rho \,
\f{f(\cos\rho)-f(-1)}{(1-\cos^2\rho)^\alpha} 
+ \int_{0}^{+1} \mathrm{d}\cos\rho \,
\f{f(\cos\rho)-f(+1)}{(1-\cos^2\rho)^\alpha}
\; .
\ee
In the case of two-to-two processes, i.e.\ processes with $n=2$, there
is an additional simplification. Indeed, we can then assume that the
resolved momenta in the Born approximation are three-dimensional,
i.e. they have two-dimensional spatial components. This corresponds to
scattering on a plane. In such a case, the necessary dimension of
the unresolved momenta drops by one. This implies that for $n=2$,
five-dimensional unresolved momenta are sufficient.

%%%%%%%%%%%%%%%%%%%%%%%%%%%%%%%%%%%%%%%%%%%%%%%%%%%%%%%%%%%%%%%%%%%%%%%%%%%%%%%%

\section{Generation of subtraction and integrated subtraction terms}
\label{sec:subtraction}

\noindent
Using the decomposition of Section~\ref{sec:decomposition} and the
parameterizations of Section~\ref{sec:parameterization}, we have
succeeded in confining the singular phase space integrations to just
two variables in the $n+1$ case, and four variables in the $n+2$
case. For a single unresolved parton, the phase space measure contains
(see Eq.~\eqref{eq:unresolved1})
\be
\iint_0^1 \mathrm{d}\eta \, \mathrm{d}\xi \, \eta^{-\ep} \,
\xi^{1 - 2\ep} \; .
\ee
The collinear limit between the unresolved and the reference parton
is at $\eta = 0$. The relevant approximation of the tree-level matrix
element in this limit in terms of a factorization formula is given in
Eq.~\eqref{cfac}. This approximation also covers the soft-collinear
limit, where $\eta = \xi = 0$. The pure soft limit of vanishing
unresolved parton energy, $\xi = 0$, which is only singular for
gluons, is described by Eq.~\eqref{ccfact}. The behavior of the
matrix elements in the collinear and soft limits is at most as
singular as
\be
\f{1}{\eta} \f{1}{\xi^2} \; .
\ee
Tree-level matrix elements with a single unresolved parton are to be
found in $\hat{\sigma}^{\mathrm{R}}$ and
$\hat{\sigma}^{\mathrm{C1}}$. We will discuss the construction of the
subtraction scheme in the case of $\hat{\sigma}^{\mathrm{R}}$. The other
cases follow exactly the same pattern. We have
\be
\label{eq:sub1}
\hat{\sigma}^{\mathrm{R}} = \sum_{ik} \iint_0^1
\f{\mathrm{d}\eta}{\eta^{1+\ep}} \f{\mathrm{d}\xi}{\xi^{1+2\ep}}
f_{i,k}(\eta,\xi) \; ,
\ee
where we sum over the unresolved, $i$, and reference, $k$, partons.
$f_{i,k}(\eta,\xi)$ is independently regular at both $\eta = 0$ and
$\xi = 0$. Unless the unresolved parton is a gluon, it even vanishes
at $\xi = 0$. For example, in the case of an initial state reference
momentum, using Eqs.~\eqref{eq:unresolved0} and
\eqref{eq:unresolved1} with $z = 1$, we obtain
\be
f_{i,k}(\eta,\xi) = \f{E^2_{\mathrm{max}}}{16\pi^3 \hat{s} N_{ab}} \bigg(
\f{\pi \mR e^{\gamma_{\mathrm{E}}}}{4E^2_{\mathrm{max}}(1-\eta)} \bigg)^{\ep}
\int_{\mathcal{S}_1^{1-2\ep}}\mathrm{d}\bm{\Omega}(\phi,\rho_1,\dots)
\int \mathrm{d}\bm{\Phi}_n\big(p_1 + p_2 - u\big) \,
\mathcal{S}_{i,k} \, \Big[ \eta \, \xi^2 \la \cm^{(0)}_{n+1} |
  \cm^{(0)}_{n+1} \ra \Big] \, \mathrm{F}_{n+1} \; .
\ee
The phase space of the remaining partons,
$\mathrm{d}\bm{\Phi}_n\big(p_1 + p_2 - u\big)$, the selector function,
$\mathcal{S}_{i,k}$, the measurement function, $\mathrm{F}_{n+1}$, and the
matrix element, $\la \cm^{(0)}_{n+1} | \cm^{(0)}_{n+1} \ra$, depend on
$\eta$ and $\xi$ through the momentum, $u$, of the unresolved
parton. The Laurent expansion of $\hat{\sigma}^{\mathrm{R}}$ can be
derived using Eq.~\eqref{eq:sub1} and the master formula 
\be
\label{eq:master}
\f{1}{x^{1+a\ep}} = -\f{1}{a\ep} \delta(x) + \left[
  \f{1}{x^{1+a\ep}} \right]_+ \; ,
\ee
where $x$ is either $\eta$ or $\xi$. Since the limits commute, the
formula should be used recursively for both variables. Furthermore
\be
\int_0^1 \mathrm{d}x \, \left[ \f{1}{x^{1+a\ep}} \right]_+ f(x) =
\int_0^1 \mathrm{d}x \, \f{f(x)-f(0)}{x^{1+a\ep}} \; .
\ee
We will call the contribution of the delta-function, $\delta(x)$, the
{\it integrated subtraction term}, or the {\it pole term} in the
variable $x$. For this term, we will say that a {\it pole has been
  taken} in variable $x$. The end-point subtraction in the
plus-distribution will be called the {\it subtraction
term}. While the application of Eq.~\eqref{eq:master} indeed leads to
an explicit expansion in $\ep$ with numerically integrable
coefficients, the non-trivial part is the evaluation of the limits of
the matrix elements multiplied with $\eta \, \xi^2$. These are needed
to evaluate $f_{i,k}(\eta,0)$, $f_{i,k}(0,\xi)$ and $f_{i,k}(0,0)$,
and take the form
\be
\lim_{\eta \to 0} \Big[ \eta \, \xi^2 \la \cm^{(0)}_{n+1} |
  \cm^{(0)}_{n+1} \ra \Big] \; , \quad
\lim_{\xi \to 0} \Big[ \eta \, \xi^2 \la \cm^{(0)}_{n+1} |
  \cm^{(0)}_{n+1} \ra \Big] \; , \quad
\lim_{\eta \to 0} \lim_{\xi \to 0} \Big[ \eta \, \xi^2 \la
  \cm^{(0)}_{n+1} | \cm^{(0)}_{n+1} \ra \Big] \; . \quad
\ee
As we pointed out at the beginning of this section, these limits can
be obtained from factorization formulae, where the process dependent
information is contained in matrix elements with $n$ partons in
the final state, while the process independent information is to be
found in the splitting and soft functions. Therefore, in order to
implement the scheme in general, one only needs to determine the
limits of the splitting and soft functions. By construction, the
resulting integrals will be pointwise convergent, since the amplitude
limits were pointwise. These features will be present in the
subsequent more complicated cases to be found in this section.

The last contribution with $n+1$ final-state partons is
$\hat{\sigma}^{\mathrm{RV}}$. The factorization formula for one-loop
matrix elements present in $\hat{\sigma}^{\mathrm{RV}}$ in the
collinear limit can be found in \eqref{coll1LFull}, whereas the one
for the soft limit in \eqref{softlimit1L}. We notice that the scaling
of the matrix elements in these limits is not uniform. In the
collinear limit, there are two different terms proportional to
\be
\f{1}{\eta} \quad \text{and} \quad \f{1}{\eta^{1+\ep}} \; ,
\ee
respectively. The second scaling is an artifact of the virtual
integration. In the soft limit, on the other hand, we encounter
\be
\f{1}{\xi^2} \quad \text{and} \quad \f{1}{\xi^{2+2\ep}} \; .
\ee
The behavior in the soft-collinear limit is a combination of the two,
since soft and collinear limits commute. The possible scalings are
\be
\f{1}{\eta} \f{1}{\xi^2} \quad \text{and} \quad \f{1}{\eta^{1+\ep}}
\f{1}{\xi^{2+2\ep}} \; .
\ee
While we cannot simply use Eq.~\eqref{eq:master}, we only need a minor
modification. Suppose, that there is a function $f(x)$, which behaves
as
\be
f(x) \xrightarrow[x \to 0]{} f_0 + x^{-b\ep} f_\ep \; ,
\ee
then, we write
\be
\int_0^1 \f{\mathrm{d}x}{x^{1+a\ep}} f(x) = -\f{1}{a\ep} f_0 -
  \f{1}{(a+b)\ep} f_\ep + \int_0^1 \f{\mathrm{d}x}{x^{1+a\ep}} \big(
    f(x) - f_0 - x^{-b\ep} f_\ep \big) \; .
\ee
This is a generalization of Eq.~\eqref{eq:master} for
functions with non-uniform scaling. The formula should be used for
$\eta$ and $\xi$ independently. The order of the variables, in which
subtraction and integrated subtraction terms are introduced is
irrelevant. We define the terminology of subtraction and integrated
subtraction terms by analogy to the previous case.

The double-real radiation contribution, $\hat{\sigma}^{RR}$, contains
two unresolved partons. Eqs.~\eqref{eq:unresolved3},
\eqref{eq:unresolved2} and \eqref{eq:dijet} show that the phase space
always contains
\be
\iiiint_0^1 \mathrm{d}\eta_1 \mathrm{d}\eta_2 \mathrm{d}\xi_1
\mathrm{d}\xi_2  \, \eta_1^{a_1 - b_1 \ep} \, \eta_2^{a_2 - b_2 \ep} \,
\xi_1^{a_3 - b_3\ep} \, \xi_2^{a_4 - b_4\ep} \; .
\ee
The physical limits corresponding to the vanishing of the four
variables, $\eta_1, \eta_2, \xi_1$ and $\xi_2$, in the
triple-collinear parameterization can be determined using
Eq.~\eqref{eq:parameters3} and Tab.~\ref{tab:kinmap}. In the
double-collinear parameterization, it is necessary to use
Eq.~\eqref{eq:parameters2} and Eq.~\eqref{eq:parameters20}. The
required factorization formulae for the limits of tree-level matrix
elements can be found in \ref{sec:treelimits}. Multiplying the matrix
elements by appropriate powers of $\eta_1, \eta_2, \xi_1$ and $\xi_2$,
we can rewrite any contribution to $\hat{\sigma}^{RR}$ in the form
\be
\iiiint_0^1 \f{\mathrm{d}\eta_1}{\eta_1^{1 + b_1 \ep}}
\f{\mathrm{d}\eta_2}{\eta_2^{1 + b_2 \ep}}
\f{\mathrm{d}\xi_1}{\xi_1^{1 + b_3\ep}} \f{\mathrm{d}\xi_2}{\xi_2^{1 +
    b_4\ep}}  \, f(\eta_1, \eta_2, \xi_1, \xi_2)  \; .
\ee
The construction of the subtraction scheme amounts to the recursive
use of Eq.~\eqref{eq:master}. Notice that some of the
next-to-next-to-leading order limits are iterated next-to-leading
order limits. For example, if two different pairs of partons become
collinear, we would apply formula Eq.~\eqref{cfac} to each of the
pairs independently. Similarly, if two partons become collinear,
and another parton becomes soft, we would apply Eq.~\eqref{cfac} to
the collinear pair, and Eq.~\eqref{ccfact} to the soft parton. The
last limit, which can be obtained in this way, is when two
partons become collinear, and both of them become soft. This is a
single-collinear double-soft limit. We would first apply
Eq.~\eqref{cfac} to the collinear pair, and then we would take the
soft limit. The approximation to the matrix element is
\be
\label{eq:collinearsoft}
| \cm^{(0)}_{a_1,a_2,a_3\dots}(u_1,u_2,\dots) |^2 \simeq 
\big( 4 \pi \as \big)^2 \f{2}{s_{12}} \Ph^{(0)\,
  \mu\nu}_{a_1a_2}(z_{12},u_{3\perp};\ep) \, \la
\cm^{(0)}_{a_3,\dots}(p,\dots)| \bm{\mathrm{J}}_\mu(u_1+u_2)
\bm{\mathrm{J}}_\nu(u_1+u_2) |\cm^{(0)}_{a_3,\dots}(p,\dots)\ra \; ,
\ee
where $\{a_1,a_2\}$ is either $\{g,g\}$ or $\{q,\bar q\}$, and
\be
\bm{\mathrm{J}}^\mu(q) = \sum_i \cT_i \f{p_i^\mu}{p_i \cdot q} \; ,
\ee
is the soft current. 

The algorithm described in this section requires a multitude of matrix
elements. The complete list including contributions, which do not
necessitate subtraction is
\be
\begin{gathered}
\la \cm^{(0)}_{n} | \cm^{(0)}_{n} \ra \; , \quad
\la \cm^{(0)}_{n} | \cT_i \cdot \cT_j | \cm^{(0)}_{n} \ra \; , \quad
\la \cm^{(0)}_{n} | \lambda_i \ra \la \lambda'_i | \cm^{(0)}_{n} \ra
\; , \\[0.2cm]
\la \cm^{(0)}_{n} | \big\{ \cT_i \cdot \cT_j, \cT_k \cdot \cT_l \big\}
| \cm^{(0)}_{n} \ra \; , \quad
\la \cm^{(0)}_{n} | f^{abc} T^a_i T^b_j T^c_k |\cm^{(0)}_{n} \ra \; ,
\quad
\la \cm^{(0)}_{n} | \cT_i \cdot \cT_j | \lambda_k \ra \la \lambda'_k |
\cm^{(0)}_{n} \ra \; , \quad
\la \cm^{(0)}_{n} | \lambda_i \lambda_j \ra \la \lambda'_i \lambda'_j
| \cm^{(0)}_{n} \ra \; , \\[0.2cm]
\la \cm^{(0)}_{n+1} | \cm^{(0)}_{n+1} \ra \; , \quad
\la \cm^{(0)}_{n+1} | \cT_i \cdot \cT_j | \cm^{(0)}_{n+1} \ra \; ,
\quad
\la \cm^{(0)}_{n+1} | \lambda_i \ra \la \lambda'_i | \cm^{(0)}_{n+1}
\ra \; , \quad 
\la \cm^{(0)}_{n+2} | \cm^{(0)}_{n+2} \ra \; , \\[0.2cm]
\la \cm^{(0)}_{n} | \cm^{(1)}_{n} \ra \; , \quad
\la \cm^{(0)}_{n} | \cT_i \cdot \cT_j | \cm^{(1)}_{n} \ra \; , \quad
\la \cm^{(0)}_{n} | \lambda_i \ra \la \lambda'_i | \cm^{(1)}_{n} \ra
\; , \quad
\la \cm^{(0)}_{n+1} | \cm^{(1)}_{n+1} \ra \; , \\[0.2cm]
\la \cm^{(1)}_{n} | \cm^{(1)}_{n} \ra \; , \quad
\la \cm^{(0)}_{n} | \cm^{(2)}_{n} \ra \; .
\end{gathered}
\ee
The one- and two-loop matrix elements can be further decomposed into
divergent parts and finite remainders as shown in
\ref{sec:VirtualIR}. This, however, does not make the list any longer,
and amounts to a replacement of $| \cm_{n,n+1}^{(1,2)} \ra$ by $|
\cf_{n,n+1}^{(1,2)} \ra$. In the case of gluons, spin correlated
matrix elements are needed. Those are indicated by the presence of
\be
| \lambda_i \ra \la \lambda'_i | \; ,
\ee
which just means that the polarization of gluon $i$ is fixed to be
$\lambda_i$ in the matrix element and $\lambda'_i$ in the conjugated
matrix elements. The remaining polarizations are summed over. In case
there is
\be
| \lambda_i \lambda_j \ra \la \lambda'_i\lambda'_j | \; ,
\ee
the matrix element has a double spin correlation in gluons $i$ and
$j$. The spin correlators are due to the contraction of matrix
elements with transverse
vectors, which occur in the splitting functions,
e.g.\ in Eqs.~\eqref{hpggep} and \eqref{hpqqep}. Indeed, we can
decompose any transverse vector as
\be
\kper^\mu = - \sum_\lambda \big( \ep^*(k,\lambda) \cdot \kper \big) \,
\ep^\mu(k,\lambda) \, ,
\ee
as long as $\kper \cdot k = \kper \cdot \bar{k} = 0$, where
$\bar{k}^\mu = k_\mu$.

%%%%%%%%%%%%%%%%%%%%%%%%%%%%%%%%%%%%%%%%%%%%%%%%%%%%%%%%%%%%%%%%%%%%%%%%%%%%%%%%

\section{Average over azimuthal angles}
\label{sec:AzimuthalAverage}

\noindent
The construction of the subtraction scheme presented in the previous
sections relied, among others, on spin correlated splitting
functions. The latter were necessary to guarantee the pointwise
convergence of the numerical integration of the amplitudes and their
respective subtraction terms. Nevertheless, explicit divergences of
virtual amplitudes do not exhibit spin correlations. This suggests
that it should be possible to obtain the integrated subtraction terms
from azimuthally averaged splitting functions. There are two loopholes
in this argument. First, it might happen that the integrated
subtraction terms do not involve spin correlations in the coefficients
of the poles in $\ep$, while still containing them in the finite
parts. Second, the splitting functions might not be contracted with
matrix elements directly. In this section, we will demonstrate that
both of these cases indeed take place. We will also point out when
azimuthally averaged splitting functions may be safely used.

Let us first consider the single-collinear case described by the
parameterization of Section~\ref{sec:single-collinear}. In the
limit $\theta = 0$, besides the transverse vector $u_\perp$,
nothing depends on the azimuthal angles $\phi, \rho_1, \dots$. The
spin correlator can be evaluated explicitly with
\be
\left[ \int_{\mathcal{S}_1^{1-2\ep}} \mathrm{d}\bm{\Omega} \, 1
  \right]^{-1} \int_{\mathcal{S}_1^{1-2\ep}}
\mathrm{d}\bm{\Omega}(\phi, \rho_1, \rho_2, \dots) \, \f{u_\perp^\mu
  u_\perp^\nu}{u_\perp^2} = \f{1}{2(1-\ep)} \left( g^{\mu\nu} -
\f{r^\mu \bar{r}^\nu + r^\nu \bar{r}^\mu}{r \cdot \bar{r}} \right) \;
,
\ee
where $u_\perp$ is given in Eq.~(\ref{eq:uperp}), and $\bar{r}^\mu =
r_\mu$. This result follows from the vanishing of the $\mu, \nu = 0$
components, rotation invariance in the $1-2\ep$ dimensions
parameterized by $\phi, \rho_1, \dots$, orthogonality to $r$, and
normalization of the trace to unity. Transversality of the amplitudes
implies that we can simply replace the spin correlated splitting
function by its averaged counterpart. Similar arguments apply in the
double-collinear sector, Section~\ref{sec:double-collinear}, for
collinear limits of both unresolved momenta independently, and in the
triple-collinear sector, Section~\ref{sec:triple-collinear}, for the
triple-collinear limit, and for the collinear limit of the unresolved
momenta with respect to the reference momentum. The situation is
slightly complicated by the fact that the parameterization of the
second unresolved momentum depends on the first. In case the latter is
collinear to the reference momentum, we have to use rotation
invariance to decouple the momenta, and only then perform the
azimuthal average.

We will now be interested in the possibility to perform an azimuthal
average of the splitting function generating the collinear pole
related to the two unresolved momenta becoming collinear to each
other, but not to the reference momentum. This case occurs in sectors
$\mathcal{S}_4$ and $\mathcal{S}_5$ of the triple-collinear
parameterization. We need to consider the integration over $\phi_2$ in
the triple-collinear sector parameterization with a general parameter
$\zeta$, Eq.~(\ref{eq:azimuthalparameterization}). We thus study the
integration measure at $\theta_2 \approx 
\theta_1 \neq 0$. The relevant transverse vector $u_{3\perp}$,
Eq.~(\ref{eq:transverse3}), defines the azimuthal angle
$\tilde{\phi}_2$, Eq.~(\ref{eq:azimuthalangle}),
\be
\tan \tilde{\phi}^\pm_2(\theta_1, \zeta) = \pm \sin\theta_1 \,
\partial^+_{\theta_2}\phi_2(\theta_1, \zeta) \; , \quad
\tilde{\phi}^+_2 \in \left[0, \f{\pi}{2} \right[ \; , \quad
\tilde{\phi}^-_2 \in \left[\f{\pi}{2}, \pi \right[ \; .
\ee
The subsequent angular parameters are integrated over with the correct
$d$-dimensional measure. Thus, if the integration measure over
$\tilde{\phi}_2$, which is obtained by a non-linear remapping is
correct, we can replace the spin correlated splitting function by its
averaged counterpart. Let us proceed by expanding the relevant
expressions around $\theta_2 \approx \theta_1$, where
\be
 \phi_2(\theta_1, \theta_2, \zeta) =
 \partial^+_{\theta_2}\phi_2(\theta_1, \zeta) \, \big|
\theta_2 - \theta_1 \big| + \mathcal{O}\big( (\theta_2 - \theta_1)^2
\big) \; .
\ee
We first examine the contribution from the region $\theta_2
> \theta_1$. There is
\be
\int \mathrm{d}\phi_2 \sin^{-2\ep}\phi_2 = \int_0^{\pi/2}
\mathrm{d}\tilde{\phi}_2 \, \tan^{-2\ep}\tilde{\phi}_2 \left(
\f{\theta_2 - \theta_1}{\sin\theta_1} \right)^{1-2\ep}
\f{1}{\cos^2\tilde{\phi}_2} + \mathcal{O}\big( (\theta_2 - \theta_1)^2
\big) \; .
\ee
The singularity in the limit is generated by the invariant $s_{12}$
\be
\f{1}{s_{12}} =
\f{1}{2 \, u_1 \cdot u_2} =
\f{\cos^2\tilde{\phi}_2}{(\theta_2 - \theta_1)^2} \f{1}{u_1^0 u_2^0} +
\mathcal{O}\left( \f{1}{\theta_2 - \theta_1} \right) \; ,
\ee
which leads to
\be
\int \mathrm{d}\phi_2 \, \sin^{-2\ep}\phi_2 \, \f{1}{s_{12}} =
\int_0^{\pi/2} \mathrm{d}\tilde{\phi}_2 \,
\tan^{-2\ep}\tilde{\phi}_2\, \f{1}{\sin^2\theta_1} \,
\left( \f{\sin\theta_1}{\theta_2-\theta_1} \right)^{1+2\ep}
\f{1}{u_1^0 u_2^0} + \mathcal{O}\left( (\theta_2 - \theta_1)^0
\right)
\; .
\ee
This expression contains a regulated logarithmic singularity in
the integration over $\theta_2$ at $\theta_2 =
\theta_1$, which will result in a single pole in $\ep$. Once the
second region with $\theta_1 > \theta_2$ is added, the pole
contribution is proportional to the integral
\be
\int_0^{\pi} \mathrm{d}\tilde{\phi}_2 \, |\tan\tilde{\phi}_2|^{-2\ep}
\; .
\ee
Unfortunately, the azimuthal integration measure over $\tilde{\phi}_2$
contains $|\tan\tilde{\phi}_2|^{-2\ep}$ instead of
$\sin^{-2\ep}\tilde{\phi}_2$. In consequence, replacing the spin
correlated splitting function by the averaged one will only be correct
at $\ep = 0$. On the other hand, the integrals can still be performed
exactly
\begin{multline}
\left[ \int_0^{\pi} \mathrm{d}\tilde{\phi}_2 \,
  |\tan\tilde{\phi}_2|^{-2\ep} \int_{\mathcal{S}_1^{-2\ep}} \mathrm{d}
  \bm{\Omega} \, 1 \right]^{-1} \int_0^{\pi} \mathrm{d}\tilde{\phi}_2
\, |\tan\tilde{\phi}_2|^{-2\ep} \int_{\mathcal{S}_1^{-2\ep}}
\mathrm{d}\bm{\Omega}(\sigma_1, \sigma_2, \dots) \, \f{u_{3\perp}^\mu
  u_{3\perp}^\nu}{u_{3\perp}^2} = \\
\f{1}{2} \left( g^{\mu\nu} - \f{u_1^\mu \bar{u}_1^\nu + u_1^\nu
  \bar{u}_1^\mu}{u_1 \cdot \bar{u}_1} \right) - \ep \,
u_{3\perp}^\mu(\tilde{\phi}_2 = 0) \, u_{3\perp}^\nu(\tilde{\phi}_2 = 0)
\; ,
\end{multline}
where $u_{3\perp}$, given in Eq.~(\ref{eq:transverse3}), does not
depend on $\sigma_1, \sigma_2, \dots$ at $\tilde{\phi}_2 =
0$. Moreover, $\bar{u}_1^\mu = u_{1\, \mu}$, with $u_1^\mu$ given in
Eq.~(\ref{eq:tripleparameterization}). The result follows from the
vanishing of the $\mu,\nu = 0$ components, orthogonality to $u_1$,
rotation invariance in the space parameterized by $\sigma_1, \sigma_2,
\dots$, normalization of the trace to unity, and finally from the
explicit value of the integral when contracted with
$u_{3\perp}^\mu(\tilde{\phi}_2 = 0) \, u_{3\perp}^\nu(\tilde{\phi}_2 =
0)$. In practice, we can thus replace the spin correlated splitting
functions Eqs.~(\ref{hpggep}) and (\ref{hpqqep}) as follows
\begin{align}
\label{eq:skewedavg}
\Ph^{(0) \, \mu\nu}_{gg}(z,u_{3\perp};\ep) \, &\longrightarrow \,
-g^{\mu\nu} \left[ 2C_A \left( \f{z}{1-z} + \f{1-z}{z} + (1-\ep) z
  (1-z) \right) \right] + 4C_A \ep(1-\ep) z(1-z) \,
u_{3\perp}^\mu(\tilde{\phi}_2 = 0) \, u_{3\perp}^\nu(\tilde{\phi}_2 =
0) \; , \nn \\[0.2cm]
\Ph^{(0) \, \mu\nu}_{q\bar q}(z,u_{3\perp};\ep) \, &\longrightarrow \,
-g^{\mu\nu} \Big[ T_F \big( 1-2z(1-z) \big) \Big] - 4T_F \ep z(1-z) \,
u_{3\perp}^\mu(\tilde{\phi}_2 = 0) \, u_{3\perp}^\nu(\tilde{\phi}_2 =
0) \; .
\end{align}
The terms in the square brackets should be compared with the averaged
splitting functions Eqs.~(\ref{avhpgg}) and (\ref{avhpqq}). As
expected, the functions are different at order $\ep$.

This discussion is relevant first and foremost to the single-collinear
pole generated by $\eta_2$ in sector 4, and by $\eta_1$ in sector 5 of
the triple-collinear sector parameterization, as can be verified using
Tab.~\ref{tab:kinmap}. Clearly, there will be no spin correlations in
the coefficient of the pole itself, but the matrix element of the
finite part will be contracted with $u_{3\perp}(\tilde{\phi}_2 =
0)$. Note that the subtraction terms to this contribution in the
collinear limit of $u_1$ with respect to the reference momentum $r$
should be derived by using an iteration of splitting functions,
instead of the triple-collinear splitting function. The relevant
approximation of the azimuthally averaged matrix elements takes the
general form
\be
\label{eq:azimuthalcc}
\overline{| \cm^{(0)}_{a_r,a_1,a_2,\dots}(r,u_1,u_2,\dots) |^2} \simeq 
\f{\big( 8 \pi \as \big)^2}{s_{12} \, s_{r12}} \, \la
\sP^{(0)}_{a_1a_2}(z_{12};\ep) \ra \, \la \cm^{(0)}_{a,\dots}(p,\dots)|
\sP^{(0)}_{a_ra_{12}}(z_{r12},u_{1\perp};\ep)
|\cm^{(0)}_{a,\dots}(p,\dots)\ra \; ,
\ee
where the variables with subscript ``$12$'' describe the first limit
$u_1 || u_2$, whereas the variables with subscript ``$r12$'', the
second limit $r || u_1 + u_2$. If spin correlations are present in the
first limit, then
\begin{multline}
\overline{| \cm^{(0)}_{a_r,g,g,\dots}(r,u_1,u_2,\dots) |^2} \simeq \\
\f{\big( 8 \pi \as \big)^2}{s_{12} \, s_{r12}} \Biggl[ 2C_A \left(
  \f{z_{12}}{1-z_{12}} + \f{1-z_{12}}{z_{12}} + (1-\ep) z_{12}
  (1-z_{12}) \right)  \, \la \cm^{(0)}_{a_r,\dots}(p,\dots)|
  \sP^{(0)}_{a_rg}(z_{r12},u_{1\perp};\ep)
  |\cm^{(0)}_{a_r,\dots}(p,\dots)\ra \\
  + 4C_A \ep(1-\ep) z_{12}(1-z_{12}) \, \la \cm^{(0)}_{a_r,\dots}(p,\dots)|
  \sP^{(0)}_{Pa_rg}\big(z_{r12},u_{1\perp},u_{3\perp}^\mu(\tilde{\phi}_2 =
  0)\big) |\cm^{(0)}_{a_r,\dots}(p,\dots)\ra \Biggr]
\; ,
\end{multline}
for gluons, and similarly
\begin{multline}
\overline{| \cm^{(0)}_{a_r,q,\bar q,\dots}(r,u_1,u_2,\dots) |^2} \simeq \\
\f{\big( 8 \pi \as \big)^2}{s_{12} \, s_{r12}} \Biggl[ T_F \Big(
  1 - 2 z_{12} (1-z_{12}) \Big)  \, \la \cm^{(0)}_{a_r,\dots}(p,\dots)|
  \sP^{(0)}_{a_rg}(z_{r12},u_{1\perp};\ep)
  |\cm^{(0)}_{a_r,\dots}(p,\dots)\ra \\
  - 4T_F \ep z_{12}(1-z_{12}) \, \la \cm^{(0)}_{a_r,\dots}(p,\dots)|
  \sP^{(0)}_{Pa_rg}\big(z_{r12},u_{1\perp},u_{3\perp}^\mu(\tilde{\phi}_2
  = 0)\big) |\cm^{(0)}_{a_r,\dots}(p,\dots)\ra \Biggr]
\; ,
\end{multline}
for quarks. Both expressions contain the polarized splitting functions
Eqs.~(\ref{polhpgg}) and (\ref{polhpgq}).

The second contribution, which is affected by the non-trivial averages
of the spin correlated splitting functions, is the single-collinear
double-soft double pole, generated by the pair of variables
$\{\xi_2,\eta_2\}$ in sector 4, and $\{\xi_2,\eta_1\}$ in sector 5 of
the triple-collinear sector parameterization (see
Tab.~\ref{tab:kinmap}). In both cases, as long as $\xi_1$ is
non-vanishing, the collinear limit of the two unresolved partons will
generate spin correlations. The double-pole contribution can be
obtained from the matrix element factorization (see
Eq.~\eqref{eq:collinearsoft})
\be
\overline{| \cm^{(0)}_{a_1,a_2,a_3\dots}(u_1,u_2,\dots) |^2} \simeq 
\big( 4 \pi \as \big)^2 \f{2}{s_{12}} \Ph^{(0)\,
  \mu\nu}_{a_1a_2}(z_{12},u_{3\perp};\ep) \, \la
\cm^{(0)}_{a_3,\dots}(p,\dots)| \bm{\mathrm{J}}_\mu(u_1+u_2)
\bm{\mathrm{J}}_\nu(u_1+u_2) |\cm^{(0)}_{a_3,\dots}(p,\dots)\ra \; ,
\ee
The splitting functions should be replaced according to
Eq.~(\ref{eq:skewedavg}). In this case, the difference between the
correct replacement and the averaged splitting functions affects the
single pole. Thus, using averaged splitting functions would lead to an
incomplete cancellation of the divergences in the cross section. There
are no other instances, where Eq.~(\ref{eq:skewedavg}) must be used,
because all other subtraction and integrated subtraction terms
contributing to the cases just discussed do not involve spin
correlations.

%%%%%%%%%%%%%%%%%%%%%%%%%%%%%%%%%%%%%%%%%%%%%%%%%%%%%%%%%%%%%%%%%%%%%%%%%%%%%%%%

\section{Separation of finite contributions}
\label{sec:FiniteContributions}

\noindent
In order to define the subtraction scheme in 't Hooft-Veltman
regularization, which will be done in the next section, it is
necessary to understand which cross section contributions are
separately finite. Clearly, the leading order cross section is finite
and can be directly evaluated in four dimensions. At next-to-leading
order, the situation is slightly more complicated. From the three
contributions, $\hat{\sigma}^{\mathrm{R}}, \hat{\sigma}^{\mathrm{V}},
\hat{\sigma}^{\mathrm{C}}$, we can decompose the first two as follows
\be
\hat{\sigma}^{\mathrm{R}} = \hat{\sigma}^{\mathrm{R}}_{\mathrm{F}} +
\hat{\sigma}^{\mathrm{R}}_{\mathrm{U}} \; , \quad
\hat{\sigma}^{\mathrm{V}} = \hat{\sigma}^{\mathrm{V}}_{\mathrm{F}} +
\hat{\sigma}^{\mathrm{V}}_{\mathrm{U}} \; .
\ee
The subscript ``F'' stands for ``finite'' here and below, while
``U'' stands for ``unresolved''. For real radiation we have
\be
\hat{\sigma}^{\mathrm{R}}_{\mathrm{F}} =
\f{1}{2\hat{s}} \f{1}{N} \int \mathrm{d} \bm{\Phi}_{n+1} \, \Big[
  \la \cm_{n+1}^{(0)} | \cm_{n+1}^{(0)} \ra \, \mathrm{F}_{n+1} +
\text{subtraction terms} \Big] \; ,
\ee
with the notation of Section~\ref{sec:outline}. The subtraction
terms are generated using the algorithm of
Section~\ref{sec:subtraction}. Any integrated subtraction terms
generated along the way belong to
$\hat{\sigma}^{\mathrm{R}}_{\mathrm{U}}$. Thus,
$\hat{\sigma}^{\mathrm{R}}_{\mathrm{U}}$ only involves tree-level
matrix elements with $n$ final-state particles. For virtual
corrections we have
\be
\hat{\sigma}^{\mathrm{V}}_{\mathrm{F}} = \f{1}{2\hat{s}} \f{1}{N} \int
\mathrm{d} \bm{\Phi}_n \, 2 \R \, \la \cm_n^{(0)} | \cf_n^{(1)} \ra \,
\mathrm{F}_n \; , \quad
\hat{\sigma}^{\mathrm{V}}_{\mathrm{U}} = \f{1}{2\hat{s}} \f{1}{N} \int
\mathrm{d} \bm{\Phi}_n \, 2 \R \, \la \cm_n^{(0)} |
\bm{\mathrm{Z}}^{(1)} | \cm_n^{(0)} \ra \, \mathrm{F}_n \; ,
\ee
where the finite remainder $|\cf_n^{(1)} \ra$ and the singular
color-space operator $\bm{\mathrm{Z}}^{(1)}$ are defined in
\ref{sec:VirtualIR}. The finiteness of the next-to-leading order cross
section implies that the following contributions are separately finite
\be
\hat{\sigma}^{\mathrm{R}}_{\mathrm{F}} \; , \quad
\hat{\sigma}^{\mathrm{V}}_{\mathrm{F}} \; , \quad
\hat{\sigma}_{\mathrm{U}} =
\hat{\sigma}^{\mathrm{R}}_{\mathrm{U}} +
\hat{\sigma}^{\mathrm{V}}_{\mathrm{U}} +
\hat{\sigma}^{\mathrm{C}} \; .
\ee
At next-to-next-to-leading order, we begin by decomposing the
double-real radiation as follows
\be
\hat{\sigma}^{\mathrm{RR}} =
\hat{\sigma}^{\mathrm{RR}}_{\mathrm{F}} +
\hat{\sigma}^{\mathrm{RR}}_{\mathrm{SU}} +
\hat{\sigma}^{\mathrm{RR}}_{\mathrm{DU}} \; ,
\ee
where
\be
\hat{\sigma}^{\mathrm{RR}}_{\mathrm{F}} =
\f{1}{2\hat{s}} \f{1}{N} \int \mathrm{d} \bm{\Phi}_{n+2} \, \Big[
  \la \cm_{n+2}^{(0)} | \cm_{n+2}^{(0)} \ra \, \mathrm{F}_{n+2} +
\text{subtraction terms} \Big] \; .
\ee
The single-unresolved (SU) contribution,
$\hat{\sigma}^{\mathrm{RR}}_{\mathrm{SU}}$, contains all integrated
subtraction terms with $n+1$ resolved particles including the
necessary subtraction terms. The relevant pole terms are listed in
Tab.~\ref{tab:SU}. The double-unresolved (DU) contribution,
$\hat{\sigma}^{\mathrm{RR}}_{\mathrm{DU}}$, contains all integrated
subtraction terms with $n$ resolved particles including the necessary
subtraction terms. The relevant pole terms are those not listed in
Tab.~\ref{tab:SU}.
\begin{table}[t]
  \begin{center}
    \renewcommand{\arraystretch}{1.5}
    \begin{tabular}{cC{2cm}C{2cm}C{2cm}C{2cm}C{2cm}C{2cm}}
      \hline
      & ${\cal S}_1$ & ${\cal S}_2$ & ${\cal S}_3$ & ${\cal S}_4$ &
      ${\cal S}_5$ & double-collinear
      \\ \hline
      double-pole & $(\eta_2,\xi_2)$ & & & $(\eta_2,\xi_2)$ &
      $(\eta_1,\xi_2)$ & $(\eta_2,\xi_2)$ \\
      single-pole & $\eta_2,\xi_2$ & $\xi_2$ & $\eta_1$ &
      $\eta_2,\xi_2$ & $\eta_1,\xi_2$ & $\eta_1,\eta_2,\xi_2$
      \\ \hline
    \end{tabular}
  \end{center}
  \caption{\label{tab:SU} Single-unresolved contributions to
    double-real radiation, $\hat{\sigma}^{\mathrm{RR}}_{\mathrm{SU}}$.
    Listed are the variables leading to pole (integrated
    subtraction) terms after
    application of the algorithm of
    Section~\ref{sec:subtraction}. Each pole is accompanied by all
    possible subtraction terms. Sectors $\mathcal{S}_1, \dots,
    \mathcal{S}_5$ belong to the triple-collinear sector
    parameterization of Section~\ref{sec:triple-collinear}. The
    variables of the double-collinear sector parameterization are
    defined in Section~\ref{sec:double-collinear}.
  }
\end{table}
The real-virtual cross section can be decomposed as follows
\be
\hat{\sigma}^{\mathrm{RV}} = \hat{\sigma}^{\mathrm{RV}}_{\mathrm{F}} +
\hat{\sigma}^{\mathrm{RV}}_{\mathrm{SU}} +
\hat{\sigma}^{\mathrm{RV}}_{\mathrm{FR}} +
\hat{\sigma}^{\mathrm{RV}}_{\mathrm{DU}} \; ,
\ee
where
\begin{align}
\hat{\sigma}^{\mathrm{RV}}_{\mathrm{F}} &= \f{1}{2\hat{s}} \f{1}{N} \int
\mathrm{d} \bm{\Phi}_{n+1} \, \Big[ 2 \R \, \la \cm_{n+1}^{(0)} |
  \cf_{n+1}^{(1)} \ra \, \mathrm{F}_{n+1} + \text{subtraction terms}
  \Big] \; , \\[0.2cm]
\hat{\sigma}^{\mathrm{RV}}_{\mathrm{SU}} &= \f{1}{2\hat{s}} \f{1}{N} \int
\mathrm{d} \bm{\Phi}_{n+1} \, \Big[ 2 \R \, \la \cm_{n+1}^{(0)} |
\bm{\mathrm{Z}}^{(1)} | \cm_{n+1}^{(0)} \ra \, \mathrm{F}_{n+1} +
\text{subtraction terms} \Big] \; .
\end{align}
The limits required to generate the subtraction terms of $2 \R \, \la
\cm_{n+1}^{(0)} | \bm{\mathrm{Z}}^{(1)} | \cm_{n+1}^{(0)} \ra$ can be
found in \ref{sec:limitsZ1}, while those of $\la \cm_{n+1}^{(0)} |
  \cf_{n+1}^{(1)} \ra$ in \ref{sec:limitsF1}. The other two
contributions, $\hat{\sigma}^{\mathrm{RV}}_{\mathrm{FR}}$ and
$\hat{\sigma}^{\mathrm{RV}}_{\mathrm{DU}}$, contain all the
integrated subtraction terms. The latter are distributed such that
$\hat{\sigma}^{\mathrm{RV}}_{\mathrm{FR}}$ involves finite remainders (FR)
of one-loop amplitudes only, whereas
$\hat{\sigma}^{\mathrm{RV}}_{\mathrm{DU}}$ the left-over Born
matrix elements. Both types of matrix elements correspond to
amplitudes with $n$ final-state particles. In order to obtain the
relevant expressions, it is necessary to use the formulae of
\ref{sec:collinearM1} and \ref{sec:softM1}.
The double-virtual cross section is decomposed as follows
\be
\hat{\sigma}^{\mathrm{VV}} = \hat{\sigma}^{\mathrm{VV}}_{\mathrm{F}} +
\hat{\sigma}^{\mathrm{VV}}_{\mathrm{FR}} +
\hat{\sigma}^{\mathrm{VV}}_{\mathrm{DU}} \; ,
\ee
where
\begin{align}
\hat{\sigma}^{\mathrm{VV}}_{\mathrm{F}} &= \f{1}{2\hat{s}} \f{1}{N} \int
\mathrm{d} \bm{\Phi}_n \, \Big[ 2 \R \, \la \cm_n^{(0)} | \cf_n^{(2)}
  \ra + \la \cf_n^{(1)} | \cf_n^{(1)} \ra \Big] \,
\mathrm{F}_n \; , \\
\hat{\sigma}^{\mathrm{VV}}_{\mathrm{FR}} &= \f{1}{2\hat{s}} \f{1}{N} \int
\mathrm{d} \bm{\Phi}_n \, 2 \R \, \la \cm_n^{(0)} | \Big(
\bm{\mathrm{Z}}^{(1)\,\dagger} + \bm{\mathrm{Z}}^{(1)} \Big) |
\cf_n^{(1)} \ra \, \mathrm{F}_n \; , \\
\hat{\sigma}^{\mathrm{VV}}_{\mathrm{DU}} &= \f{1}{2\hat{s}} \f{1}{N} \int
\mathrm{d} \bm{\Phi}_n \, \Big[ 2 \R \, \la \cm_n^{(0)} |
  \bm{\mathrm{Z}}^{(2)} | \cm_n^{(0)} \ra + \la \cm_n^{(0)} |
\bm{\mathrm{Z}}^{(1)\,\dagger} \bm{\mathrm{Z}}^{(1)} |
\cm_n^{(0)} \ra \Big] \, \mathrm{F}_n \; .
\end{align}
Finally, the factorization contributions are decomposed as follows
\be
\hat{\sigma}^{\mathrm{C1}} = \hat{\sigma}^{\mathrm{C1}}_{\mathrm{SU}} +
\hat{\sigma}^{\mathrm{C1}}_{\mathrm{DU}} \; , \quad
\hat{\sigma}^{\mathrm{C2}} = \hat{\sigma}^{\mathrm{C2}}_{\mathrm{FR}} +
\hat{\sigma}^{\mathrm{C2}}_{\mathrm{DU}} \; ,
\ee
where $\hat{\sigma}^{\mathrm{C1}}_{\mathrm{SU}}$ is obtained from
Eq.~(\ref{eq:C1andC2}) by replacing $\hat{\sigma}^{\mathrm{R}}$ with
$\hat{\sigma}^{\mathrm{R}}_{\mathrm{F}}$, and
$\hat{\sigma}^{\mathrm{C1}}_{\mathrm{DU}}$ by replacing
$\hat{\sigma}^{\mathrm{R}}$ with
$\hat{\sigma}^{\mathrm{R}}_{\mathrm{U}}$. Similarly,
$\hat{\sigma}^{\mathrm{C2}}_{\mathrm{FR}}$ corresponds to keeping only
the virtual contribution with the replacement
$\hat{\sigma}^{\mathrm{V}} \to
\hat{\sigma}^{\mathrm{V}}_{\mathrm{F}}$, while the 
$\hat{\sigma}^{\mathrm{C2}}_{\mathrm{DU}}$ contains the rest of the
convolutions together with
$\hat{\sigma}^{\mathrm{V}} \to
\hat{\sigma}^{\mathrm{V}}_{\mathrm{U}}$.
Gathering the different contributions, we define
\be
\hat{\sigma}_{\mathrm{FR}} = \hat{\sigma}^{\mathrm{RV}}_{\mathrm{FR}}
+ \hat{\sigma}^{\mathrm{VV}}_{\mathrm{FR}} +
\hat{\sigma}^{\mathrm{C2}}_{\mathrm{FR}} \; , \quad
\hat{\sigma}_{\mathrm{SU}} = \hat{\sigma}^{\mathrm{RR}}_{\mathrm{SU}}
+ \hat{\sigma}^{\mathrm{RV}}_{\mathrm{SU}} +
\hat{\sigma}^{\mathrm{C1}}_{\mathrm{SU}} \; , \quad
\hat{\sigma}_{\mathrm{DU}} = \hat{\sigma}^{\mathrm{RR}}_{\mathrm{DU}}
+ \hat{\sigma}^{\mathrm{RV}}_{\mathrm{DU}} +
\hat{\sigma}^{\mathrm{VV}}_{\mathrm{DU}} + 
\hat{\sigma}^{\mathrm{C1}}_{\mathrm{DU}} + 
\hat{\sigma}^{\mathrm{C2}}_{\mathrm{DU}} \; .
\ee
The separately finite contributions are
\be
\hat{\sigma}^{\mathrm{RR}}_{\mathrm{F}} \; , \quad
\hat{\sigma}^{\mathrm{RV}}_{\mathrm{F}} \; , \quad
\hat{\sigma}^{\mathrm{VV}}_{\mathrm{F}} \; , \quad
\hat{\sigma}_{\mathrm{FR}} \; , \quad
\hat{\sigma}_{\mathrm{SU}} + \hat{\sigma}_{\mathrm{DU}} \; .
\ee
The finiteness of $\hat{\sigma}_{\mathrm{FR}}$ can be proven by
noticing that the sum $\hat{\sigma}_{\mathrm{FR}} +
\hat{\sigma}_{\mathrm{SU}} + \hat{\sigma}_{\mathrm{DU}}$ is finite by
the finiteness of the next-to-next-to-leading order cross section, and
the analytic structure of the matrix elements in
$\hat{\sigma}_{\mathrm{FR}}$ and $\hat{\sigma}_{\mathrm{SU}} +
  \hat{\sigma}_{\mathrm{DU}}$ is different. Indeed,
$\hat{\sigma}_{\mathrm{FR}}$ may have thresholds due to virtual
integrations, which cannot be present in $\hat{\sigma}_{\mathrm{SU}} +
\hat{\sigma}_{\mathrm{DU}}$, as the latter only involves tree-level
matrix elements. Thus, $\hat{\sigma}_{\mathrm{FR}}$ must be
separately finite. Another way to conduct the proof is to notice that
$\hat{\sigma}_{\mathrm{FR}}$ is generated from leading order splitting
and soft functions (compare with \ref{sec:collinearM1} and
\ref{sec:softM1}). It is thus finite by the finiteness of the
next-to-leading order cross section, since the unresolved
contributions to the latter are derived from exactly the same
splitting and soft functions \cite{Weinzierl:2011uz}. The finiteness
of $\hat{\sigma}_{\mathrm{FR}}$ implies, of course, the finiteness of
$\hat{\sigma}_{\mathrm{SU}} + \hat{\sigma}_{\mathrm{DU}}$.

Using a suitable measurement function, it is possible to obtain the
next-to-leading order cross section for $n+1$ well-separated partons
from a next-to-next-to-leading order cross section calculation for $n$
well-separated partons. This measurement function would satistify
$\mathrm{F}_n = 0$. It would set the contributions
$\hat{\sigma}^{\mathrm{VV}}_{\mathrm{F}}, 
\hat{\sigma}_{\mathrm{FR}}$, and $\hat{\sigma}_{\mathrm{DU}}$ to
zero. On the other hand, $\hat{\sigma}^{\mathrm{RR}}_{\mathrm{F}},
\hat{\sigma}^{\mathrm{RV}}_{\mathrm{F}}$ and
$\hat{\sigma}_{\mathrm{SU}}$ would correspond to
$\hat{\sigma}^{\mathrm{R}}_{\mathrm{F}},
\hat{\sigma}^{\mathrm{V}}_{\mathrm{F}}$ and
$\hat{\sigma}_{\mathrm{U}}$ as follows
\be
\hat{\sigma}^{\mathrm{RR}}_{\mathrm{F}} \to
\hat{\sigma}^{\mathrm{R}}_{\mathrm{F}} \; , \quad
\hat{\sigma}^{\mathrm{RV}}_{\mathrm{F}} \to
\hat{\sigma}^{\mathrm{V}}_{\mathrm{F}} \; , \quad
\hat{\sigma}_{\mathrm{SU}} \to \hat{\sigma}_{\mathrm{U}} \; .
\ee

At this point, there are two remaining contributions, the single- and
double-unresolved, which are not finite by themselves. Both involve
tree-level matrix elements only. Nevertheless, they have a different
physical interpretation. The single-unresolved contribution
corresponds to an inclusive phase space integral over
$n+1$-resolved-particles kinematics. The phase space singularities are
regulated with suitable subtraction terms. The double-unresolved
contribution, on the other hand, corresponds to a phase space integral
over $n$-resolved-particles kinematics. Due to appropriate separation
cuts inherent in the measurement function, there are no further phase
space singularities. For the purpose of a four-dimensional formulation
of the subtraction scheme, we must render both contributions
separately finite. Since their sum is finite, we can concentrate on
one of them. Once it is rendered finite by specifically designed
counterterms, it is sufficient to subtract the same counterterm from
the other. Clearly, the single-unresolved contribution is
substantially less complex than the double-unresolved. We will thus
proceed with its analysis. Once our task is finished, we will have the
separately finite contributions summarized in Tab.~\ref{tab:finite}.

\begin{table}[t]
  \begin{center}
    \renewcommand{\arraystretch}{1.5}
    \begin{tabular}{ll}
      \hline
      LO & $\hat{\sigma}^B$ \\
      NLO & $\hat{\sigma}^{\mathrm{R}}_{\mathrm{F}} \; , \quad
      \hat{\sigma}^{\mathrm{V}}_{\mathrm{F}} \; , \quad
      \hat{\sigma}_{\mathrm{U}} =
      \hat{\sigma}^{\mathrm{R}}_{\mathrm{U}} +
      \hat{\sigma}^{\mathrm{V}}_{\mathrm{U}} +
      \hat{\sigma}^{\mathrm{C}}$ \\
      NNLO & $\hat{\sigma}^{\mathrm{RR}}_{\mathrm{F}} \; , \quad
      \hat{\sigma}^{\mathrm{RV}}_{\mathrm{F}} \; , \quad
      \hat{\sigma}^{\mathrm{VV}}_{\mathrm{F}} \; , \quad
      \hat{\sigma}_{\mathrm{FR}} =
      \hat{\sigma}^{\mathrm{RV}}_{\mathrm{FR}} +
      \hat{\sigma}^{\mathrm{VV}}_{\mathrm{FR}} +
      \hat{\sigma}^{\mathrm{C2}}_{\mathrm{FR}} \; , $ \\ %\quad
      & $\hat{\sigma}_{\mathrm{SU}} =
      \hat{\sigma}^{\mathrm{RR}}_{\mathrm{SU}} +
      \hat{\sigma}^{\mathrm{RV}}_{\mathrm{SU}} +
      \hat{\sigma}^{\mathrm{C1}}_{\mathrm{SU}}\; , \quad
      \hat{\sigma}_{\mathrm{DU}} =
      \hat{\sigma}^{\mathrm{RR}}_{\mathrm{DU}} +
      \hat{\sigma}^{\mathrm{RV}}_{\mathrm{DU}} + 
      \hat{\sigma}^{\mathrm{VV}}_{\mathrm{DU}} + 
      \hat{\sigma}^{\mathrm{C1}}_{\mathrm{DU}} + 
      \hat{\sigma}^{\mathrm{C2}}_{\mathrm{DU}}$ 
      \\ \hline
    \end{tabular}
  \end{center}
  \caption{\label{tab:finite} Separately finite parts of a
    next-to-next-to-leading order cross section divided into: finite
    (F), unresolved (U), finite-remainder (FR), single-unresolved
    (SU), and double-unresolved (DU) contributions. Precise
    definitions can be found in the text.
  }
\end{table}

The single-unresolved contribution consists of three parts:
double-real radiation, $\hat{\sigma}^{\mathrm{RR}}_{\mathrm{SU}}$,
real-virtual radiation, $\hat{\sigma}^{\mathrm{RV}}_{\mathrm{SU}}$,
and factorization, $\hat{\sigma}^{\mathrm{C1}}_{\mathrm{SU}}$. In a
first step, we notice that if, instead of a next-to-next-to-leading
order measurement function, we would use a next-to-leading order one,
which would require $n+1$ well separated particles, the double-unresolved
contribution would vanish. Thus, the single-unresolved contribution
would be finite. Once we return to the original measurement
function, divergences do not cancel in the sum
$\hat{\sigma}^{\mathrm{RR}}_{\mathrm{SU}} +
\hat{\sigma}^{\mathrm{RV}}_{\mathrm{SU}} +
\hat{\sigma}^{\mathrm{C1}}_{\mathrm{SU}}$ anymore. This lack of
cancellation is, therefore, due to the subtraction terms. Let us now
concentrate on the real-virtual and double-real
corrections causing final state divergences. Neglecting irrelevant
integration variables, the essential part of any contribution to
$\hat{\sigma}^{\mathrm{RV}}_{\mathrm{SU}}$ in any phase space sector
has the following form in the parameterization of
Section~\ref{sec:single-collinear} (see Section~\ref{sec:subtraction})
\be
\label{eq:RVf}
\iint_0^1 \f{\mathrm{d}\eta}{\eta^{1+a\ep}}
\f{\mathrm{d}\xi}{\xi^{1+b\ep}} \, \Big( 
  f(\eta, \xi) - f(0, \xi) - f(\eta, 0) + f(0, 0) \Big) \; .
\ee
The function $f(\eta,\xi)$ is a product of a selector function, the
matrix element of the $\bm{\mathrm{Z}}^{(1)}$ operator, and the
measurement function. The terms with $\eta$ and/or $\xi$
vanishing are the subtraction terms. The essential point is that they
are integrated over the full range of variation of the resolved
particle kinematics given by $\eta$ and $\xi$, i.e.\ over the unit
interval.

Contributions to $\hat{\sigma}^{\mathrm{RR}}_{\mathrm{SU}}$ can be
classified according to the number of poles taken. Double-pole
contributions always correspond to the disappearance of the second
unresolved parton with momentum $u_2$, as can be checked by direct
inspection of the parameterizations of
Section~\ref{sec:parameterization}. They take the form
\be
\iint_0^1 \f{\mathrm{d}\eta_1}{\eta_1^{1+a\ep}}
\f{\mathrm{d}\xi_1}{\xi_1^{1+b\ep}} \, \Big( g(\eta_1, \xi_1) - g(0,
\xi_1) - g(\eta_1, 0) + g(0, 0) \Big) \; ,
\ee
where $g(\eta_1,\xi_1)$ is a product of a selector function, splitting
or soft function, matrix element, and the measurement
function. $\eta_1$ and $\xi_1$ can be identified directly with $\eta$
and $\xi$ from (\ref{eq:RVf}) due to the way they enter the matrix
element. $g(\eta_1,\xi_1)$ gives a contribution to the divergences of
$-f(\eta,\xi)$. In the same way, any of the subtraction terms
corresponding to vanishing $\eta_1$ and/or $\xi_1$ constitutes a
contribution to the subtraction terms of the divergences of $-f$. If
we could write all contributions to
$\hat{\sigma}^{\mathrm{RR}}_{\mathrm{SU}}$ in the same way, not only
the divergences of the integrals of $f(\eta,\xi)$ and
$g(\eta_1,\xi_1)$ would cancel, but the same would also be true of
their subtraction terms. Unfortunately, the situation is complicated
by the left-over integration present in the single-pole contributions
to double-real radiation. We have to deal with integrals of the form
\begin{multline}
\label{eq:RRg}
  \iiint_0^1 \f{\mathrm{d}y}{y^{1+a\ep}} \f{\mathrm{d}x_1}{x_1^{1+b_1\ep}}
  \f{\mathrm{d}x_2}{x_2^{1+b_2\ep}} \, \Biggl\{ \Biggl[
    \Big( g(y, x_1, x_2) - g(y, x_1, 0) \Big)
    - \Big( g(0, x_1, x_2) - g(0, x_1, 0) \Big) \Biggr] \\
  - \Biggl[ \Big( g(y, 0, x_2) - g(0, 0, x_2) \Big)
    - \Big( g(y, 0, 0)- g(0, 0, 0) \Big) \Biggr] \Biggr\} \; ,
\end{multline}
where $\{y,x_1,x_2\} \subset \{\eta_1, \eta_2, \xi_1, \xi_2\}$, and
the pole has been taken in the variable $\{\eta_1, \eta_2, \xi_1,
\xi_2\} \setminus \{y,x_1,x_2\}$. It can be checked, that in all
contributions, there is always one variable, which can be directly
identified with either $\eta$ or $\xi$ from (\ref{eq:RVf}). We denote
this variable by $y$. The second kinematic variable, denoted by $x$
(e.g.\ if $y$ corresponds to $\eta$, then $x$ corresponds to $\xi$), is
a function of $x_1$ and $x_2$
\be
x = x(x_1,x_2) \; .
\ee
Using the same arguments as in the double-pole case, one can convince
one-self that there is nothing to correct, if $x = x_1$ or $x =
x_2$. However, if the transformation between these variables is
non-linear, the divergences of the subtraction terms built of $g$ may
not correspond to the divergences of the subtraction terms built of
$f$. Alternatively, it may turn out that they are integrated over a
different range, i.e.\ not over the unit interval. In each of these
cases, we must introduce counterterms to compensate for the
difference. It is important that when $x \to 0$, the divergences of
the subtraction terms must constitute correct contributions to the
divergences of the subtraction terms of  $-f$. The reason is that the
unsubtracted integral gives a correct contribution to the divergences
of the integral of $-f(\eta,\xi)$, and the subtraction terms are
defined as its limits at vanishing $y$ and/or $x$. This allows to
derive the counterterms by inspection of the limit $x \to 0$. We now
discuss different cases separately. Initial state divergences will be
treated last.

\subsection{Case I}

\noindent
We assume that $x$ is related to $x_1, x_2$ by
\be
x = x_1 \big( 1 + c(x_1) x_2 \big) \; , \quad c'(0) = 0 \; ,
\ee
where $c$ is some function. Notice that it is then the difference
\be
g(y,x_1,x_2) - g(y,x_1,0) \; ,
\ee
in (\ref{eq:RRg}) that gives a contribution to the unsubtracted
divergences, since neither of the two terms vanishes in the
presence of a next-to-leading order measurement
function. Nevertheless, each of them has a different relation between
$x_1, x_2$ and $x$. Indeed, for $g(y,x_1,0)$ there is $x = x_1$. In
consequence, subtraction terms with $x_2 = 0$ will not require
counterterms. Let us then consider terms with a non-trivial dependence
on $x_2$ and define
\be
h(x_2) = - \Big( g(y,0,x_2) - g(0,0,x_2) \Big) \; .
\ee
Omitting the irrelevant integration over $y$, the subtraction terms
take the form
\be
\label{eq:caseI1}
  \iint_0^1 \f{\mathrm{d}x_1}{x_1^{1+b_1\ep}}
  \f{\mathrm{d}x_2}{x_2^{1+b_2\ep}} \, h(x_2) = \int_0^1
  \f{\mathrm{d}x}{x^{1+b_1\ep}} \int_0^{x_{2\, \mathrm{max}}(x)}
    \f{\mathrm{d}x_2}{x_2^{1+b_2\ep}} \, \Biggl[
    \f{\mathrm{d}x_1}{\mathrm{d}x}(x,x_2) \, \Big( \f{x}{x_1(x,x_2)}
    \Big)^{1+b_1\ep} \Biggr] \, h(x_2) \; ,
\ee
where
\be
x_{2\, \mathrm{max}}(0) = 1 \; .
\ee
If the second integral has any dependence on $x$, it must be
corrected. Such a dependence may be induced by a non-trivial
dependence on $x$ of $x_{2\, \mathrm{max}}(x)$ or of the integrand. As
discussed above, the correct behavior is defined by the limit $x \to
0$
\be
\lim_{x \to 0} \int_0^{x_{2\, \mathrm{max}}(x)}
\f{\mathrm{d}x_2}{x_2^{1+b_2\ep}} \, \Biggl[
  \f{\mathrm{d}x_1}{\mathrm{d}x}(x,x_2) \, \Big( \f{x}{x_1(x,x_2)}
  \Big)^{1+b_1\ep} \Biggr] \, h(x_2) = \int_0^1
\f{\mathrm{d}x_2}{x_2^{1+b_2\ep}} \, \big( 1 + c(0) x_2 \big)^{b_1\ep} \,
h(x_2) \; .
\ee
The counterterm to be added to the single-unresolved contributions is,
therefore
\be
\label{eq:caseI2}
\int_0^1 \f{\mathrm{d}x}{x^{1+b_1\ep}} \Biggl\{ \int_0^1
  \f{\mathrm{d}x_2}{x_2^{1+b_2\ep}} \, \big( 1 + c(0) x_2 \big)^{b_1\ep} \,
  h(x_2) -\int_0^{x_{2\, \mathrm{max}}(x)}
  \f{\mathrm{d}x_2}{x_2^{1+b_2\ep}} \, \Biggl[
    \f{\mathrm{d}x_1}{\mathrm{d}x}(x,x_2) \, \Big( \f{x}{x_1(x,x_2)}
    \Big)^{1+b_1\ep} \Biggr] \, h(x_2) \Biggr\} \; .
\ee
The integral over $x$ can be performed explicitly by a change of the
order of integration, with the result
\be
\int_0^1 \f{\mathrm{d}x_2}{x_2^{1+b_2\ep}} \, \Delta(x_2) \, h(x_2) \; .
\ee
Thus finally, the counterterm takes the form
\be
- \iint_0^1 \f{\mathrm{d}y}{y^{1+a\ep}}
\f{\mathrm{d}x_2}{x_2^{1+b_2\ep}} \, \Delta(x_2) \, \big( g(y, 0, x_2)
- g(0, 0, x_2) \big) \; .
\ee
The same counterterm will have to be subtracted from the
double-unresolved contributions in order to render them finite. Below,
we give specific values for three different cases. They depend
on the scaling of $x_2$, i.e.\ on the exponent $b_2$. We provide
the latter for CDR and HV regularizations. The latter is given in
anticipation of the discussion of the next section. Notice that
double-unresolved contributions will always use the CDR values, while
it will only be the single-unresolved contributions that will be
affected by the regularization scheme change.

\subsubsection{Triple-collinear parameterization sector
  $\mathcal{S}_4$, collinear pole in $\eta_2$}
\label{sec:caseIS4}

\noindent
The variable assignments are
\be
y = \eta_1 \; , \quad x_1 = \xi_1 \; , \quad x_2 = \xi_2 \; ,
\ee
with
\be
x = x_1 \big( 1 + x_2 \min( 1, 1/x_1 - 1 ) \big) \; .
\ee
In this case, the subtraction term (\ref{eq:caseI1}) takes the form
\be
\int_0^1 \f{\mathrm{d}x}{x^{1+b_1\ep}} \int_{\max(0, 2x-1)}^1
\f{\mathrm{d}x_2}{x_2^{1+b_2\ep}} \, ( 1 + x_2 )^{b_1\ep}
h(x_2) + \int_{1/2}^1 \mathrm{d}x \int_0^{2x-1}
\f{\mathrm{d}x_2}{x_2^{1+b_2\ep}} \f{( 1 - x_2 )^{b_1\ep}}{( x -
  x_2 )^{1+b_1\ep}} \, h(x_2) \; .
\ee
We notice that the integrand of the $x_2$ integration depends on $x$
in the second term, while the integration range depends on $x$ in both
terms. This leads to the necessity of compensation. The counterterm
(\ref{eq:caseI2}) is
\begin{multline}
\int_{1/2}^1 \mathrm{d}x \int_0^{2x-1}
\f{\mathrm{d}x_2}{x_2^{1+b_2\ep}} \, \Biggl[ \f{( 1 + x_2
  )^{b_1\ep}}{x^{1+b_1\ep}} - \f{( 1 - x_2 )^{b_1\ep}}{( x - x_2
  )^{1+b_1\ep}} \Biggr] \, h(x_2) = \\
\int_0^1 \f{\mathrm{d}x_2}{x_2^{1+b_2\ep}} \int_{(1+x_2)/2}^1
\mathrm{d}x \, \Biggl[ \f{( 1 + x_2 )^{b_1\ep}}{x^{1+b_1\ep}} - \f{(
    1 - x_2 )^{b_1\ep}}{( x - x_2 )^{1+b_1\ep}} \Biggr] \, h(x_2)
= \\ \int_0^1 \f{\mathrm{d}x_2}{x_2^{1+b_2\ep}} \, \Biggl[ \f{1}{b_1\ep}
  \Big( 1 - ( 1 + x_2 )^{b_1\ep} \Big) \Biggr] \, h(x_2)
\; .
\end{multline}
Returning to the original variables, the counterterm is parameterized
by
\be
\Delta^{\eta_2}_{\mathcal{S}_4} = \f{1 - ( 1 + \xi_2 )^{b_1\ep}}{b_1\ep}
\; ,
\ee
with the scaling
\be
b_1^{\mathrm{CDR}} = 4 \; , \quad b_1^{\mathrm{HV}} = 2 \; .
\ee

\subsubsection{Triple-collinear parameterization sector
  $\mathcal{S}_5$, collinear pole in $\eta_1$}
\label{sec:caseIS5}

\noindent
This case is identical to the previous upon the interchange $\eta_1
\leftrightarrow \eta_2$. Thus
\be
\Delta^{\eta_1}_{\mathcal{S}_5} = \Delta^{\eta_2}_{\mathcal{S}_4} \; .
\ee

\subsubsection{Triple-collinear parameterization sector
  $\mathcal{S}_5$, soft pole in $\xi_2$}
\label{sec:caseIS5soft}

\noindent
The variable assignments are
\be
y = \xi_1 \; , \quad x_1 = \eta_2 \; , \quad x_2 = \eta_1 \; ,
\ee
with
\be
x = x_1 \Big( 1 - \f{x_2}{2} \Big) \; .
\ee
The subtraction term (\ref{eq:caseI1}) takes the form
\be
\int_0^1 \f{\mathrm{d}x}{x^{1+b_1\ep}} \int_0^{\min(1, 2-2x)}
\f{\mathrm{d}x_2}{x_2^{1+b_2\ep}} \, \Big( 1 - \f{x_2}{2}
\Big)^{b_1\ep} h(x_2) \; .
\ee
We notice that, while the integrand of the second integral is
independent of $x$, the integration range is not. In order to
compensate for this fact, we introduce the counterterm
\begin{multline}
\int_{1/2}^1 \f{\mathrm{d}x}{x^{1+b_1\ep}}
\int_{2-2x}^1 \f{\mathrm{d}x_2}{x_2^{1+b_2\ep}} \, \Big( 1 -
\f{x_2}{2} \Big)^{b_1\ep} h(x_2) = \\
\int_0^1 \f{\mathrm{d}x_2}{x_2^{1+b_2\ep}} \int_{1-x_2/2}^1
\f{\mathrm{d}x}{x^{1+b_1\ep}} \, \Big( 1 - \f{x_2}{2} \Big)^{b_1\ep}
h(x_2) = \\
\int_0^1 \f{\mathrm{d}x_2}{x_2^{1+b_2\ep}} \Biggl[ \f{1}{b_1 \ep}
  \Biggl( 1 - \Big( 1 - \f{x_2}{2} \Big)^{b_1\ep} \Biggr) \Biggr] \,
h(x_2) \; .
\end{multline}
Returning to the original variables, we obtain
\be
\Delta^{\xi_2}_{\mathcal{S}_5} = \f{1 - \Big( 1 - \f{\eta_1}{2}
  \Big)^{b_1\ep}}{b_1\ep} \; ,
\ee
with the scaling
\be
b_1^{\mathrm{CDR}} = 2 \; , \quad b_1^{\mathrm{HV}} = 1 \; .
\ee

\subsection{Case II: triple-collinear parameterization sector
  $\mathcal{S}_2$, soft pole in $\xi_2$}
\label{sec:caseII}

\noindent
The variable assignments are
\be
y = \xi_1 \; , \quad x_1 = \eta_1 \; , \quad x_2 = \eta_2 \; .
\ee
The reason for separate treatment is that the resolved parton momentum
is parameterized in a symmetric way by the sector variables
\be
x = \f{1}{2} x_1 x_2 \; .
\ee
It is, therefore, only $g(y,x_1,x_2)$ that contributes to the
unsubtracted divergences. The symmetry requires to consider
counterterms with a dependence on either variable, $x_1$ and $x_2$. We
treat explicitly the $x_2$ case with a generic function $h(x_2)$. The
relevant contribution to the subtraction term can be rewritten as
\be
\label{eq:S2}
\iint_0^1 \f{\mathrm{d}x_1}{x_1^{1+b_1\ep}}
\f{\mathrm{d}x_2}{x_2^{1+b_2\ep}} \, h(x_2) = \int_0^{1/2}
\f{\mathrm{d}x}{x^{1+b_1\ep}} \int_{2x}^1
  \f{\mathrm{d}x_2}{x_2^{1+b_2\ep}} \, \Big( \f{x_2}{2} \Big)^{b_1\ep}
  \, h(x_2) \; .
\ee
We notice that the range of the resolved parameter $x$ is restricted
to $[0,1/2]$ although it should be $[0,1]$. At the same time the
integration over $x_2$ has an $x$ dependent range. We first determine
the correct behavior at $x \to 0$. This can be achieved by writing the
right-hand side of Eq.~(\ref{eq:S2}) as
\be
 \int_0^{1/2} \f{\mathrm{d}x}{x^{1+b_1\ep}} \, \Biggl[
    \int_{2x}^1 \f{\mathrm{d}x_2}{x_2^{1+b_2\ep}} \, \Big( \f{x_2}{2}
    \Big)^{b_1\ep} \, h(0) + \int_0^1
    \f{\mathrm{d}x_2}{x_2^{1+b_2\ep}} \, \Big( \f{x_2}{2}
    \Big)^{b_1\ep} \, \big( h(x_2) - h(0) \big) - \int_0^{2x}
    \f{\mathrm{d}x_2}{x_2^{1+b_2\ep}} \, \Big( \f{x_2}{2}
    \Big)^{b_1\ep} \, \big( h(x_2) - h(0) \big) \Biggr] \; .
\ee
The first two terms in the square bracket have a uniform scaling in
$x$ as required of subtraction contributions matching the divergences
of the real-virtual single-unresolved cross section. The last term
must be removed by a counterterm. It is also necessary to extend the
integration range of $x$. In consequence, the complete counterterm has
the form
\be
\label{eq:S22}
\int_{1/2}^1 \f{\mathrm{d}x}{x^{1+b_1\ep}} \, \Biggl[ \int_{2x}^1
  \f{\mathrm{d}x_2}{x_2^{1+b_2\ep}} \, \Big( \f{x_2}{2} \Big)^{b_1\ep}
  \, h(0) + \int_0^1 \f{\mathrm{d}x_2}{x_2^{1+b_2\ep}} \, \Big(
  \f{x_2}{2} \Big)^{b_1\ep} \, \big( h(x_2) - h(0) \big) \Biggr] \, +
\int_0^{1/2} \f{\mathrm{d}x}{x^{1+b_1\ep}} \int_0^{2x}
\f{\mathrm{d}x_2}{x_2^{1+b_2\ep}} \, \Big( \f{x_2}{2} \Big)^{b_1\ep}
\, \big( h(x_2) - h(0) \big) \; .
\ee
The first term in the square bracket can be integrated over both
variables with the result
\be
\int_{1/2}^1 \f{\mathrm{d}x}{x^{1+b_1\ep}} \int_{2x}^1
\f{\mathrm{d}x_2}{x_2^{1+b_2\ep}} \, \Big( \f{x_2}{2} \Big)^{b_1\ep}
\, h(0) = \f{1}{b_1 b_2 \ep^2} \f{\big( 2^{-b_2\ep} - 1 \big) \, b_1 -
  \big( 2^{-b_1\ep} - 1 \big) \, b_2 }{b_1 - b_2} h(0) \; .
\ee
The second and third terms in (\ref{eq:S22}) can be combined together
with the result
\begin{multline}
\int_{1/2}^1 \f{\mathrm{d}x}{x^{1+b_1\ep}} \int_0^1
\f{\mathrm{d}x_2}{x_2^{1+b_2\ep}} \, \Big( \f{x_2}{2} \Big)^{b_1\ep}
\, \big( h(x_2) - h(0) \big) + \int_0^{1/2}
\f{\mathrm{d}x}{x^{1+b_1\ep}} \int_0^{2x}
\f{\mathrm{d}x_2}{x_2^{1+b_2\ep}} \, \Big( \f{x_2}{2} \Big)^{b_1\ep}
\, \big( h(x_2) - h(0) \big) = \\
\int_0^1 \f{\mathrm{d}x_2}{x_2^{1+b_2\ep}} \int_{x_2/2}^1
\f{\mathrm{d}x}{x^{1+b_1\ep}} \, \Big( \f{x_2}{2} \Big)^{b_1\ep}
\, \big( h(x_2) - h(0) \big) = \int_0^1
\f{\mathrm{d}x_2}{x_2^{1+b_2\ep}} \, \Biggl[ \f{1}{b_1
    \ep} \Biggl( 1 - \Big( \f{x_2}{2} \Big)^{b_1 \ep} \Biggr) \Biggr] \,
\big( h(x_2) - h(0) \big) \; .
\end{multline}
Using the above result and symmetry of the expressions with respect to
the interchange $x_1 \leftrightarrow x_2$, we obtain the complete
counterterm for sector $\mathcal{S}_2$ expressed through the original
variables
\be
\begin{split}
- \f{1}{b_1 b_2 \ep^2} &\f{\big( 2^{-b_2\ep} - 1 \big) \, b_1 - \big(
  2^{-b_1\ep} - 1 \big) \, b_2}{b_1 - b_2} \int_0^1
\f{\mathrm{d}\xi_1}{\xi_1^{1+a\ep}} \, \big( g(\xi_1, 0, 0) - g(0, 0,
0) \big) \\ - &\iint_0^1 \f{\mathrm{d}\xi_1}{\xi_1^{1+a\ep}}
\f{\mathrm{d}\eta_1}{\eta_1^{1+b_1\ep}}
\f{1 - \big( \f{\eta_1}{2} \big)^{b_2\ep}}{b_2\ep} \big( g(\xi_1,
\eta_1, 0) - g(\xi_1, 0, 0) - g(0, \eta_1, 0) + g(0, 0, 0) \big)
\\ - &\iint_0^1 \f{\mathrm{d}\xi_1}{\xi_1^{1+a\ep}}
\f{\mathrm{d}\eta_2}{\eta_2^{1+b_2\ep}}
\f{1 - \big( \f{\eta_2}{2} \big)^{b_1\ep}}{b_1\ep} \big( g(\xi_1, 0,
\eta_2) - g(\xi_1, 0, 0) - g(0, 0, \eta_2) + g(0, 0, 0) \big) \; ,
\end{split}
\ee
with the scaling
\be
b_1^{\mathrm{CDR}} = 3 \; , \quad
b_2^{\mathrm{CDR}} = 2 \; , \quad
b_1^{\mathrm{HV}} = 2 \; , \quad
b_2^{\mathrm{HV}} = 1 \; .
\ee

\subsection{Case III: double-collinear parameterization and
  triple-collinear parameterization sector $\mathcal{S}_3$, collinear
  pole in $\eta_1$}
\label{sec:caseIII}

\noindent
The variable assignments are
\be
y = \eta_2 \; , \quad x_1 = \xi_2 \; , \quad x_2 = \xi_1 \; .
\ee
This case is different from those treated until now, because at
$\eta_1 = 0$ the unresolved parton with momentum $u_1$ is collinear to
the reference parton, which is assumed to be in the final state. In
the notation of Section~\ref{sec:parameterization}, the reference
momentum is $r_1$ in the double-collinear parameterization, and $r$ in
the triple-collinear parameterization. Here, we will denote it by
$r$ irrespective of the case. As far as the double-collinear
parameterization is concerned, we will treat the general case of
subsection~\ref{sec:double-collinear-general} throughout, and only
return to the special case of
subsection~\ref{sec:double-collinear-special} at
the end. Since both partons, the unresolved and
reference, are moving in the same direction, the reference momentum
for comparison with the real-virtual corrections is not $r$, but
rather
\be
r' = r + u_1 \; .
\ee
Let us introduce rescaled variables for the energy of the reference
parton: $\xi_r$ for the original one, and $\xi_{r'}$ for the composite
one. Taking into account the different variation ranges, we write
\be
r^0 = E_{\mathrm{max}} \, \xi_r \, \xi_{r\,\mathrm{max}} \; , \quad
r^{\prime 0} = E_{\mathrm{max}} \, \xi_{r'} \,
\xi_{r'\,\mathrm{max}} \; ,
\ee
where $\xi_{r\,\mathrm{max}}$ denotes the maximum of
$r^0/E_{\mathrm{max}}$, and similarly for $\xi_{r'\,\mathrm{max}}$. By
definition, $\xi_r, \xi_{r'} \in [0,1]$. The variation range of
$r^{\prime 0}$ is the same as in the respective real-virtual
contributions. Since the comparison between
$\hat{\sigma}^{\mathrm{RV}}_{\mathrm{SU}}$ and
$\hat{\sigma}^{\mathrm{RR}}_{\mathrm{SU}}$ involves the reference
momentum, we define a new function $h$ through
\be
g(y,x_1,x_2) = \int_0^1 \mathrm{d}\xi_r \, h(\xi_r,y,x_1,x_2) \; .
\ee
By inspection of the limits of the double-real radiation matrix
element, we note that
\be
\label{eq:h0}
h(\xi_r,y,x_1,0) - h(\xi_r,y,0,0) = 0 \; .
\ee
Therefore, we only need to consider the subtraction terms related to
the functions
\be
h(\xi_r,y,0,x_2) \; , \quad h(\xi_r,0,0,x_2) \; .
\ee
The result for the correction to the second one must be the limit at
$y = 0$ of the result for the correction to the first one. Thus, we
concentrate on $h(\xi_r,y,0,x_2)$. Due to the special kinematics
\be
\xi_{r\,\mathrm{max}} \big|_{x_1 = 0} = 1 - x_2 \; , \quad
\xi_{r'\,\mathrm{max}} \big|_{x_1 = 0} = 1 \; .
\ee
We can now specify the relationship between the variables of the
double-real contribution, and those of the real-virtual contribution
\be
x = x_1 x_2 x_{2\,\mathrm{max}}(x_2) \; , \quad \xi_{r'} = \xi_r +
(1-\xi_r) x_2 \; ,
\ee
where
\be
x_{2\,\mathrm{max}}(x_2) = \min \left[ 1, \, \f{1}{x_2}
  \f{1-x_2}{1-\f{E_{\mathrm{max}}}{\sqrt{\hat{s}}} \, \big( \hat{u}_1
    \cdot \hat{u}_2 \big) \, x_2} \right] \; .
\ee
Furthermore, we introduce
\be
\bar{z} = \f{x_2}{\xi_r + (1-\xi_r) x_2} \; .
\ee
The variable $\bar{z}$ is integrated over in order to obtain a
contribution, which cancels the poles of the matrix element of
$\bm{\mathrm{Z}}^{(1)}$ present  in
$\hat{\sigma}^{\mathrm{RV}}_{\mathrm{SU}}$ for fixed kinematics
specified by $x,y$ and $\xi_{r'}$. The soft limit in the integration
over the phase space of the unresolved parton with momentum $u_1$
corresponds to $\bar{z} = 0$ (this is the reason for the bar in the
notation, since usually the soft limit is at $z = 1$). The function
$x_{2\,\mathrm{max}}$ depends on $y$ through the scalar product
$\hat{u}_1 \cdot \hat{u}_2$. In the triple-collinear case, this
dependence is simple, since $\hat{u}_1 \cdot \hat{u}_2 = 2y$. On the
other hand, the dependence in the double-collinear case is only
indirect and involves the angles of the other reference momentum. For
convenience, we define
\be
y' = \f{E_{\mathrm{max}}}{\sqrt{\hat{s}}} \, \big( \hat{u}_1 \cdot
\hat{u}_2 \big) \in [0,1] \; , \quad x_{\mathrm{max}} =
\f{1}{1+\sqrt{1-y'}} \; .
\ee
It turns out that
\be
\label{eq:range3}
x_2 \in [0, x_{\mathrm{max}}] \; \Rightarrow \; x_{2\,\mathrm{max}}(x_2)
= 1 \; , \quad x_2 \in [x_{\mathrm{max}}, 1] \; \Rightarrow \;
x_{2\,\mathrm{max}}(x_2) = \f{1}{x_2} \f{1-x_2}{1-y' x_2} \; .
\ee
The integral of the subtraction term takes the form
\begin{multline}
\iiiint_0^1 \mathrm{d}\xi_r \f{\mathrm{d}y}{y^{1+a\ep}}
\f{\mathrm{d}x_1}{x_1^{1+b_1\ep}} \f{\mathrm{d}x_2}{x_2^{1+b_2\ep}} \,
h(\xi_r,y,0,x_2) = \\ \iiint_0^1 \f{\mathrm{d}y}{y^{1+a\ep}} \,
\mathrm{d}\xi_{r'} \f{\mathrm{d}\bar{z}}{\bar{z}^{1+(b_2-b_1)\ep}}
\int_0^{\xi_{r'}\bar{z}\,x_{2\,\mathrm{max}}(\xi_{r'}\bar{z})}
\f{\mathrm{d}x}{x^{1+b_1\ep}} \, \xi_{r'}^{(b_1-b_2)\ep} \big(
x_{2\,\mathrm{max}}(\xi_{r'}\bar{z}) \big)^{b_1\ep} \,
\f{1}{1-\xi_{r'}\bar{z}} \, h\bigg(
\f{\xi_{r'}(1-\bar{z})}{1-\xi_{r'}\bar{z}}, y, 0, \xi_{r'}\bar{z}
\bigg) = \\
 \iiint_0^1 \f{\mathrm{d}y}{y^{1+a\ep}} \,
\mathrm{d}\xi_{r'} \f{\mathrm{d}\bar{z}}{\bar{z}^{1+(b_2-b_1)\ep}}
\int_0^{\xi_{r'}\bar{z}\,x_{2\,\mathrm{max}}(\xi_{r'}\bar{z})}
\f{\mathrm{d}x}{x^{1+b_1\ep}} \, \tilde{h}(\xi_{r'},y,\bar{z})
\; ,
\end{multline}
where we have introduced a shorthand notation, $\tilde{h}$, for the
integrand.
To derive the behavior of the subtraction term at small $x$, we must
work with the appropriate order of the integration variables. For now,
we shall neglect $y$, which is irrelevant to this problem. In view of
the parameterization of the contributions to
$\hat{\sigma}^{\mathrm{RV}}_{\mathrm{SU}}$, the integration order must be
\be
\int \mathrm{d}x \int \mathrm{d}\xi_{r'} \int \mathrm{d}\bar{z} \; .
\ee
Using Eq.~\eqref{eq:range3}, we obtain
\begin{multline}
\label{eq:case32}
\iint_0^1 \mathrm{d}\xi_{r'}
\f{\mathrm{d}\bar{z}}{\bar{z}^{1+(b_2-b_1)\ep}}
\int_0^{\xi_{r'}\bar{z}\,x_{2\,\mathrm{max}}(\xi_{r'}\bar{z})}
\f{\mathrm{d}x}{x^{1+b_1\ep}} \, \tilde{h}(\xi_{r'},y,\bar{z}) = \\
\int_0^{x_{\mathrm{max}}} \f{\mathrm{d}x}{x^{1+b_1\ep}} \, \Bigg[
  \int_x^1 \mathrm{d}\xi_{r'}
  \int_{x/\xi_{r'}}^{\min(1,x_{\mathrm{max}}/\xi_{r'})}
  \f{\mathrm{d}\bar{z}}{\bar{z}^{1+(b_2-b_1)\ep}} +
  \int_{x_{\mathrm{max}}}^1 \mathrm{d}\xi_{r'}
  \int_{x_{\mathrm{max}}/\xi_{r'}}^{\min\big(1, \f{1}{\xi_{r'}} \f{1 -
      x}{1 - y' x}\big)}
  \f{\mathrm{d}\bar{z}}{\bar{z}^{1+(b_2-b_1)\ep}} \Bigg]
\, \tilde{h}(\xi_{r'},y,\bar{z}) \; .
\end{multline}
The correct subtraction term matching the poles of the respective
subtraction term of the real-virtual contribution, is obtained from
this expression by taking the limit $x \to 0$ of the square bracket,
and extending the integration range over $x$. It reads
\begin{multline}
\label{eq:case32corr}
\int_0^1 \f{\mathrm{d}x}{x^{1+b_1\ep}} \int_0^1 \mathrm{d}\xi_{r'}
\, \Bigg[
  \int_{x/\xi_{r'}}^{\min(1,x_{\mathrm{max}}/\xi_{r'})}
  \f{\mathrm{d}\bar{z}}{\bar{z}^{1+(b_2-b_1)\ep}} \,
  \tilde{h}(\xi_{r'},y,0) \\ +
  \int_0^{\min(1,x_{\mathrm{max}}/\xi_{r'})}
  \f{\mathrm{d}\bar{z}}{\bar{z}^{1+(b_2-b_1)\ep}} \,
  \Big( \tilde{h}(\xi_{r'},y,\bar{z}) - \tilde{h}(\xi_{r'},y,0)
  \Big) + \int_{\min(1,x_{\mathrm{max}}/\xi_{r'})}^1
  \f{\mathrm{d}\bar{z}}{\bar{z}^{1+(b_2-b_1)\ep}} \,
  \tilde{h}(\xi_{r'},y,\bar{z}) \Bigg]
 \; ,
\end{multline}
where we have made use of the fact that a potential singularity at
$\xi_{r'} = 0$ is regulated by the selector function. The counterterm
we are seeking is the difference between Eq.~\eqref{eq:case32corr} and
Eq.~\eqref{eq:case32}. The contribution containing
$\tilde{h}(\xi_{r'},y,\bar{z})$ is easily obtained by returning to the
original order of integration variables, and noticing that it is only
necessary to extend the integration range over $x$
\be
\iint_0^1 \mathrm{d}\xi_{r'}
\f{\mathrm{d}\bar{z}}{\bar{z}^{1+(b_2-b_1)\ep}}
\int_{\xi_{r'}\bar{z}\,x_{2\,\mathrm{max}}(\xi_{r'}\bar{z})}^1
\f{\mathrm{d}x}{x^{1+b_1\ep}} \, \tilde{h}(\xi_{r'},y,\bar{z}) =
\\ \iint_0^1 \mathrm{d}\xi_r \f{\mathrm{d}x_2}{x_2^{1+b_2\ep}} \,
\Bigg[ \f{1 - \big( x_2 x_{2\,\mathrm{max}}(x_2)
    \big)^{b_1\ep}}{b_1\ep} \Bigg] \, h(\xi_r,y,0,x_2) \; .
\ee
This expression is not integrable at $\ep = 0$, but we can extract the
explicit pole in $\ep$ by a subtraction
\begin{multline}
\label{eq:case32p1}
\iint_0^1 \mathrm{d}\xi_r \f{\mathrm{d}x_2}{x_2^{1+b_2\ep}} \, \Bigg[
  \f{1 - \big( x_2 x_{2\,\mathrm{max}}(x_2) \big)^{b_1\ep}}{b_1\ep}
  \Bigg] \, h(\xi_r,y,0,x_2) = \f{1}{(b_2-b_1)b_2\ep^2} \int_0^1
\mathrm{d}\xi_r\, h(\xi_r,y,0,0) \\ + \iint_0^1 \mathrm{d}\xi_r
\f{\mathrm{d}x_2}{x_2^{1+b_2\ep}} \, \f{1}{b_1\ep} \Bigg[ \Big( 1 -
  \big( x_2 x_{2\,\mathrm{max}}(x_2) \big)^{b_1\ep} \Big) \,
  h(\xi_r,y,0,x_2) - ( 1 - x_2^{b_1\ep} ) \, h(\xi_r,y,0,0) \Big]
\; .
\end{multline}
The second contribution to the counterterm is obtained from
Eq.~\eqref{eq:case32corr}, by taking only the terms involving
$\tilde{h}(\xi_r,y,0,0)$
\be
\label{eq:case32p2}
\int_0^1 \f{\mathrm{d}x}{x^{1+b_1\ep}} \int_0^1 \mathrm{d}\xi_{r'}
\int_{x/\xi_{r'}}^0 \f{\mathrm{d}\bar{z}}{\bar{z}^{1+(b_2-b_1)\ep}} \,
  \tilde{h}(\xi_{r'},y,0) = -\f{1}{(b_2-b_1)b_2\ep^2} \int_0^1
\mathrm{d}\xi_r\, h(\xi_r,y,0,0) \; .
\ee
As expected, the divergence cancels in the sum of
Eqs.~\eqref{eq:case32p1} and \eqref{eq:case32p2}. The counterterm is
thus finite and integrable. Returning to the original variables, the
counterterm to be added to the single-unresolved double-real radiation
contribution, and subtracted from the double-unresolved contribution is
\begin{multline}
- \iint_0^1 \f{\mathrm{d}\eta_2}{\eta_2^{1+a\ep}}
\f{\mathrm{d}\xi_1}{\xi_1^{1+b_2\ep}} \, \f{1}{b_1\ep}
\Big[ \Big( 1 - \big( \xi_1 \xi_{2\,\mathrm{max}}(\xi_1,\eta_2)
  \big)^{b_1\ep} \Big) \, g(\eta_2,0,\xi_1) -\big( 1 - \xi_1^{b_1\ep}
  \big) \, g(\eta_2,0,0) \\ - \Big( 1 - \big( \xi_1
  \xi_{2\,\mathrm{max}}(\xi_1,0) \big)^{b_1\ep} \Big) \, g(0,0,\xi_1)
  + \big( 1 - \xi_1^{b_1\ep} \big) \, g(0,0,0) \Big] \; ,
\end{multline}
with the scaling
\be
b_1^{\mathrm{CDR}} = 3 \; , \quad b_1^{\mathrm{HV}} = 1 \; .
\ee
Finally, we note that the counterterm for the special case of the
double-collinear parameterization of
subsection~\ref{sec:double-collinear-special} is the same as above
with the replacement
\be
\xi_{2\,\mathrm{max}} = 1 \; .
\ee

\subsection{Case IV: initial state divergences}
\label{sec:caseIV}

\noindent
The corrections needed in the case of initial state divergences have
their origin in the same pole as Case III. We will thus consider the
{\it collinear pole in $\eta_1$ in the double-collinear
parameterization, and in sector $\mathcal{S}_3$ of the
triple-collinear parameterization}. However, the reference momentum
will now be in the initial state. Furthermore, we must compare the
contributions to $\hat{\sigma}^{\mathrm{RR}}_{\mathrm{SU}}$ with those to
$\hat{\sigma}^{\mathrm{C1}}_{\mathrm{SU}}$, and not those to
$\hat{\sigma}^{\mathrm{RV}}_{\mathrm{SU}}$. Instead of
Eq.~\eqref{eq:RVf}, we have
\be
\iiint_0^1 \mathrm{d}z \f{\mathrm{d}\eta}{\eta^{1+a\ep}}
\f{\mathrm{d}\xi}{\xi^{1+b\ep}} \, \Big( 
f(z, \eta, \xi) - f(z, 0, \xi) - f(z, \eta, 0) + f(z, 0, 0) \Big) \; ,
\ee
where $z$ is the variable used in the convolution with the splitting
functions in Eq.~\eqref{eq:C1andC2}, and the latter have been included
in $f$. In order to facilitate the comparison, we assume that the
kinematics in the calculation of
$\hat{\sigma}^{\mathrm{C1}}_{\mathrm{SU}}$ is the same as in
$\hat{\sigma}^{\mathrm{RR}}_{\mathrm{SU}}$, which means that $z$ is
defined by the emission of a fictitious parton from the initial state,
such that the parton momentum entering the matrix element is $z r$,
$r$ being the same reference momentum as for the unresolved parton
with momentum $u_1$ in $\hat{\sigma}^{\mathrm{RR}}_{\mathrm{SU}}$ (one
of the initial state momenta by assumption). Of course,
$\hat{\sigma}^{\mathrm{C1}}$ is boost invariant,  and we can evaluate
it in any system we like. The question whether its partial
contribution $\hat{\sigma}^{\mathrm{C1}}_{\mathrm{SU}}$ also has this
property will be discussed at the end of this subsection. The relevant
contribution to $\hat{\sigma}^{\mathrm{RR}}_{\mathrm{SU}}$ has the
form of Eq.~\eqref{eq:RRg}. We rewrite it through the original
variables of the contribution
\begin{multline}
  \iiint_0^1 \f{\mathrm{d}\eta_2}{\eta_2^{1+a\ep}}
  \f{\mathrm{d}\xi_1}{\xi_1^{1+b_1\ep}}
  \f{\mathrm{d}\xi_2}{\xi_2^{1+b_2\ep}} \, \Biggl\{ \Biggl[ \Big(
    g(\eta_2, \xi_1, \xi_2) - g(\eta_2, \xi_1, 0) \Big) - \Big( g(0,
    \xi_1, \xi_2) - g(0, \xi_1, 0) \Big) \Biggr] \\
  - \Biggl[ \Big( g(\eta_2, 0, \xi_2) - g(0, 0, \xi_2) \Big)
    - \Big( g(\eta_2, 0, 0)- g(0, 0, 0) \Big) \Biggr] \Biggr\} \; .
\end{multline}
$\xi_1$ and $\xi_2$ are related to $z$ and $\xi$ through
\be
z = 1 - \f{2E_{\mathrm{max}}}{\sqrt{\hat{s}}} \xi_1 \; , \quad
\xi = \f{\xi_{2\,\mathrm{max}}}{\xi'_{2 \, \mathrm{max}}} \xi_2 \; ,
\quad \xi_{2 \, \mathrm{max}} = \min \left[ 1, \, \xi'_{2 \,
    \mathrm{max}} \right] \; , \quad \xi'_{2 \, \mathrm{max}} =
\f{1}{\xi_1} \f{1-\xi_1}{1-\f{E_{\mathrm{max}}}{\sqrt{\hat{s}}} \,
  \xi_1 \, \hat{r} \cdot \hat{u}_2} \; ,
\ee
Just as in Case III (compare to Eq.~\eqref{eq:h0}), we note that
\be
g(\eta_2,0,\xi_2) - g(\eta_2,0,0) = 0 \; .
\ee
We thus only need to consider
\be
g(\eta_2,\xi_1,0) \; , \quad g(0,\xi_1,0) \; .
\ee
We will work with $g(\eta_2,\xi_1,0)$, as the correction to $g(0,\xi_1,0)$
can be derived as the limit to that of $g(\eta_2,\xi_1,0)$. Neglecting
the irrelevant integration over $\eta_2$, the integral of the
subtraction term is
\begin{multline}
\label{eq:subini}
\iint_0^1 \f{\mathrm{d}\xi_1}{\xi_1^{1+b_1\ep}}
\f{\mathrm{d}\xi_2}{\xi_2^{1+b_2\ep}} \, g(\eta_2, \xi_1, 0) = \\
\bigg( \f{2E_{\mathrm{max}}}{\sqrt{\hat{s}}} \bigg)^{(b_1-b_2)\ep}
\int_{1-2E_{\mathrm{max}}/\sqrt{\hat{s}}}^1
\f{\mathrm{d}z}{(1-z)^{1+(b_1-b_2)\ep}}
\int_0^{\xi_{2\,\mathrm{max}}/\xi'_{2\,\mathrm{max}}}
\f{\mathrm{d}\xi}{\xi^{1+b_2\ep}} \, \bigg(
\f{\xi_{2\,\mathrm{max}}}{\xi'_{2\,\mathrm{max}}} \bigg)^{b_2\ep} \,
  g( \eta_2, \xi_1(z), 0) \; .
\end{multline}
The necessary counterterm amounts to the extension of the integration
range over $\xi$. It is thus
\begin{multline}
\bigg( \f{2E_{\mathrm{max}}}{\sqrt{\hat{s}}} \bigg)^{(b_1-b_2)\ep}
\int_{1-2E_{\mathrm{max}}/\sqrt{\hat{s}}}^1
\f{\mathrm{d}z}{(1-z)^{1+(b_1-b_2)\ep}}
\int_{\xi_{2\,\mathrm{max}}/\xi'_{2\,\mathrm{max}}}^1
\f{\mathrm{d}\xi}{\xi^{1+b_2\ep}} \, \bigg(
\f{\xi_{2\,\mathrm{max}}}{\xi'_{2\,\mathrm{max}}} \bigg)^{b_2\ep} \,
g( \eta_2, \xi_1(z), 0) = \\ \iint_0^1
\f{\mathrm{d}\xi_1}{\xi_1^{1+b_1\ep}} \, \Bigg[ \f{1}{b_2\ep} \Bigg( 1
  - \bigg( \f{\xi_{2\,\mathrm{max}}}{\xi'_{2\,\mathrm{max}}} \bigg)^{b_2\ep}
    \Bigg) \Bigg] \, g(\eta_2, \xi_1, 0) \; .
\end{multline}
As it stands, this correction is not integrable in $\xi_1$ if
$g(\eta_2,0,0) \neq 0$. The reason is that
$\xi_{2\,\mathrm{max}}/\xi'_{2\,\mathrm{max}} = \xi_1 +
\mathcal{O}(\xi_1^2)$. The singularity at $\xi_1 = 0$ corresponds to
$z = 1$. There is a subtraction at $z = 1$ in the
convolution with the splitting functions in the factorization
contribution, $\hat{\sigma}^{\mathrm{C1}}_{\mathrm{SU}}$. In the
double-real contributions we are considering, an analoguous
subtraction is not necessary, since the integration range of $\xi$
vanishes at $z = 1$, as seen in Eq.~\eqref{eq:subini}. The
counterterm we have derived reintroduces the singularity at the
endpoint. This singularity must be subtracted in order to match the
factorization contributions. This amounts to replacing
\be
\f{1}{(1-z)^{1+c\ep}} \, \longrightarrow \, \bigg[
  \f{1}{(1-z)^{1+c\ep}} \bigg]_+ \; .
\ee
We thus finally arrive at the complete counterterm
\begin{multline}
-\iiint_0^1 \f{\mathrm{d}\eta_2}{\eta_2^{1+a\ep}}
\f{\mathrm{d}\xi_1}{\xi_1^{1+b_1\ep}} \, \f{1}{b_2\ep} \Bigg[ \Bigg( 1
  - \bigg(
  \f{\xi_{2\,\mathrm{max}}(\eta_2,\xi_1)}{\xi'_{2\,\mathrm{max}}(\eta_2,\xi_1)}
  \bigg)^{b_2\ep} \Bigg) \, g(\eta_2, \xi_1, 0) - \big( 1 -
  \xi_1^{b_2\ep} \big) \, g(\eta_2, 0, 0) \\ - \Bigg( 1 - \bigg(
  \f{\xi_{2\,\mathrm{max}}(0,\xi_1)}{\xi'_{2\,\mathrm{max}}(0,\xi_1)}
  \bigg)^{b_2\ep} \Bigg) \, g(0, \xi_1, 0) + \big( 1 -
  \xi_1^{b_2\ep} \big) \, g(0, 0, 0) \Bigg] \; ,
\end{multline}
with the scaling
\be
b_2^{\mathrm{CDR}} = 3 \; , \quad b_2^{\mathrm{HV}} = 1 \; .
\ee
We have derived the counterterm for this case by comparing the
single-unresolved double-real radiation contribution to the
single-unresolved factorization contribution in a boosted
frame. Unfortunately, the separation into single- and
double-unresolved contributions is not Lorentz invariant, because it
is based on taking poles in energy and angle variables. In the next
section, we will manipulate differently both types of
contributions. In consequence, the construction will only be correct,
if the factorization contributions,
$\hat{\sigma}^{\mathrm{C1}}_{\mathrm{SU}}$ and
$\hat{\sigma}^{\mathrm{C1}}_{\mathrm{DU}}$, will be evaluated in the
boosted frame as assumed here. This is the reason for the necessity of
the unusual parameterization of Section~\ref{sec:single-collinear}.

%%%%%%%%%%%%%%%%%%%%%%%%%%%%%%%%%%%%%%%%%%%%%%%%%%%%%%%%%%%%%%%%%%%%%%%%%%%%%%%%

\section{'t Hooft-Veltman regularization of separately finite contributions}
\label{sec:HVregularization}

\noindent
The final stage of our construction is the introduction of the 't
Hooft-Veltman regularization. The latter differs from the conventional
dimensional regularization in the description of the resolved
partons. In HV, their momenta and polarizations are four-dimensional,
while in CDR they are $d$-dimensional. The differences in the number
of spin degrees-of-freedom only affect the gluons. In HV, tree-level
matrix elements do not have any expansion in $\ep$. Indeed, the $\ep$
dependence is due to spin sums over squared matrix elements. If these
sums are only restricted to four-dimensional degrees-of-freedom, there
can be no dependence on $\ep$. Due to the virtual integrations, one-
and two-loop matrix elements do have a non-trivial dependence on
$\ep$. Nevertheless, our goal will be to only work with the first
term of the $\ep$-expansion of the finite remainders. These can also
be viewed as four-dimensional loop corrections. We note that the two
features of HV regularization, four-dimensional momenta and
four-dimensional polarizations, are quite different. Since our
subtraction scheme makes extensive use of phase space
parameterizations, the restriction to four-dimensional momenta will be
achieved by modifying the phase spaces. On the other hand, the
restriction to four-dimensional polarizations amounts to the removal
of higher order terms of the $\ep$-expansion of the matrix
elements. We now consider the different contributions described in
Section~\ref{sec:FiniteContributions} in increasing level of
complexity.

The simplest contributions do not involve any singularities in
$\ep$. These are
\be
\hat{\sigma}^B \; , \quad
\hat{\sigma}^{\mathrm{R}}_{\mathrm{F}} \; , \quad
\hat{\sigma}^{\mathrm{V}}_{\mathrm{F}} \; , \quad
\hat{\sigma}^{\mathrm{RR}}_{\mathrm{F}} \; , \quad
\hat{\sigma}^{\mathrm{RV}}_{\mathrm{F}} \; , \quad
\hat{\sigma}^{\mathrm{VV}}_{\mathrm{F}} \; .
\ee
We can simply set $\ep = 0$ in the phase space integrals and in the
matrix elements. We thus directly obtain HV regularized
contributions, as if the calculation were done completely in
four-dimensions without ever making reference to dimensional
regularization.

The second group of contributions only has $n$ resolved partons. These
are
\be
\hat{\sigma}_{\mathrm{U}} =
      \hat{\sigma}^{\mathrm{R}}_{\mathrm{U}} +
      \hat{\sigma}^{\mathrm{V}}_{\mathrm{U}} +
      \hat{\sigma}^{\mathrm{C}} \; , \quad
\hat{\sigma}_{\mathrm{FR}} =
      \hat{\sigma}^{\mathrm{RV}}_{\mathrm{FR}} +
      \hat{\sigma}^{\mathrm{VV}}_{\mathrm{FR}} +
      \hat{\sigma}^{\mathrm{C2}}_{\mathrm{FR}} \; , \quad
\hat{\sigma}_{\mathrm{DU}} =
      \hat{\sigma}^{\mathrm{RR}}_{\mathrm{DU}} +
      \hat{\sigma}^{\mathrm{RV}}_{\mathrm{DU}} + 
      \hat{\sigma}^{\mathrm{VV}}_{\mathrm{DU}} + 
      \hat{\sigma}^{\mathrm{C1}}_{\mathrm{DU}} + 
      \hat{\sigma}^{\mathrm{C2}}_{\mathrm{DU}} \; .
\ee
We first note that since these contributions are separately finite, we
can drop the higher order terms in the $\ep$-expansion of the matrix
elements. Indeed, the cancellation of divergences is due to the
unresolved phase space integrals of the soft and splitting functions,
and the form of the divergences of the virtual amplitudes, which are
contained in the $\bm{\mathrm{Z}}^{(1,2)}$ operators. The cancellation
is not related to any particular functional dependence of the
amplitudes on the kinematics. Thus, it occurs separately for any order
of the expansion in $\ep$ of the matrix elements. However, while the
leading order will give a finite contribution, the $\ep$-suppressed
subleading orders will give vanishing contributions in the limit
$\ep \to 0$. In consequence, we can remove them from the very
beginning. We stress that this mechanism only works after azimuthal
averages have been performed. Otherwise, spin correlators would mix
different orders of the $\ep$-expansion, and at least the finite parts
of the results would not be correct. We point out that the
cancellation of divergences happens even at the level of independent
color correlated amplitudes. We do not see, however, much use of this
fact for our purposes, besides testing of course.

At this point, we have four-dimensional polarizations. We now note
that the contributions under consideration are finite for any
measurement function, as long as it is infrared safe. $\mathrm{F}_n$
is the only measurement function present, and we can use it to
restrict the resolved momenta to be four-dimensional with the
replacement
\be
\label{eq:restriction}
\mathrm{F}_n \, \longrightarrow \, \mathrm{F}_n \, \Biggl( \f{\mu_R^2
  e^{\gamma_{\rm{E}}}}{4\pi} \Biggr)^{-(n-1)\ep} \Bigg[ \prod_{i=1}^{n-1}
(2\pi)^{-2\ep} \delta^{(-2\ep)}(q_i) \Bigg] \; ,
\ee
where $q_i$ are the momenta of the final-state resolved partons, and
the delta-functions remove their $\ep$-dimensional components. We note
that it is only possible to explicitly restrict the momenta of all but
one of the final-state particles. This is due to momentum
conservation. On the other  hand, the same momentum conservation will
always provide a restriction on the  last final-state momentum. Thus,
the result is independent of the subset of momenta chosen in
\eqref{eq:restriction}. In the simplest case, where there are either no 
reference momenta or they are in the initial state, the replacement
amounts to the following change of the resolved parton phase spaces
\be
\label{eq:remainder}
\int \mathrm{d}\bm{\Phi}_n\Big( p_1 + p_2 \to \sum_{i=1}^n q_i \Big)
\, \longrightarrow \, \int \prod_{i=1}^n \f{\mathrm{d}^3q_i}{(2\pi)^32q_i^0} \,
(2\pi)^4\delta^{(4)}\Big( \sum_{i=1}^n q_i - p_1 - p_2 \Big) \; .
\ee
In other words, the calculation can be performed with a
four-dimensional phase space for the resolved partons. The situation
is more complicated, if reference momenta are in the final state. The
rule \eqref{eq:remainder} is still valid for all the final-state
particles not explicitely parameterized in the given sector. There
will appear, however, delta-functions of the form
\be
\begin{split}
\delta^{(-2\ep)}(r+u)  = (r^0 + u^0)^{2\ep} \, \delta^{(-2\ep)}(\hat{r})
\quad &\text{for the single-collinear sector} \\
\delta^{(-2\ep)}(r+u_1+u_2) = (r^0+u_1^0+u_2^0)^{2\ep} \,
\delta^{(-2\ep)}(\hat{r}) \quad &\text{for the triple-collinear
  sector,} \\
\delta^{(-2\ep)}(r_1+u_1) \, \delta^{(-2\ep)}(r_2+u_2) = \big[
  (r_1^0+u_1^0) (r_2^0+u_2^0) \big]^{2\ep} \,
\delta^{(-2\ep)}(\hat{r}_1) \, \delta^{(-2\ep)}(\hat{r}_2) \quad
&\text{for the double-collinear sector.}
\end{split}
\ee
These delta-functions will restrict the angular integrations for the
reference momenta to be four-dimensional. The energy factors are,
nevertheless, non-trivial. In the case of collinear poles, the
integration over the reference momentum energy is not simply
four-dimensional, since the actual resolved momentum is a sum of the
reference and unresolved momenta.

It remains to argue that the restriction \eqref{eq:restriction} does
not affect the finite part of the result. This is a true assertion,
since the result is finite independently of the measurement
function. The missing $\ep$-dimensional integrations are finite, but
evaluated over a volume of the phase space, which vanishes at $\ep =
0$. Thus the missing contribution vanishes as well and can be
neglected from the start. We have thus shown that the HV
regularization for this group of contributions amounts to evaluating
matrix elements and resolved-parton phase spaces in four
dimensions. The integration over the unresolved partons is
nevertheless performed in $d$ dimensions. In particular, if there is
one unresolved parton and the contribution is due to a soft limit, the
parton kinematics will be integrated non-trivially over the fifth
dimension. Similarly, if there are two unresolved partons and the
contribution is due to a double-soft limit, one parton will be
effectively integrated in five dimensions, while the other in
six. This is, however, the upper bound on the number of required
dimensions independently of the multiplicity.

The last contribution to consider is the single-unresolved cross
section
\be
\hat{\sigma}_{\mathrm{SU}} =
      \hat{\sigma}^{\mathrm{RR}}_{\mathrm{SU}} +
      \hat{\sigma}^{\mathrm{RV}}_{\mathrm{SU}} +
      \hat{\sigma}^{\mathrm{C1}}_{\mathrm{SU}} \; .
\ee
It differs from the previous case in one aspect. It involves both
$n+1$- and $n$-parton matrix elements. Let us first assume that the
calculation is performed with a next-to-leading order measurement
function, i.e. $\mathrm{F}_n = 0$. In this case, we can just repeat the
previous discussion with $n \to n+1$, and simply evaluate $n+1$-parton
matrix elements and $n+1$-parton phase spaces in
four-dimensions. However, once we turn to the next-to-next-to-leading
order measurement function, i.e.\ $\mathrm{F}_n \neq 0$, the
subtraction terms will not match the singularities of the amplitudes
anymore. The reason is that the limits originally used in the
construction of the subtraction terms in Section~\ref{sec:subtraction}
apply to $d$-dimensional amplitudes. For
$\hat{\sigma}^{\mathrm{RV}}_{\mathrm{SU}}$ and
$\hat{\sigma}^{\mathrm{C1}}_{\mathrm{SU}}$, this problem is easily
resolved. In the subtraction terms, we take the splitting functions at
$\ep = 0$. The soft functions do not depend on $\ep$ and do not
require modifications. The calculation of
$\hat{\sigma}^{\mathrm{RV}}_{\mathrm{SU}}$ and
$\hat{\sigma}^{\mathrm{C1}}_{\mathrm{SU}}$ can now be performed with a
four-dimensional phase space and four-dimensional matrix elements.

The double-real single-unresolved contribution,
$\hat{\sigma}^{\mathrm{RR}}_{\mathrm{SU}}$, requires more care. The
pole contributions must be evaluated with complete splitting
functions, i.e.\ at $\ep \neq 0$. However, when subtraction terms to
these pole terms are derived, we take into account that the
limits are iterated. This means, for example, that instead of the
triple-collinear splitting function, Eq.~\eqref{ccfacm}, we can use
a product of two splitting functions. The first one will generate the
pole, and it must be taken azimuthally averaged and at $\ep \neq 0$. The
second one corresponding to the subtraction term must be taken
spin-correlated and at $\ep = 0$. An example of this procedure is the
pole at $\eta_2$ due to a quark and a gluon in the triple-collinear
parameterization sector $\mathcal{S}_4$, to which we would generate a
collinear subtraction term with a gluon as reference parton. The
corresponding limit would be a special case of
Eq.~\eqref{eq:azimuthalcc}
\be
\overline{| \cm^{(0)}_{g,q,g,\dots}(r,u_1,u_2,\dots) |^2} \simeq 
\f{\big( 8 \pi \as \big)^2}{s_{12} \, s_{r12}} \, \la
\sP^{(0)}_{qg}(z_{12};\ep \neq 0) \ra \, \la \cm^{(0)}_{q,\dots}(p,\dots)|
\sP^{(0)}_{gq}(z_{r12},u_{1\perp};\ep = 0)
|\cm^{(0)}_{q,\dots}(p,\dots)\ra \; .
\ee
Since next-to-leading order soft functions do not depend on $\ep$,
there is nothing special to do in their case. This concerns the cases
of a soft pole and collinear subtraction, and of a collinear pole and
soft subtraction. We note, however, that the iterated
soft-pole-soft-subtraction limit is easier to obtain form the
double-soft limit. The double-soft function does depend on
$\ep$. However, this dependence does not contribute to the iterated
limit.  After each contribution listed in Tab.~\ref{tab:SU} has been
treated this way, we may evaluate all tree-level matrix elements, both
with $n+1$- and $n$-partons in the final state, in four
dimensions. The derivation of the subtraction and integrated
subtraction terms is performed using the algorithm of
Section~\ref{sec:subtraction}, but taking into account that the phase
space restriction \eqref{eq:restriction} now acts on the unresolved
partons $u_1$ and $u_2$. Since we use \eqref{eq:restriction} with $n
\to n+1$, the following delta-functions will appear in the
triple-collinear parameterization
\be
\begin{split}
\delta^{(-2\ep)}(r+u_1) \, \delta^{(-2\ep)}(u_2) \quad &\text{for the
  collinear pole in $\eta_1$ in sector $\mathcal{S}_3$,} \\
\delta^{(-2\ep)}(r+u_2) \, \delta^{(-2\ep)}(u_1) \quad &\text{for the
  collinear pole in $\eta_2$ in sector $\mathcal{S}_1$,} \\
\delta^{(-2\ep)}(r) \, \delta^{(-2\ep)}(u_1+u_2) \quad &\text{for the
  collinear pole in $\eta_1$ in sector $\mathcal{S}_5$, and in
  $\eta_2$ in sector $\mathcal{S}_4$,} \\
\delta^{(-2\ep)}(r) \, \delta^{(-2\ep)}(u_1) \quad &\text{for the soft
  pole in $\xi_2$ in sectors $\mathcal{S}_1$, $\mathcal{S}_2$,
  $\mathcal{S}_4$ and $\mathcal{S}_5$.}
\end{split}
\ee
In the double-collinear parameterization, on the other hand, there
will be
\be
\begin{split}
\delta^{(-2\ep)}(r_1+u_1) \, \delta^{(-2\ep)}(r_2) \,
\delta^{(-2\ep)}(u_2) \quad &\text{for the collinear pole in
  $\eta_1$,} \\
\delta^{(-2\ep)}(r_2+u_2) \, \delta^{(-2\ep)}(r_1) \,
\delta^{(-2\ep)}(u_1) \quad &\text{for the collinear pole in
  $\eta_2$,} \\
\delta^{(-2\ep)}(r_1) \, \delta^{(-2\ep)}(r_2) \,
\delta^{(-2\ep)}(u_1) \quad &\text{for the soft pole in $\xi_2$.}
\end{split}
\ee
Delta-functions involving a reference momentum are only present, if
it is in the final state. Notice that in all collinear pole cases,
there are no $d$-dimensional integrations left. However, there is a
$d$-dimensional integration over the unrestricted 
direction of $u_2$ in the soft-pole case. In practice, most
integrations can be performed analytically, because nothing depends on
the respective angles. There will, however, remain a single
integration beyond the dimensions of the resolved momenta. Thus, in
the general case, there will be a five-dimensional integration. The
delta-functions, which do not involve the reference momentum, influence
the scaling in the singular variables. This, in turn, has consequences
for the counterterms derived in
Section~\ref{sec:FiniteContributions}. There, we gave the necessary
scaling exponents for both CDR and HV cases, the latter following from the
application of the above listed delta-functions. Whenever the
exponents concerned energy variables (as for collinear poles of
subsections~\ref{sec:caseIS4}, \ref{sec:caseIS5}, \ref{sec:caseIII}
and \ref{sec:caseIV}), $\xi_{1,2}$, the difference
between the exponent in CDR and that in HV was 2, simply because the
delta-functions provide an additional factor of $\xi_{1,2}^{2\ep}$. On
the other hand, the scaling of the angular variables in the soft-pole
cases of subsections~\ref{sec:caseIS5soft} and \ref{sec:caseII} was
affected by the change of the angular measure. Indeed, the
delta-function $\delta^{(-2\ep)}(u_1)$ introduces a factor of
\be
\big[ 4(1-\hat{\eta}_1)\hat{\eta}_1 \big]^\ep \; ,
\ee
which changes the exponent of the relevant angular variables by 1
($\eta_2$ in sector $\mathcal{S}_5$, and both $\eta_1$ and $\eta_2$ is
sector $\mathcal{S}_2$).

Notice that the difference in scaling only influences the finite parts
of cross sections. In consequence, if one would proceed along the
algorithm of this  section, but without separating the single- and
double-unresolved contributions, and without applying the counterterms
of Section~\ref{sec:FiniteContributions}, the result would be finite,
but incorrect. This remark completes the discussion of the 't
Hooft-Veltman regularization of the single-unresolved contributions,
and by the same of the complete cross section.

Finally, we remind that the factorization contributions
$\hat{\sigma}^{\mathrm{C1}}_{\mathrm{SU}}$ and
$\hat{\sigma}^{\mathrm{C1}}_{\mathrm{DU}}$ must be calculated in a
boosted frame as discussed in subsection~\ref{sec:caseIV}. One could
wonder, if this means that the complete calculation is frame
dependent. This is not the case, since after all the modifications,
the result for the $\mathcal{O}(\ep^0)$ cross section is exactly the
same as in CDR. In CDR, however, each cross section contribution
listed in Section~\ref{sec:outline} is separately Lorentz
invariant.

%%%%%%%%%%%%%%%%%%%%%%%%%%%%%%%%%%%%%%%%%%%%%%%%%%%%%%%%%%%%%%%%%%%%%%%%%%%%%%%%

\section{Example: $gg \to t \bar t + n g$, $n = 0,1,2$}
\label{sec:example}

\noindent
In this section, we present a comparison between results obtained
using conventional dimensional regularization and 't Hooft-Veltman
regularization for an example cross section at next-to-next-to-leading
order. The aim is to demonstrate that the modifications described in
Sections~\ref{sec:FiniteContributions} and \ref{sec:HVregularization}
lead to correct results in a realistic calculation. We select
inclusive top-quark pair production in the gluon fusion channel with
up to two gluons in the final state. This choice is motivated by the
fact that collinear limits involving gluons have the most involved
structure of spin correlations. We are thus able to verify the
azimuthal average corrections we have provided in
Section~\ref{sec:AzimuthalAverage}. We neglect all contributions
involving finite remainders of one- and two-loop amplitudes. As
explained before, their independence of the regularization scheme can
be proven explicitly by noticing that they enter the calculation in
exactly the same manner as Born amplitudes in a next-to-leading order
calculation. Finally, we do not include the contribution of the
subtracted six-point Born amplitude, which is trivially independent of
the regularization, as it is finite by definition.

The relevant partonic cross section is rendered dimensionless and
independent of the value of the strong coupling with the normalization
\be
\tilde{\sigma}^{(2)} = \f{m_t^2}{\as^4} \hat{\sigma}^{(2)} \; ,
\ee
where $\hat{\sigma}^{(2)}$ is the total cross section contribution at
$\mathcal{O}(\as^4)$. Furthermore, we set
\be
\mu_R = \mu_F = m_t \; .
\ee
Our results are obtained at the point
\be
\beta =\sqrt{1-\f{4 m_t^2}{\hat{s}}}= 0.5 \; .
\ee

Tables \ref{tab:DoubleUnresolvedCDR} and \ref{tab:DoubleUnresolvedHV}
contain results for partial double-unresolved contributions calculated
in CDR and HV regularizations respectively. In each case, the last row
gives the total double-unresolved contribution. Whenever errors are
quoted, they are due to Monte Carlo integration. Contributions
involving two-parton kinematics have been computed with a
deterministic integration method, which implies that their error is
negligible, and, therefore, not specified in the tables. By
construction, double-unresolved contributions should be finite in both
regularizations. Indeed, coefficients of the poles in $\ep$ are
consistent with $0$ within one standard deviation for all but the
leading singularity in CDR, where consistency at a level below two
standard deviations is observed. The calculation has been performed
with optimization based on the behavior of the finite part, which is
one reason for the slightly lower quality of the leading pole
contribution. Notice, nevertheless, that there is no difference
between CDR and HV at the leading pole of each contribution as far as
the actual integrand is concerned. Therefore, divergence cancellation
in the HV case within one sigma is sufficient to claim divergence
cancellation in both regularizations. On the other hand, analytic
cancellation of the coefficient of the $1/\ep^4$ pole has already been
shown in \cite{Czakon:2011ve}. Due to severe cancellations between the
different partial contributions, it was necessary to use large Monte
Carlo samples. For instance, the quoted precision of the double-real
contributions required nearly $10^{11}$ points. In all cases, the
convergence in the HV regularization was noticeably better. As far as
the finite parts of the results are concerned, we aimed at about 1\%
precision for the total contribution. Nevertheless, the agreement
between evaluations in CDR and HV regularizations is at the level of
one permille.

\begin{table}[h]
\begin{center}
\renewcommand{\arraystretch}{1.5}
\begin{tabular}{l.....}
\hline
 & \multicolumn{1}{c}{$1/\ep^{4}$} & \multicolumn{1}{c}{$1/\ep^{3}$} & \multicolumn{1}{c}{$1/\ep^{2}$} & \multicolumn{1}{c}{$1/\ep$} & \multicolumn{1}{c}{$\ep^0$} \\ 
\hline
$\tilde{\sigma}^{\mathrm{VV}}_{\mathrm{DU}}$ & 0.0321959 & 0.135003 & 0.177418 & 0.04517 & -0.1242 \\ 
$\tilde{\sigma}^{\mathrm{RV}}_{\mathrm{DU}}$ & -0.0724423(9) & -0.456495(4) & -1.196150(11) & -1.81962(4) & -2.8562(1) \\ 
$\tilde{\sigma}^{\mathrm{RR}}_{\mathrm{DU}}$ & 0.0402448(2) & 0.321486(1) & 1.045064(6) & 1.61821(4) & 1.3065(3) \\
$\tilde{\sigma}^{\mathrm{C1}}_{\mathrm{DU}}$ &  & -0.154649(4) & -0.447655(20) & 0.09385(8) & 1.8313(2) \\ 
$\tilde{\sigma}^{\mathrm{C2}}_{\mathrm{DU}}$ &  & 0.154650 & 0.421336 & 0.06247 & -0.1878 \\ 
\hline
$\tilde{\sigma}^{\mathrm{CDR}}_{\mathrm{DU}}$ & -0.0000016(9) & -0.000005(6) & 0.000013(24) & 0.00007(9) & -0.0304(4) \\
\hline
\end{tabular}
\caption{\label{tab:DoubleUnresolvedCDR} Double-unresolved (DU)
  contributions to the partonic cross section $gg \to t\bar t + X$,
  with $X$ consisting of up to two gluons, evaluated in conventional
  dimensional regularization (CDR). The error estimates quoted in
  parentheses are due to Monte Carlo integration. The definition of
  partial contributions is given in the text.}
\end{center}
\end{table}

\begin{table}[h]
\begin{center}
\renewcommand{\arraystretch}{1.5}
\begin{tabular}{l.....}
\hline
 & \multicolumn{1}{c}{$1/\ep^{4}$} & \multicolumn{1}{c}{$1/\ep^{3}$} & \multicolumn{1}{c}{$1/\ep^{2}$} & \multicolumn{1}{c}{$1/\ep$} & \multicolumn{1}{c}{$\ep^0$} \\ 
\hline
$\tilde{\sigma}^{\mathrm{VV}}_{\mathrm{DU}}$ & 0.0321959 & 0.086177 & 0.021985 & -0.03200 & 0\\ 
$\tilde{\sigma}^{\mathrm{RV}}_{\mathrm{DU}}$ & -0.0724415(9) & -0.346630(3) & -0.702124(8) & -1.04640(3) & -2.3910(1)\\ 
$\tilde{\sigma}^{\mathrm{RR}}_{\mathrm{DU}}$ & 0.0402447(2) & 0.260452(1) & 0.706469(6) & 1.06119(3) & 1.8461(2)\\  
$\tilde{\sigma}^{\mathrm{C1}}_{\mathrm{DU}}$ &  & -0.154646(4) & -0.283008(15) & 0.08326(5) & 0.5144(1)\\ 
$\tilde{\sigma}^{\mathrm{C2}}_{\mathrm{DU}}$ &  & 0.154650 & 0.256668 & -0.06603 & 0\\ 
\hline
$\tilde{\sigma}^{\mathrm{HV}}_{\mathrm{DU}}$ & -0.0000009(9) & 0.000003(6) & -0.000010(17) & 0.00002(6) & -0.0304(2)\\
\hline 
\end{tabular}
\caption{\label{tab:DoubleUnresolvedHV} Double-unresolved (DU)
  contributions to the partonic cross section $gg \to t\bar t + X$,
  with $X$ consisting of up to two gluons, evaluated in
  't Hooft-Veltman regularization (HV). The error estimates quoted in
  parentheses are due to Monte Carlo integration. The definition of
  partial contributions is given in the text.}
\end{center}
\end{table}

The partial results quoted in the tables have been obtained by
integrating tree-level amplitudes only. They are:
\begin{itemize}

\item[$\tilde{\sigma}^\mathrm{VV}_{\mathrm{DU}}$ :] Double-virtual
  contributions obtained by integrating the two-loop and one-loop
  squared amplitudes for $gg \to t\bar t$ without their finite
  remainders.

\item[$\tilde{\sigma}^\mathrm{RV}_{\mathrm{DU}}$ :] Real-virtual
  contributions obtained from the integrated subtraction terms of the
  one-loop amplitude for $gg \to t\bar t + g$, without the
  contribution of the finite remainder of the one-loop amplitude for
  $gg \to t\bar t$.

\item[$\tilde{\sigma}^\mathrm{RR}_{\mathrm{DU}}$ :] Double-real
  contributions obtained from the double-unresolved integrated
  subtraction terms of the Born amplitude for $gg \to t\bar t + gg$,
  including corrections described in
  Section~\ref{sec:HVregularization}, which make the total
  double-unresolved contribution finite.

\item[$\tilde{\sigma}^\mathrm{C1}_{\mathrm{DU}}$ :] Factorization
  contributions obtained from the convolution of the leading order
  splitting function with the cross section contribution of the
  integrated subtraction terms of the Born amplitude for $gg \to t\bar
  t + g$.

\item[$\tilde{\sigma}^\mathrm{C2}_{\mathrm{DU}}$ :] Factorization
  contributions obtained from the convolution of the leading order
  splitting function with the cross section contribution of the
  one-loop amplitude for $gg \to t\bar t$ without its finite
  remainder, and the convolution of the next-to-leading order
  splitting function as well as two leading-order splitting functions
  with the Born cross section for $gg \to t\bar t$. 

\end{itemize}
The results in CDR have been obtained without azimuthal averaging,
i.e.\ with splitting functions containing full spin correlations. The
results in HV, on the other hand, have been obtained with azimuthal
averaging, i.e.\ with averaged splitting functions whenever
possible. Details can be found in
Section~\ref{sec:AzimuthalAverage}. The counterterms of
Section~\ref{sec:FiniteContributions} have been applied to
averaged splitting functions. We finally note that in the HV
regularization, the finite parts receive non-vanishing contributions
from integrated subtraction terms only. Indeed, there is no
contribution from $\tilde{\sigma}^\mathrm{VV}_{\mathrm{DU}}$ and
$\tilde{\sigma}^\mathrm{C2}_{\mathrm{DU}}$. The latter will, however,
contribute if $\mu_R \neq \mu_F$.

\begin{table}[h]
\begin{center}
\renewcommand{\arraystretch}{1.5}
\begin{tabular}{l...}
\hline
 & \multicolumn{1}{c}{$1/\ep^{2}$} & \multicolumn{1}{c}{$1/\ep$} & \multicolumn{1}{c}{$\ep^0$} \\ 
\hline 
$\tilde{\sigma}^{\mathrm{RR}}_{\mathrm{SU}}$ & 0.064772(4) & 0.42742(3) & 1.0623(3)\\ 
$\tilde{\sigma}^{\mathrm{RV}}_{\mathrm{SU}}$ & -0.064780(6) & -0.31419(4) & -0.6044(2)\\ 
$\tilde{\sigma}^{\mathrm{C1}}_{\mathrm{SU}}$ &  & -0.11329(3) & -0.1999(1)\\  
$\tilde{\sigma}^{\mathrm{A}}_{\mathrm{SU}}$ &  &  & -0.00737(2)\\  
\hline 
$\tilde{\sigma}^{\mathrm{CDR}}_{\mathrm{SU}}$ & -0.000008(8) & -0.00006(6) & 0.2506(3)\\  
\hline
\end{tabular}
\caption{\label{tab:SingleUnresolvedCDR} Single-unresolved (SU)
  contributions to the partonic cross section $gg \to t\bar t + X$,
  with $X$ consisting of up to two gluons, evaluated in conventional
  dimensional regularization (CDR). The error estimates quoted in
  parentheses are due to Monte Carlo integration. The definition of
  partial contributions is given in the text.}
\end{center}
\end{table}

\begin{table}[h]
\begin{center}
\renewcommand{\arraystretch}{1.5}
\begin{tabular}{l...}
\hline
 & \multicolumn{1}{c}{$1/\ep^{2}$} & \multicolumn{1}{c}{$1/\ep$} & \multicolumn{1}{c}{$\ep^0$} \\ 
\hline 
$\tilde{\sigma}^{\mathrm{RR}}_{\mathrm{SU}}$ & 0.064780(5) & 0.25429(3) & 0.2584(2) \\ 
$\tilde{\sigma}^{\mathrm{RV}}_{\mathrm{SU}}$ & -0.064770(7) & -0.14096(2) & 0 \\ 
$\tilde{\sigma}^{\mathrm{C1}}_{\mathrm{SU}}$ &  & -0.11329(2) & 0 \\  
$\tilde{\sigma}^{\mathrm{A}}_{\mathrm{SU}}$ &  &  & -0.00734(1) \\  
\hline 
$\tilde{\sigma}^{\mathrm{HV}}_{\mathrm{SU}}$ & 0.000011(8) & 0.00004(4) & 0.2511(2) \\  
\hline
\end{tabular}
\caption{\label{tab:SingleUnresolvedHV} Single-unresolved (SU)
  contributions to the partonic cross section $gg \to t\bar t + X$,
  with $X$ consisting of up to two gluons, evaluated in
  't Hooft-Veltman regularization (HV). The error estimates quoted in
  parentheses are due to Monte Carlo integration. The definition of
  partial contributions is given in the text.}
\end{center}
\end{table}

Results for the single-unresolved partial contributions are displayed
in the Tabs.~\ref{tab:SingleUnresolvedCDR} and
\ref{tab:SingleUnresolvedHV} for CDR and HV regularizations
respectively. We again observe finiteness of the total contributions
given in the last row of each table. The coefficients of the poles are
consistent with 0 at the one sigma level in all cases but the leading
singularity evaluated in HV regularization. There, consistency is only
observed at a level better than two standard deviations. In this
respect, the same comments apply as in the discussion of the
double-unresolved contribution. The single-unresolved contributions in
both regularizations have a precision of better than two permille
relative error. The agreement between them is at the same level, and
does only slightly exceed one standard deviation.

The partial cross section contributions in the single-unresolved case
are:
\begin{itemize}

\item[$\tilde{\sigma}^\mathrm{RR}_{\mathrm{SU}}$ :] Double-real
  contributions obtained from the single-unresolved integrated
  subtraction terms of the Born amplitude for $gg \to t\bar t + gg$,
  including corrections described in
  Section~\ref{sec:HVregularization}, which make the total
  single-unresolved contribution finite. The splitting functions used
  in the derivation of the integrated subtraction terms are given by
  the azimuthally averaged expression Eqs~\eqref{avhpgg}. The correct
  result is obtained after adding
  $\tilde{\sigma}^\mathrm{A}_{\mathrm{SU}}$.

\item[$\tilde{\sigma}^\mathrm{RV}_{\mathrm{SU}}$ :] Real-virtual
  contributions obtained by integrating the one-loop amplitude for $gg
  \to t\bar t + g$ together with its subtraction terms, after removal
  of all finite remainders.

\item[$\tilde{\sigma}^\mathrm{C1}_{\mathrm{SU}}$ :] Factorization
  contributions obtained from the convolution of the leading order
  splitting function with the cross section contribution of the Born
  amplitude for $gg \to t\bar t + g$ together with its subtraction
  terms. 

\item[$\tilde{\sigma}^\mathrm{A}_{\mathrm{SU}}$ :] Difference between
  the single-unresolved contributions obtained with spin-correlated
  and azimuthally-averaged splitting functions in integrated
  subtraction terms as explained in
  Section~\ref{sec:AzimuthalAverage}.

\end{itemize}
We notice that $\tilde{\sigma}^\mathrm{RV}_{\mathrm{SU}}$ and
$\tilde{\sigma}^\mathrm{C1}_{\mathrm{SU}}$ only contain poles in HV
regularization. However, $\tilde{\sigma}^\mathrm{C1}_{\mathrm{SU}}$
would develop a finite part if $\mu_R \neq \mu_F$.

Finally, we can compare the single- and double-unresolved contributions
to the full partonic NNLO cross section, which was calculated in
Ref.~\cite{Czakon:2013goa}. At the chosen value of $\beta$, there is
\begin{equation}
\tilde{\sigma}= 1.223 \pm 0.003 \; .
\end{equation}
The single-unresolved part contributes $20\%$ to the full cross
section, whereas the double-unresolved part only contributes $2\%$ in
this specific case.

%%%%%%%%%%%%%%%%%%%%%%%%%%%%%%%%%%%%%%%%%%%%%%%%%%%%%%%%%%%%%%%%%%%%%%%%%%%%%%%%

\section{Concluding remarks}

\noindent
We have presented a complete construction of the sector-improved
residue subtraction scheme in four dimensions. It is now possible to
evaluate next-to-next-to-leading order cross sections using ordinary
tree-level matrix elements without higher order terms of the
$\ep$-expansion. This is crucial, since it allows to use the myriad of
publicly available software designed for efficient tree-level
calculations. Of course, it is also necessary to have access to
one- and two-loop amplitudes. Fortunately, at least the former are
also available from open access packages. The problem is thus
currently reduced to the two-loop virtual corrections. Recent progress
shows that a breakthrough may be possible on the scale of the next
few years.

At present, our construction is an algorithm, which leads directly
from various soft and splitting functions collected in appendices of
this paper, to process independent subtraction and integrated
subtraction terms necessary for a Monte Carlo implementation of the
phase space integration. In the future, we intend to provide an
optimized code that will handle the burden of bookkeeping and will
contain built-in tree-level Standard Model amplitudes, similarly to
the next-to-leading software packages of Refs.~\cite{Czakon:2009ss}
and \cite{Bevilacqua:2013iha}. Virtual amplitudes will still have to
be provided by the user.

Of course, the work on the scheme does not stop here. There are
several possible improvements. For instance, we imagine that it
would be quite advantageous to allow for random polarization in the
integration of the most computationally intensive $n+2$ tree-level,
and $n+1$ one-loop amplitudes. This can definitely be achieved by
polarized splitting functions. Another issue to consider is the
introduction of cutoffs on the subtraction phase space. In any case,
more experience has to be accumulated in order to decide, which
modifications to include first. We leave this to future work.

%%%%%%%%%%%%%%%%%%%%%%%%%%%%%%%%%%%%%%%%%%%%%%%%%%%%%%%%%%%%%%%%%%%%%%%%%%%%%%%%

\section*{Acknowledgments}

\noindent
This research was supported by the German Research Foundation (DFG)
via the Sonderforschungsbereich/Transregio SFB/TR-9 ``Computational
Particle Physics''. The work of M.C. was supported by the DFG
Heisenberg programme.

%%%%%%%%%%%%%%%%%%%%%%%%%%%%%%%%%%%%%%%%%%%%%%%%%%%%%%%%%%%%%%%%%%%%%%%%%%%%%%%%

\appendix

%%%%%%%%%%%%%%%%%%%%%%%%%%%%%%%%%%%%%%%%%%%%%%%%%%%%%%%%%%%%%%%%%%%%%%%%%%%%%%%%

\section{Notation}
\label{sec:Notation}

\noindent
\begin{description}

\item[{\it Spacetime dimension}]
\be
d = 4-2\ep \; .
\ee

\item[{\it Bare strong coupling}]
\be
\alpha_s^0 = \Biggl( \f{\mR e^{\gamma_{\mathrm{E}}}}{4\pi}
\Biggr)^\ep \, Z_\as \zeta_\as \as \; ,
\ee
\begin{eqnarray*}
\mu_R \; &-& \; \mbox{renormalization scale} \; , \\
Z_\as \; &-& \; \mbox{\msbar renormalization constant} \; , \\
\zeta_\as \; &-& \; \mbox{heavy-quark decoupling constant
  \cite{Chetyrkin:1997un}} \; .
\end{eqnarray*}

\item[{\it Matrix elements}]
\bea
\label{meldef}
\cm^{c_1,\dots,c_n;s_1,\dots,s_n}_{a_1,\dots,a_n}(p_1,\dots,p_n)
&=& \Big( \la c_1,\dots,c_n| \otimes \la s_1,\dots,s_n| \Big)
\, |\cm_{a_1,\dots,a_n} \left(p_1,\dots,p_n\right) \ra
\;, \\[.6cm]
\label{meldefC}
|\cm_{n}\ra &=& |\cm_{a_1,\dots,a_n}\left(p_1,\dots,p_n\right)\ra
\; , \quad \sum\limits_{\substack{\mathrm{color}\\ \mathrm{spin}}}
|\cm_{n}|^2 = \la\cm_{n}|\cm_{n}\ra\; , \\
|\cm_n\ra &=& \Biggl( \f{\mu_R^2 e^{\gamma_{\rm{E}}}}{4\pi}
\Biggr)^{-l\ep} \Bigl( |\cm_n^{(0)}\ra+|\cm_n^{(1)}\ra+|\cm_n^{(2)}\ra
+\dots \Bigr) \; .
\eea
\begin{center}
\begin{tabular}{lll}
$c_i \;-$ color of parton $i$, & $a_i \;-$ flavor of parton $i$, &
  $|c_1,\dots,c_n \ra \;-$ color basis vectors, \\
$s_i \;-$ spin of parton $i$, & $p_i \;-$ momentum of parton $i$, & 
  $| s_1,\dots,s_n \ra \;-$ spin basis vectors, \\
$l \;-$ $\as$ power of Born approximation. &&
\end{tabular}
\end{center}

\item[{\it Phase spaces}]
\be
\int \mathrm{d}\bm{\Phi}_n\Big( p_1 + p_2 \to \sum_{i=1}^n q_i \Big) =
\Biggl( \f{\mu_R^2 e^{\gamma_{\rm{E}}}}{4\pi} \Biggr)^{(n-1)\ep} \int
\prod_{i=1}^n \f{\mathrm{d}^{d-1}q_i}{(2\pi)^{d-1}2q_i^0} \,
(2\pi)^d\delta^{(d)}\Big( \sum_{i=1}^n q_i - p_1 - p_2 \Big) \; .
\ee

\item[{\it Sums over partons}]
\begin{equation*}
\sum_{ij\dots} \; - \; \mbox{sum over all indices $i,j,\dots$} \; , \quad
\sum_{(i,j,\dots)} \; - \; \mbox{sum over distinct indices $i,j,\dots$} \; .
\end{equation*}
\begin{eqnarray*}
i,j,k,\dots &-& \mbox{indices for arbitrary partons, both massless and
  massive,} \\
i_0,j_0,k_0,\dots &-& \mbox{indices for massless partons,} \\
I,J,K,\dots &-& \mbox{indices for massive partons.} \\
\end{eqnarray*}

\item[{\it Kinematic invariants}]
\bea
p_I^2 = m_I^2 \; , & v_I = p_I/m_I \; , & v_{IJ} =
\sqrt{1-\f{m_I^2m_J^2}{(p_Ip_J)^2}} \; , \\
&s_{ij} = 2\sigma_{ij} p_i \cdot p_j + i0^+ \; .&
\eea
\begin{equation*}
\sigma_{ij} = +1 \; - \; \mbox{if the momenta $p_i$ and $p_j$ are both
  incoming or outgoing} \; , \quad
\sigma_{ij} = -1 \; - \; \mbox{otherwise.}
\end{equation*}

\item[{\it Color charge operators \cite{Catani:1996vz}}]
\bea
\la c_1,\dots,c_i,\dots,c_n,c|\cT_i|b_1,\dots,b_i,\dots,b_n\ra &=&
\la c_1,\dots,c_i,\dots,c_n|T^c_i|b_1,\dots,b_i,\dots,b_n\ra
\nn\\ &=& \delta_{c_1b_1} \dots T^c_{c_i b_i}\dots\delta_{c_nb_n} \; .
\eea
\be
\sum_{i}\cT_i|\cm_{n}\ra=0 \; , \quad
T^c_iT^c_j= \cT_i \cdot \cT_j=\cT_j\cdot\cT_i, \quad
\cT_i \cdot \cT_i = \cT_i^2=C_i=C_{a_i} \; , 
\ee
\be
C_g = C_A \; , \quad C_q = C_{\bar{q}} = C_F \; .
\ee
\begin{eqnarray*}
T^c_{c_1c_2} &=& if^{c_1cc_2} \; - \; \mbox{emitter is a gluon} \; , \\
T^c_{c_1 c_2} &=& t^c_{c_1 c_2} (= -t^c_{c_2 c_1}) \; - \;
\mbox{emitter is an outgoing quark (anti-quark)} \; , \\
T^c_{c_1 c_2} &=& -t^c_{c_2 c_1} (= t^c_{c_1 c_2}) \; - \;
\mbox{emitter is an ingoing quark (anti-quark)} \; . \\
\end{eqnarray*}
\be
\mathrm{Tr}\left[t^at^b\right] = T_F\delta^{ab} =
\f{1}{2}\delta^{ab} \; .
\ee

\end{description}

%%%%%%%%%%%%%%%%%%%%%%%%%%%%%%%%%%%%%%%%%%%%%%%%%%%%%%%%%%%%%%%%%%%%%%%%%%%%%%%%

\section{Spherical coordinates in $d$ dimensions}
\label{sec:spherical}

\noindent
Let $\mathrm{d}^d \bm{r}$ be the Euclidean integration measure in
$\mathbb{R}^d$. We can decompose it into a radial and an angular part
with the help of a $\delta$-function insertion, if we rescale the
$\bm{r}$ vector as $\bm{r} = r \, \bm{\hat{n}}$
\be
\label{eq:sphericalcoordinates}
\int_{\mathbb{R}^d} \mathrm{d}^d \bm{r}  = \int_0^\infty \mathrm{d}r
\, r^{d-1} \int_{\mathbb{R}^d} \mathrm{d}^d \bm{\hat{n}} \, \delta(1 -
\lVert \bm{\hat{n}} \rVert) = \int_0^\infty \mathrm{d}r \,
r^{d-1} \int_{\mathcal{S}_1^{d-1}} \mathrm{d} \bm{\Omega}
\; .
\ee
We have thus defined a rotationally invariant measure,
$\mathrm{d}\bm{\Omega}$, on the unit $(d-1)$-sphere,
$\mathcal{S}_1^{d-1}$. Notice that we will, from now on, include the
dimensionality in the notation of the versors $\bm{\hat{n}}$. Let us
introduce a recursive parameterization in terms of angles
\be
\bm{\hat{n}}^{(d)}(\theta_1, \theta_2, \dots, \theta_{d-1}) =
\begin{pmatrix*}[l]
\cos\theta_1 \\ \\
\sin\theta_1 \; \bm{\hat{n}}^{(d-1)}(\theta_2, \dots, \theta_{d-1})
\end{pmatrix*}
\; , \quad \bm{\hat{n}}^{(1)} = 1
\; ,
\ee
where
\be
\theta_1, \dots, \theta_{d-2} \in [0, \pi] \; , \quad
\theta_{d-1} \in [0, 2\pi ]
\; .
\ee
An important property of this parameterization is
\bea
\bm{\hat{n}}^{(d)}(\theta_1, \dots, \theta_{n-1}, 0, \theta_{n+1},
\dots, \theta_{d-1}) &=& \bm{\hat{n}}^{(d)}(\theta_1, \dots,
\theta_{n-1}, 0, 0, \dots) \; ,
\nn \\
\bm{\hat{n}}^{(d)}(\theta_1, \dots, \theta_{n-1}, \pi, \theta_{n+1},
\dots, \theta_{d-1}) &=& \bm{\hat{n}}^{(d)}(\theta_1, \dots,
\theta_{n-1}, \pi, 0, 0, \dots)
\; .
\eea
The recursive definition of the versor can be implemented in the
integration measure
\be
\label{eq:recursiveintegration}
\int_{\mathcal{S}_1^{d-1}} \mathrm{d}\bm{\Omega}(\theta_1, \theta_2,
\dots, \theta_{d-1}) =
\int_0^\pi \mathrm{d}\theta_1 \, \sin^{d-2}\theta_1
\int_{\mathcal{S}_1^{d-2}} \mathrm{d}\bm{\Omega}( \theta_2, \dots,
\theta_{d-1})
\; .
\ee
The volume of the unit $(d-1)$-sphere is
\be
\int_{\mathcal{S}_1^{d-1}} \mathrm{d}\bm{\Omega} \, 1 = 
\f{2\pi^{\f{d}{2}}}{\Gamma\left( \f{d}{2} \right)}
\; .
\ee
We will also need the following result
\be
\int_{\mathcal{S}_1^{d-1}} \mathrm{d}\bm{\Omega} \, \delta^{(d)}\left(
\alpha \bm{\hat{n}}^{(d)} \right) =
\alpha^{1-d} \int_{\mathbb{R}^d} \mathrm{d}^d \left(\alpha
\bm{\hat{n}}^{(d)}\right) \, \delta\left(\alpha - \left\lVert \alpha
\bm{\hat{n}}^{(d)} \right\rVert\right) \, \delta^{(d)}\left(\alpha
\bm{\hat{n}}^{(d)}\right) = \f{1}{\alpha^{d-1}} \, \delta(\alpha)
\; ,
\ee
which implies the correct reduction of the
dimensionality of space
\be
\int_{\mathcal{S}_1^{d-1}} \mathrm{d}\bm{\Omega} \,
\delta^{(d-n)}\left( \bm{\hat{n}}^{(d)} \right) =
\int_{\mathcal{S}_1^{n-1}} \mathrm{d}\bm{\Omega}
\; ,
\ee
We further introduce a representation of the angular versor
parameterization through rotations of a basis vector. To this
end, we define
\be
\bm{\hat{n}_0}^{(d)} = \begin{pmatrix} 1 \\ 0 \\ 0
  \\ \vdots \end{pmatrix}
\; ,
\ee
and a $d \times d$ rotation matrix transforming the coordinates $i$
and $j$
\be
\bm{R}^{(d)}_{ij}(\theta) =
\begin{blockarray}{cccccc}
& i && j && \\
\begin{block}{(ccccc)c}
\begin{matrix} 1 && \\ & \ddots & \\ && 1 \end{matrix} &&&& & \\
& \cos\theta && \sin\theta & & i \\
&& \begin{matrix} 1 && \\ & \ddots & \\ && 1 \end{matrix} && & \\
& -\sin\theta && \cos\theta & & j \\
&&&& \begin{matrix} 1 && \\ & \ddots & \\ && 1 \end{matrix} & \\
\end{block}
\end{blockarray}
\; ,
\ee
where the unspecified entries are null. If the rotations act in
different planes, then the respective rotation matrices commute
\be
\{ i, j \} \cap \{k, l \} = \emptyset \; \Longrightarrow \; \left[
  \bm{R}^{(d)}_{ij}(\theta_1) , \, \bm{R}^{(d)}_{kl}(\theta_2) \right] = 0
\; .
\ee
The versor parameterization can be expressed through rotations as
\be
\bm{\hat{n}}^{(d)}(\theta_1, \dots, \theta_{d-1}) =
\bm{R}^{(d)}_1(\theta_1, \dots, \theta_{d-1})
\bm{\hat{n}_0}^{(d)}
\; ,
\ee
where we have introduced the shorthand notation
\be
\bm{R}^{(d)}_n(\theta_1, \dots, \theta_{d-n}) =
\bm{R}^{(d)}_{d,d-1}(\theta_{d-n}) {\dots}
\bm{R}^{(d)}_{n+1,n}(\theta_1)
\; .
\ee
Due to the commutation properties of the rotation matrices, there is
\be
\left[ \bm{R}^{(d)}_1(\theta_1, \dots, \theta_{n-1}, 0, 0, \dots) , \,
\bm{R}^{(d)}_{n+1}(\theta_{n+1}, \dots, \theta_{d-1})
\right] = 0 \; .
\ee
%

%%%%%%%%%%%%%%%%%%%%%%%%%%%%%%%%%%%%%%%%%%%%%%%%%%%%%%%%%%%%%%%%%%%%%%%%%%%%%%%%

\section{Infrared divergences of virtual amplitudes}
\label{sec:VirtualIR}

\noindent
We consider renormalized on-shell virtual amplitudes including wave
function renormalization factors, which are non-trivial for external
massive quarks. The strong coupling is assumed to be renormalized in
the \msbar scheme with decoupling of massive quarks. Infrared
divergences can be factorized from virtual amplitudes as follows
\be
\label{FinRem}
|\cm_n \ra = \bm{\mathrm{Z}}(\ep,\{p_i\},\{m_i\},\mu_R) \, |\cf_n \ra \; ,
\ee
where the infrared (IR) renormalization constant $\bm{\mathrm{Z}}$ is an
operator in color space, and depends on the momenta $\{p_i\}=\{p_1,
..., p_n\}$ and masses $\{m_i\}=\{m_1,...,m_n\}$ of the external
partons. The finite remainder, $|\cf_n \ra$, has a well-defined limit
when $\ep \rightarrow 0$. Expanding equation (\ref{FinRem}) in a
series in $\as$ we obtain for the first three terms
\begin{align}
 |\mathcal{M}_n^{(0)}\ra
&=|\mathcal{F}_n^{(0)}\ra \;, \label{eq:ExpandZM0} \\ 
|\mathcal{M}_n^{(1)}\ra &=
\bm{\mathrm{Z}}^{(1)}|\mathcal{M}_n^{(0)}\ra
+|\mathcal{F}_n^{(1)}\ra \;,\label{eq:ExpandZM1}\\ 
|\mathcal{M}_n^{\left(2\right)}\ra &=
\bm{\mathrm{Z}}^{\left(2\right)}|\mathcal{M}_n^{(0)}\ra +
\bm{\mathrm{Z}}^{(1)}|\mathcal{F}_n^{(1)}\ra +
|\mathcal{F}_n^{\left(2\right)}\ra \nn\\ 
&=\left(
\bm{\mathrm{Z}}^{\left(2\right)}-\bm{\mathrm{Z}}^{(1)}
\bm{\mathrm{Z}}^{(1)}\right)|\mathcal{M}_n^{(0)}\ra
+ \bm{\mathrm{Z}}^{(1)}|\mathcal{M}_n^{(1)}\ra +
|\mathcal{F}_n^{\left(2\right)}\ra \; ,
\label{eq:ExpandZM2}
\end{align}
with $\bm{\mathrm{Z}} = \bm{1} + \bm{\mathrm{Z}}^{(1)} +
\bm{\mathrm{Z}}^{(2)} + \mathcal{O}(\alpha_s^3)$. The IR
renormalization constant satisfies the renormalization group equation
(RGE)
\be
\label{Zdef}
\f{\mathrm{d}}{\mathrm{d}\ln\mu_R}\,\bm{\mathrm{Z}}(\ep,\{p_i\},\{m_i\},\mu_R)
= - \bm{\Gamma}(\{p_i\},\{m_i\},\mu_R) \,
\bm{\mathrm{Z}}(\ep,\{p_i\},\{m_i\},\mu_R) \; ,
\ee 
where the anomalous dimension operator ${\bf \Gamma}$ is given by
\cite{Aybat:2006mz, Becher:2009kw, Czakon:2009zw, Mitov:2009sv,
  Ferroglia:2009ii, Mitov:2010xw}
\be\label{eq:Gamma}
\begin{split}
\bm{\Gamma}(\{p_i\},\{m_i\},\mu_R)\,=&\,\sum\limits_{(i_0,j_0)}\f{\cT_{i_0}\cdot
  \cT_{j_0}}{2}\,\gamma_{\text{cusp}}(\as)\,\ln \, \biggl(\f{\mu_R^2}{-s_{i_0j_0}}
\biggr) \,+ \,\sum\limits_{i_0} \gamma^{i_0}(\as)\\ 
&-\sum\limits_{(I,J)}\f{\cT_I\cdot
  \cT_J}{2}\,\gamma_{\text{cusp}}(v_{IJ},\as)\,+ \,\sum\limits_I
\gamma^I(\as)\,+\,\sum\limits_{I,j_0}\cT_I\cdot\cT_{j_0}\,
  \gamma_{\text{cusp}}(\as)\,\ln \, \biggl(\f{m_I\,\mu_R}{-s_{Ij_0}}\biggr)\\
&+\sum\limits_{(I,J,K)}i\,f^{abc}\,T_I^a\,T_J^b\,T_K^c\,F_1(v_{IJ},v_{JK},v_{KI})\\
&+\sum\limits_{(I,J)}\sum\limits_{k_0}\,i\,f^{abc}\,
  T_I^a\,T_J^b\,T_{k_0}^c\,f_2\left(v_{IJ},
  \ln\,\biggl(\f{-\sigma_{Jk_0}\,v_J\cdot p_{k_0}}{-\sigma_{Ik_0}\,v_I\cdot
  p_{k_0}}\biggr)\right)\,+\,\mathcal{O}(\as^3)
\; .
\end{split}
\ee 
The triple color correlations given in the third and fourth lines of
Eq.~(\ref{eq:Gamma}) cannot contribute to the divergences of spin and
color summed amplitudes at next-to-next-to-leading order, as 
long as the Born amplitudes do not contain complex couplings or
masses~\cite{Czakon:2013hxa}.

The explicit solution of the RGE Eq.~(\ref{Zdef}) can be found in
Ref.~\cite{Becher:2009cu}, and reads up to order $\alpha_s^2$ 
\be
\label{result}
   {\bf Z} = 1 + \f{\as}{4\pi} 
    \left( \f{\Gamma_0'}{4\ep^2}
    + \f{\bm{\Gamma}_0}{2\ep} \right) 
   + \left( \f{\as}{4\pi} \right)^2 \left[
    \f{(\Gamma_0')^2}{32\ep^4} 
    + \f{\Gamma_0'}{8\ep^3} 
    \left( \bm{\Gamma}_0 - \f32\,\beta_0 \right) 
    + \f{\bm{\Gamma}_0}{8\ep^2} 
    \left( \bm{\Gamma}_0 -2\beta_0 \right) 
    + \f{\Gamma_1'}{16\ep^2}
    + \f{\bm{\Gamma}_1}{4\ep} \right] + \mathcal{O}(\alpha_s^3) \; ,
\ee
where the leading beta-function coefficient is
\be
\beta_0=\f{11}{3}C_A-\f{4}{3}T_Fn_l \; ,
\ee
with $n_l$ the number of massless quark flavors. The expression
contains the anomalous dimension $\bm{\Gamma}$ and its derivative
\be\label{Gampr}
   \Gamma'(\as) 
   = \f{\partial}{\partial\ln\mu_R}\,
   \bm{\Gamma}(\{p_i\},\mu_R,\as) \; ,
\ee
expanded according to
\be
   \bm{\Gamma} = \sum_{n=0}^\infty\,\bm{\Gamma}_n 
    \left( \f{\as}{4\pi} \right)^{n+1} , \quad
   \Gamma' = \sum_{n=0}^\infty\,\Gamma'_n 
    \left( \f{\as}{4\pi} \right)^{n+1} \; .
\ee
$\bm{\Gamma}$ is given in terms of the anomalous dimensions
$\gamma_{\text{cusp}}$, $\gamma^q$, $\gamma^{Q}$, $\gamma^{g}$, and
two functions $F_1$ and $f_2$. We give explicit formulae for the
coefficients of the expansion in $\as$
\be
   \gamma (\as) = \sum_{n=0}^\infty\,\gamma_n 
    \left( \f{\as}{4\pi} \right)^{n+1}\; ,
\ee
which we have taken literally from Refs.~\cite{Becher:2009kw,Becher:2009cu}.
The massless cusp anomalous dimension is
\begin{align}
   \gamma_0^{\rm cusp} &= 4 \,, \nn\\
   \gamma_1^{\rm cusp} &= \left( \f{268}{9} 
    - \f{4\pi^2}{3} \right) C_A - \f{80}{9}\,T_F n_l \,.
\end{align}
In the massive case the cusp anomalous dimension can be written as 
\be
\begin{split}
   \gamma_{\rm cusp}(v,\alpha_s) =& \,\gamma_{\rm
     cusp}(\alpha_s) \, \f{1}{v} \left[ \f{1}{2} \ln
     \left(\f{1+v}{1-v}\right) - i\pi\right]\\ 
& +8C_A\left(\f{\as}{4\pi}\right)^2 \Bigg\{\zeta_3 -
   \f{5\pi^2}{6}+\f{1}{4}\ln^2\left(\f{1+v}{1-v}\right)\\
& \hspace{2cm} +\f{1}{v^2}\left[\f{1}{24}\ln^3\left(\f{1+v}{1-v}\right)+
  \ln\left(\f{1+v}{1-v}\right)\left(\f{1}{2}\Li_2\left(\f{1-v}{1+v}\right)-
  \f{5\pi^2}{12}\right)+\Li_3\left(\f{1-v}{1+v}\right)-\zeta_3\right]\\
& \hspace{2cm} +\f{1}{v} \hspace{.125cm} \left[\f{5\pi^2}{6}+
     \f{5\pi^2}{12}\ln\left(\f{1+v}{1-v}\right)
  -\ln\left(\f{2v}{1+v}\right)\ln\left(\f{1+v}{1-v}\right)\right.\\
& \hspace{2cm} \left. \hspace{.9cm} - \f{1}{4}\ln^2\left(\f{1+v}{1-v}\right)
  -\f{1}{24}\ln^3\left(\f{1+v}{1-v}\right)+\Li_2\left(\f{1-v}{1+v}\right)\right]\\
& \hspace{2cm} +i\pi\left\{\f{1}{v^2}\left[\f{\pi^2}{6} -
     \f{1}{4}\ln^2\left(\f{1+v}{1-v}\right) -
     \Li_2\left(\f{1-v}{1+v}\right)\right]\right.\\
& \hspace{2cm} \left. \hspace{.8cm}
   +\f{1}{v}\left[-\f{\pi^2}{6}+2\ln\left(\f{2v}{1+v}\right)
     + \ln\left(\f{1+v}{1-v}\right)+
     \f{1}{4}\ln^2\left(\f{1+v}{1-v}\right)\right]\right.\\
& \hspace{2cm} \left. \hspace{.8cm} -\ln\left(\f{1+v}{1-v}\right)\right\}
\Bigg\} \; .
\end{split}
\ee
For massless quarks (anti-quarks) we also have
\begin{align}
   \gamma_0^q &= -3 C_F \,, \nn\\
   \gamma_1^q &= C_F^2 \left( -\f{3}{2} + 2\pi^2
    - 24\zeta_3 \right)
    + C_F C_A \left( - \f{961}{54} - \f{11\pi^2}{6} 
    + 26\zeta_3 \right)
    + C_F T_F n_l \left( \f{130}{27} + \f{2\pi^2}{3} \right)\;,
\end{align}
whereas the massive quark (anti quark) anomalous dimension is
\be\label{gammaQ}
\begin{split}
   \gamma_0^Q &= - 2 C_F \,, \\
   \gamma_1^Q &= C_F C_A \left( \f{2\pi^2}{3} - \f{98}{9} 
    - 4\zeta_3 \right) + \f{40}{9}\,C_F T_F n_l \,.
\end{split}
\ee
The anomalous dimension for gluons reads
\begin{align}
   \gamma_0^g &= - \beta_0 
    = - \f{11}{3}\,C_A + \f43\,T_F n_l \,, \nn\\
   \gamma_1^g &= C_A^2 \left( -\f{692}{27} + \f{11\pi^2}{18}
    + 2\zeta_3 \right) 
    + C_A T_F n_l \left( \f{256}{27} - \f{2\pi^2}{9} \right)
    + 4 C_F T_F n_l \,.
\end{align}
Finally, we take the functions $F_1$ and $f_2$ from
Ref.~\cite{Ferroglia:2009ii}
\be\label{F_1f_2}
\begin{split}
   F_1(v_{12},v_{23},v_{31}) 
   &= \f{1}{3} \sum_{I,J,K=1}^{3} \ep_{IJK}\,
    \f{\as}{4\pi}\,g(v_{IJ})\,
    \gamma_{\rm cusp}(v_{KI},\as) \,, \\
   f_2\Big( v_{12}, 
    \ln\f{-\sigma_{23}\,v_2\cdot p_3}%
            {-\sigma_{13}\,v_1\cdot p_3} \Big) 
   &= - \f{\as}{4\pi}\,g(v_{12})\,
    \gamma_{\rm cusp}(\as)\,
    \ln\left(\f{-\sigma_{23}\,v_2\cdot p_3}%
            {-\sigma_{13}\,v_1\cdot p_3}\right) \,,
\end{split}
\ee
where
\be
\begin{split}
g(v)=&\,
\f{5\pi^2}{6}-\f{1}{4}\ln^2\left(\f{1+v}{1-v}\right)\\
& +
\f{1}{v}\left[-\f{5\pi^2}{6}+
  \ln\left(\f{2v}{1+v}\right)\ln\left(\f{1+v}{1-v}\right)+
  \f{1}{4}\ln^2\left(\f{1+v}{1-v}\right) -
  \Li_2\left(\f{1-v}{1+v}\right)\right]\\
&+i\pi\left\{\ln\left(\f{1+v}{1-v}\right) -
\f{1}{v}\left[2\ln\left(\f{2v}{1+v}\right) +
  \ln\left(\f{1+v}{1-v}\right)\right]\right\} \; .
\end{split}
\ee
%

%%%%%%%%%%%%%%%%%%%%%%%%%%%%%%%%%%%%%%%%%%%%%%%%%%%%%%%%%%%%%%%%%%%%%%%%%%%%%%%%

\section{Infrared limits of tree-level amplitudes}
\label{sec:treelimits}

%%%%%%%%%%%%%%%%%%%%%%%%%%%%%%%%%%%%%%%%%%%%%%%%%%%%%%%%%%%%%%%%%%%%%%%%%%%%%%%%

\subsection{Collinear limits}
\label{sec:splitting}

\noindent
Consider the collinear limit of two final state momenta $p_1$ and $p_2$
\bea
\label{clim}
p_1^\mu = z p^\mu + k_\perp^\mu - \f{k_\perp^2}{z} 
\f{n^\mu}{2 p\cdot n} \; ,
&& p_2^\mu = (1-z) p^\mu - k_\perp^\mu - \f{k_\perp^2}{1-z}
\f{n^\mu}{2 p\cdot n}
\; ,
\nn \\
s_{12} = 2 p_1 \cdot p_2 = - \f{k_\perp^2}{z(1-z)} \; ,
&& p^2 = n^2 = p\cdot k_\perp = n\cdot k_\perp = 0 \; ,
\nn \\
&k^\mu_\perp \to 0 \; .&
\eea
The matrix element factorizes as
\be
\label{cfac}
| \cm^{(0)}_{a_1,a_2,\dots}(p_1,p_2,\dots) |^2 \simeq 4 \pi \as \,
\f{2}{s_{12}} \, \la \cm^{(0)}_{a,\dots}(p,\dots)|
\sP^{(0)}_{a_1a_2}(z,\kper;\ep) |\cm^{(0)}_{a,\dots}(p,\dots)\ra
\; . 
\ee
The flavor $a$ is set by flavor conservation, i.e.\ if $a_{1,2} = g$ then
$a = a_{2,1}$, while if $a_1 = \bar{a}_2$ then $a = g$. The splitting
functions, $\sP_{a_1a_2}^{(0)}$, are operators in spin space, and act
on the spin of the parton with flavor $a$
\be
\la s | \sP_{a_1a_2}^{(0)} | s' \ra = \Ph_{a_1 a_2}^{(0), \, ss'}
\; ,
\ee
with
\begin{align}
\label{hpggep}
\Ph_{gg}^{(0), \, \mu \nu}(z,\kper;\ep) &= 2C_A
\;\left[ - g^{\mu \nu} \left( \f{z}{1-z} + \f{1-z}{z} \right)
- 2 (1-\ep) z(1-z) \f{\kper^{\mu} \kper^{\nu}}{\kper^2}
\right] \; , \\
\label{hpqqep}
\Ph_{q{\bar q}}^{(0), \, \mu \nu}(z,\kper;\ep) 
= \Ph_{{\bar q}q}^{(0), \, \mu \nu}(z,\kper;\ep)
&= T_F \left[ - g^{\mu \nu} + 4 z(1-z) \f{\kper^{\mu}
    \kper^{\nu}}{\kper^2} \right] \; , \\
\label{hpqgep}
\Ph_{qg}^{(0), \, s s'}(z,\kper;\ep) = \Ph_{{\bar
    q}g}^{(0), \, s s'}(z,\kper;\ep) &= \delta^{ss'} \, C_F
\;\left[ \f{1 + z^2}{1-z} - \ep (1-z) \right] \; , \\[.2cm]
\label{hpgqep}
\Ph_{gq}^{(0), \, s s'}(z,\kper;\ep) = \Ph_{g{\bar
    q}}^{(0), \, s s'}(z,\kper;\ep) &= \Ph_{qg}^{(0), \,s
  s'}(1-z,\kper;\ep) \; .
\end{align}
We will need the average of Eq.~(\ref{cfac}) over the transverse direction
\be
\label{cfacAv}
\overline{| \cm^{(0)}_{a_1,a_2,\dots}(p_1,p_2,\dots) |^2} \simeq 4 \pi \as \,
\f{2}{s_{12}} \, \la \sP^{(0)}_{a_1a_2}(z;\ep) \ra \,
|\cm^{(0)}_{a,\dots}(p,\dots)|^2
\; ,
\ee
where the averaged splitting functions are 
\begin{align}
\label{avhpgg}
\la \sP^{(0)}_{gg}(z;\ep) \ra &= 2C_A \, \left[ \f{z}{1-z} +
  \f{1-z}{z} + z(1-z) \right] \; , \\
\label{avhpqq}
\la \sP^{(0)}_{q{\bar q}}(z;\ep) \ra = \la \sP^{(0)}_{{\bar
    q}q}(z;\ep) \ra &= T_F \left[ 1 - \f{2 z(1-z)}{1-\ep} \right] \; ,
\\
\label{avhpqg}
\la \sP^{(0)}_{qg}(z;\ep) \ra = \la \sP^{(0)}_{{\bar q}g}(z;\ep) \ra
&= C_F \, \left[ \f{1 + z^2}{1-z} - \ep (1-z) \right] \; , \\[.2cm]
\label{avhpgq}
\la \sP^{(0)}_{gq}(z;\ep) \ra = \la \sP^{(0)}_{g{\bar q}}(z;\ep) \ra
&= \la \sP^{(0)}_{qg}(1-z;\ep) \ra \; .
\end{align}
We are also interested in the case without summation over the
polarization of the final state gluon. Let us assume that the latter
has momentum $p_1$. The splitting functions depend on the polarization
vector $\varepsilon^{\mu}_1$, which, for our purposes, may be
assumed to be real. The polarized splitting functions read
\bea
\label{polhpgg}
\Ph_{Pgg}^{(0), \, \mu
  \nu}\left(z,\kper,\varepsilon^{\mu}_1\right) &=& 2 C_A\left[g^{\mu
    \nu}\f{(\varepsilon_1\cdot\kper)^2}{\kper^2}\left(\f{1-z}{z}\right)
  + \left(\f{z}{1-z}\right)
  \varepsilon^{\mu}_1\varepsilon^{\nu}_1-z(1-z)\f{\kper^{\mu}
    \kper^{\nu}}{\kper^2}\right] \; , \\ \label{polhpgq}
\Ph_{Pgq}^{(0), \, s
  s'}\left(z,\kper,\varepsilon^{\mu}_1\right) &=&
\delta^{ss'}C_F\left[-2\f{(\varepsilon_1 \cdot \kper)^2}{\kper^2}
  \left(\f{1-z}{z}\right)  + \f{1}{2}z\right] 
\; .
\eea
We recover the unpolarized splitting functions, if we sum over the
gluon polarizations
\be
\sum_{\text{spin}}\varepsilon^{\mu}_1\varepsilon^{\nu}_1 =
-g^{\mu\nu}+\f{p^{\mu}n^{\nu}+p^{\nu}n^{\mu}}{p\cdot n} \; ,
\quad \sum_{\text{spin}}1=2(1-\ep) \; .
\ee

\begin{figure}[t]
\begin{center}
 \includegraphics[scale=.75]{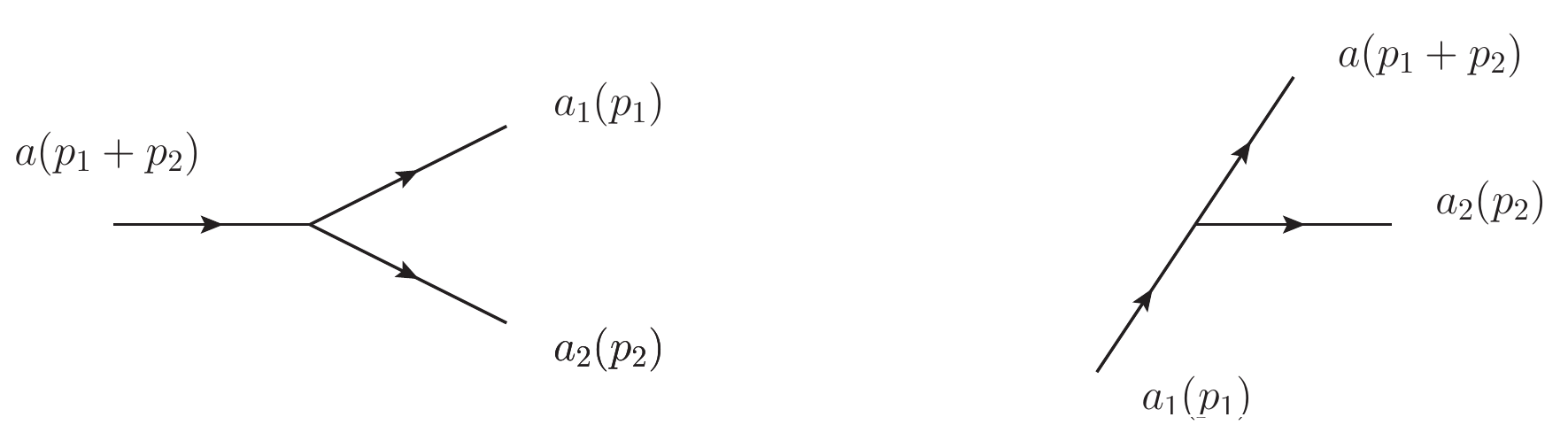}
\end{center}
\caption{\label{fig:splitting} Final state collinear splitting
  configuration (left) vs. initial state collinear splitting
  configuration (right).}
\end{figure}

\noindent
The initial state collinear limit can be recovered from the given
formulae with minor replacements. Both collinear configurations are
depicted schematically in Fig.~(\ref{fig:splitting}). The essential
difference is in the direction of the momenta. All splitting functions
require the following replacement
\be
\label{crossingrule}
\sP_{a_1a_2} \quad \longrightarrow \quad \big( - \big)^{2s_a + 2s_{a_1}}
\sP_{a_1a_2} \; ,
\ee
where $s_a$ and $s_{a_1}$ are the spins of partons $a$ and $a_1$
respectively. The splitting variable $z$ can be obtained in the
collinear limit from the energies of the involved partons. The
crossing amounts to the replacement
\be
z=\f{p_1^0}{p_1^0+p_2^0} \in [0,1] \quad \longrightarrow \quad
z=\f{p_1^0}{p_1^0-p_2^0}  \in [1, +\infty [ \; .
\ee 
Let us now turn to the triple-collinear limit. Consider the set of
three vectors
\be
\label{kin3}
p_i^\mu = x_i p^\mu +k_{\perp i}^\mu - \f{k_{\perp i}^2}{x_i} 
\f{n^\mu}{2p \cdot n} \, , \quad i=1,2,3 \; ,
\ee
where as before $p^2 = n^2 = p\cdot k_{\perp i} = n\cdot k_{\perp i} =
0$. This configuration fulfills no other constraints, but rather the
limits are expressed through derived variables
\be
z_i = \f{x_i}{\sum_{j=1}^3 \,x_j} \; , \quad
{\ktil}_i^\mu = k_{\perp i}^\mu - \f{x_i}{\sum_{k=1}^3 \,x_k} \;
\sum_{j=1}^3 k_{\perp j}^\mu \; , \quad
t_{ij,k} = 2 \;\f{z_i s_{jk}-z_j s_{ik}}{z_i+z_j} +
\f{z_i-z_j}{z_i+z_j} \,s_{ij} \; .
\ee
The factorization formula is obtained in the limit $k_{\perp i}
\rightarrow 0$ and reads
\be
\label{ccfacm}
| \cm^{(0)}_{a_1,a_2,a_3,\dots}(p_1,p_2,p_3,\dots) |^2 \simeq
\left( \f{8 \pi \as}{s_{123}}\right)^{2}
 \; \la
 \cm^{(0)}_{a,\dots}(xp,\dots)|\sP^{(0)}_{a_1a_2a_3}(z_i,k_{\perp
   i};\ep) |\cm^{(0)}_{a,\dots}(xp,\dots)\ra \; ,
\ee
with $s_{123} = (p_1+p_2+p_3)^2$ and $x = x_1+x_2+x_3$. In the
following we will drop the superscript, $(0)$, in
$\sP^{(0)}_{a_1a_2a_3}$ in order not to clutter the notation beyond
the necessary. The flavor of the parton $a$ is obtained by
flavor conservation. The complete set of splitting functions
is taken from Ref.~\cite{Catani:1999ss} (see also
\cite{Campbell:1997hg, Catani:1998nv}). In the case of spin
conservation, we only give the averaged splitting functions $\la
\sP_{a_1 a_2 a_3} \ra$
\be
\Ph_{a_1 a_2 a_3}^{ss'} = \delta^{ss'} \la \sP_{a_1 a_2 a_3} \ra \; .
\ee
We have
\be
\label{qqqprimesf}
\la \sP_{{\bar q}^\prime_1 q^\prime_2 q_3} \ra \, = \f{1}{2} \, 
C_F T_F \,\f{s_{123}}{s_{12}} \left[ - \f{t_{12,3}^2}{s_{12}s_{123}}
+\f{4z_3+(z_1-z_2)^2}{z_1+z_2} 
+ (1-2\ep) \left(z_1+z_2-\f{s_{12}}{s_{123}}\right)
\right] \; .
\ee
The function for identical quark flavors reads
\be
\label{qqqsf}
\la \sP_{{\bar q}_1q_2q_3} \ra \, =
\left[ \la \sP_{{\bar q}^\prime_1q^\prime_2q_3} \ra \, + \,(2\lra 3) \,\right]
+ \la \sP^{({\rm id})}_{{\bar q}_1q_2q_3} \ra \; ,
\ee 
where
\bea
\label{idensf}
\la \sP^{({\rm id})}_{{\bar q}_1q_2q_3} \ra \,
&=& C_F \left( C_F-\f{1}{2} C_A \right)
 \Biggl\{ (1-\ep)\left( \f{2s_{23}}{s_{12}} - \ep \right)\nn\\
&&+ \f{s_{123}}{s_{12}}\Biggl[\f{1+z_1^2}{1-z_2}-\f{2z_2}{1-z_3}
    -\ep\left(\f{(1-z_3)^2}{1-z_2}+1+z_1-\f{2z_2}{1-z_3}\right) 
- \ep^2(1-z_3)\Biggr] \nn\\
&&- \f{s_{123}^2}{s_{12}s_{13}}\f{z_1}{2}\left[\f{1+z_1^2}{(1-z_2)(1-z_3)}-\ep
    \left(1+2\f{1-z_2}{1-z_3}\right)
    -\ep^2\right] \Biggr\} + (2\lra 3) \; .
\eea
The remaining functions are
\be
\label{qggsf}
\la \sP_{g_1 g_2 q_3} \ra \, =
C_F^2 \, \la \sP_{g_1 g_2 q_3}^{({\rm ab})} \ra \,
+ \, C_F C_A \, \la \sP_{g_1 g_2 q_3}^{({\rm nab})} \ra  \;,
\ee
with
\bea
\label{qggabsf}
\la \sP_{g_1 g_2 q_3}^{({\rm ab})} \ra \, 
&=&\Biggl\{\f{s_{123}^2}{2s_{13}s_{23}}
z_3\left[\f{1+z_3^2}{z_1z_2}-\ep\f{z_1^2+z_2^2}{z_1z_2}-\ep(1+\ep)\right]\nn\\
&&+\f{s_{123}}{s_{13}}\Biggl[\f{z_3(1-z_1)+(1-z_2)^3}{z_1z_2}+\ep^2(1+z_3)
-\ep (z_1^2+z_1z_2+z_2^2)\f{1-z_2}{z_1z_2}\Biggr]\nn\\
&&+(1-\ep)\left[\ep-(1-\ep)\f{s_{23}}{s_{13}}\right]
\Biggr\}+(1\lra 2) \;, \\
\label{qggnabsf}
\la \sP_{g_1 g_2 q_3}^{({\rm nab})} \ra \,
&=&\Biggl\{(1-\ep)\left(\f{t_{12,3}^2}{4s_{12}^2}+\f{1}{4}
-\f{\ep}{2}\right)+\f{s_{123}^2}{2s_{12}s_{13}}
\Biggl[\f{(1-z_3)^2(1-\ep)+2z_3}{z_2}\nn\\
&&+\f{z_2^2(1-\ep)+2(1-z_2)}{1-z_3}\Biggr]
-\f{s_{123}^2}{4s_{13}s_{23}}z_3\Biggl[\f{(1-z_3)^2(1-\ep)+2z_3}{z_1z_2}
+\ep(1-\ep)\Biggr]\nn\\
&&+\f{s_{123}}{2s_{12}}\Biggl[(1-\ep)
\f{z_1(2-2z_1+z_1^2) - z_2(6 -6 z_2+ z_2^2)}{z_2(1-z_3)}
+2\ep\f{z_3(z_1-2z_2)-z_2}{z_2(1-z_3)}\Biggr]\nn\\
&&+\f{s_{123}}{2s_{13}}\Biggl[(1-\ep)\f{(1-z_2)^3
+z_3^2-z_2}{z_2(1-z_3)}
-\ep\left(\f{2(1-z_2)(z_2-z_3)}{z_2(1-z_3)}-z_1 + z_2\right)\nn\\
&&-\f{z_3(1-z_1)+(1-z_2)^3}{z_1z_2}
+\ep(1-z_2)\left(\f{z_1^2+z_2^2}{z_1z_2}-\ep\right)\Biggr]\Biggr\}
+(1\lra 2) \;.
\eea
Similarly
\be
\label{gqqsf}
\Ph^{\mu\nu}_{g_1 q_2 {\bar q}_3}  \, =
C_F T_F \, \Ph_{g_1 q_2 {\bar q}_3}^{\mu\nu \,({\rm ab})} \,
+ \, C_A T_F\, \Ph_{g_1 q_2 {\bar q}_3}^{\mu\nu \,({\rm nab})}  \;,
\ee
with
\bea
\label{gqqabsf}
\Ph^{\mu\nu \,({\rm ab})}_{g_1q_2{\bar q}_3} &=&
%\Biggl\{
-g^{\mu\nu}\Biggl[ -2 
+ \f{2 s_{123} s_{23} + (1-\ep) (s_{123} - s_{23})^2}{s_{12}s_{13}}\Biggr]\nn\\
&&+ \f{4s_{123}}{s_{12}s_{13}}\left({\ktil}_{3}^\mu
    {\ktil}_{2}^\nu+{\ktil}_{\hs 2}^\mu
    {\ktil}_{3}^\nu-(1-\ep){\ktil}_{\hs 1}^\mu 
    {\ktil}_{1}^\nu \right)
%    \Biggr\} 
\;, \\
\label{gqqnabsf}
\Ph^{\mu\nu \,({\rm nab})}_{g_1q_2{\bar q}_3} &=& \f{1}{4}
\,\Biggl\{ \f{s_{123}}{s_{23}^2}
\Biggl[ g^{\mu\nu} \f{t_{23,1}^2}{s_{123}}-16\f{z_2^2z_3^2}{z_1(1-z_1)}
\left(\f{{\ktil}_2}{z_2}-\f{{\ktil}_3}{z_3}\right)^\mu
\left(\f{{\ktil}_2}{z_2}-\f{{\ktil}_3}{z_3}\right)^\nu \,\Biggr]\nn\\
&&+ \f{s_{123}}{s_{12}s_{13}} \Biggl[ 2 s_{123} g^{\mu\nu}
- 4 ( {\ktil}_2^\mu {\ktil}_3^\nu + {\ktil}_3^\mu {\ktil}_2^\nu
- (1-\ep) {\ktil}_1^\mu {\ktil}_1^\nu ) \Biggr] \nn\\
&&- g^{\mu\nu} \Biggl[ - ( 1 -2 \ep) + 2\f{s_{123}}{s_{12}} 
\f{1-z_3}{z_1(1-z_1)} + 2\f{s_{123}}{s_{23}} 
\f{1-z_1 + 2 z_1^2}{z_1(1-z_1)}\Biggr]\nn\\
&&+ \f{s_{123}}{s_{12}s_{23}} \Biggl[ - 2 s_{123} g^{\mu\nu}
\f{z_2(1-2z_1)}{z_1(1-z_1)} - 16 {\ktil}_3^\mu {\ktil}_3^\nu 
\f{z_2^2}{z_1(1-z_1)} 
+ 8(1-\ep) {\ktil}_2^\mu {\ktil}_2^\nu \nn\\
&&+ 4 ({\ktil}_2^\mu {\ktil}_3^\nu  + {\ktil}_3^\mu {\ktil}_2^\nu )
%\left(\f{2 z_2 z_3-z_1(z_2 -z_3)}{z_1(1-z_1)}-\ep\right) 
\left(\f{2 z_2 (z_3-z_1)}{z_1(1-z_1)}+ (1-\ep) \right)
%+ 8(1-\ep) {\ktil}_2^\mu {\ktil}_2^\nu
\Biggr] \Biggr\} + \left( 2 \leftrightarrow 3 \right)  \;.
\eea
Finally
\bea
\label{gggsf}
\Ph^{\mu\nu}_{g_1g_2g_3} &=& C_A^2 
\,\Biggl\{\f{(1-\ep)}{4s_{12}^2}
\Biggl[-g^{\mu\nu} t_{12,3}^2+16s_{123}\f{z_1^2z_2^2}{z_3(1-z_3)}
\left(\f{{\ktil}_2}{z_2}-\f{{\ktil}_1}{z_1}\right)^\mu
\left(\f{{\ktil}_2}{z_2}-\f{{\ktil}_1}{z_1}\right)^\nu \;\Biggr]\nn\\
&&- \f{3}{4}(1-\ep)g^{\mu\nu}+\f{s_{123}}{s_{12}}g^{\mu\nu}\f{1}{z_3}
    \Biggl[\f{2(1-z_3)+4z_3^2}{1-z_3}-\f{1-2z_3(1-z_3)}{z_1(1-z_1)}\Biggr]\nn\\
&&+ \f{s_{123}(1-\ep)}{s_{12}s_{13}}\Biggl[2z_1\left({\ktil}^\mu_2
    {\ktil}^\nu_2\hs\f{1-2z_3}{z_3(1-z_3)}+
    {\ktil}^\mu_3{\ktil}^\nu_3\hs
    \f{1-2z_2}{z_2(1-z_2)}\right)\nn\\
&&+ \f{s_{123}}{2(1-\ep)} g^{\mu\nu}
    \left(\f{4z_2z_3+2z_1(1-z_1)-1}{(1-z_2)(1-z_3)}
    - \f{1-2z_1(1-z_1)}{z_2z_3}\right)\nn\\
&&+ \left({\ktil}_2^\mu{\ktil}_3^\nu
   +{\ktil}_3^\mu{\ktil}_2^\nu\right)
    \left(\f{2z_2(1-z_2)}{z_3(1-z_3)}-3\right)\Biggr]\Biggr\}
    + (5\mbox{ permutations}) \;.
\eea
The averaged splitting functions are
\bea
\label{gqqabsfav}
\la \sP^{({\rm ab})}_{g_1q_2{\bar q}_3} \ra \,&=&
%\Biggl\{
-2-(1-\ep)s_{23}\left(\f{1}{s_{12}}+\f{1}{s_{13}}\right)
+ 2\f{s_{123}^2}{s_{12}s_{13}}\left(1+z_1^2-\f{z_1+2z_2 z_3}{1-\ep}\right) 
\nn\\
&&-\f{s_{123}}{s_{12}}\left(1+2z_1+\ep-2\f{z_1+z_2}{1-\ep}\right)
- \f{s_{123}}{s_{13}}\left(1+2z_1+\ep-2\f{z_1+z_3}{1-\ep}\right)
%\Biggr\} 
\;, \\
\label{gqqnabsfav}
\la \sP^{({\rm nab})}_{g_1q_2{\bar q}_3} \ra
\,&=&\Biggl\{-\f{t^2_{23,1}}{4s_{23}^2}
+\f{s_{123}^2}{2s_{13}s_{23}} z_3
\Biggl[\f{(1-z_1)^3-z_1^3}{z_1(1-z_1)}
-\f{2z_3\left(1-z_3 -2z_1z_2\right)}{(1-\ep)z_1(1-z_1)}\Biggr]\nn\\
&&+\f{s_{123}}{2s_{13}}(1-z_2)\Biggl[1
+\f{1}{z_1(1-z_1)}-\f{2z_2(1-z_2)}{(1-\ep)z_1(1-z_1)}\Biggr]\nn\\
&&+\f{s_{123}}{2s_{23}}\Biggl[\f{1+z_1^3}{z_1(1-z_1)}
+\f{z_1(z_3-z_2)^2-2z_2z_3(1+z_1)}
{(1-\ep)z_1(1-z_1)}\Biggr] \nn\\
&&-\f{1}{4}+\f{\ep}{2}
-\f{s_{123}^2}{2s_{12}s_{13}}\Biggl(1+z_1^2-\f{z_1+2z_2z_3}{1-\ep}
\Biggr) \Biggr\}
+ (2\lra  3) \;, \\
\label{gggsfav}
\la \sP_{g_1g_2g_3} \ra \,&=& C_A^2\Biggl\{\f{(1-\ep)}{4s_{12}^2}
t_{12,3}^2+\f{3}{4}(1-\ep)+\f{s_{123}}{s_{12}}\Biggl[4\f{z_1z_2-1}{1-z_3}
+\f{z_1z_2-2}{z_3}+\f{3}{2} +\f{5}{2}z_3\nn\\
&&+\f{\left(1-z_3(1-z_3)\right)^2}{z_3z_1(1-z_1)}\Biggr]
+\f{s_{123}^2}{s_{12}s_{13}}\Biggl[\f{z_1z_2(1-z_2)(1-2z_3)}{z_3(1-z_3)}
+z_2z_3 -2 +\f{z_1(1+2z_1)}{2}\nn\\
&&+\f{1+2z_1(1+z_1)}{2(1-z_2)(1-z_3)}
+\f{1-2z_1(1-z_1)}{2z_2z_3}\Biggr]\Biggr\}
+ (5\mbox{ permutations}) \;.
\eea
Initial state collinear limits are recovered by crossing (\ref{crossingrule}).

%%%%%%%%%%%%%%%%%%%%%%%%%%%%%%%%%%%%%%%%%%%%%%%%%%%%%%%%%%%%%%%%%%%%%%%%%%%%%%%%

\subsection{Soft limits}
\label{sec:soft}

\noindent
Consider the limit of vanishing gluon momentum, $q \to 0$. The matrix
element factorizes as
\be
\begin{split}
\label{ccfact}
| \cm^{(0)}_{g,a_1,\dots}(q,p_1,\dots) |^2 &\simeq
- 4 \pi \as \sum_{ij}\, {\cal S}_{ij}(q)
\;\la \cm^{(0)}_{a_1,\dots}(p_1,\dots)|\cT_i \cdot
\cT_j|\cm^{(0)}_{a_1,\dots}(p_1,\dots)\ra \\ &=
- 4 \pi \as \sum_{(i,j)}\, \Big( {\cal S}_{ij}(q) - {\cal S}_{ii}(q)
\Big) \;\la \cm^{(0)}_{a_1,\dots}(p_1,\dots)|\cT_i \cdot
\cT_j|\cm^{(0)}_{a_1,\dots}(p_1,\dots)\ra
 \;,
\end{split}
\ee
where
\be
\label{eikfun}
{\cal S}_{ij}(q) = \f{p_i \cdot p_j}{(p_i \cdot q)\, (p_j\cdot q)}
\;.
\ee
The double-soft limit is defined by rescaling uniformly the momenta,
$q_1$ and $q_2$, of two gluons or of a quark and an anti-quark of the
same flavor
\be
q_1\rightarrow\lambda q_1\,,\quad q_2\rightarrow\lambda q_2\,,\quad
\lambda \rightarrow 0\;.
\ee
In the case of a final state $q{\bar q}$-pair, the matrix element
factorizes as
\be
\label{qqsoftfac}
| \cm^{(0)}_{q,{\bar q},a_1,\dots}(q_1,q_2,p_1,\dots) |^2 \simeq
 \left( 4 \pi \as \right)^2 T_F
\sum_{ij} {\cal I}_{ij}(q_1,q_2) \,
\la \cm^{(0)}_{a_1,\dots}(p_1,\dots)|\cT_i \cdot
\cT_j|\cm^{(0)}_{a_1,\dots}(p_1,\dots)\ra\; ,
\ee
where the function $ {\cal I}_{ij}(q_1,q_2)$ has the form
\be
\label{Iij1}
{\cal I}_{ij}(q_1,q_2) = \f{(p_i \cdot q_1)\, (p_j \cdot q_2)
+ (p_j \cdot q_1)\, (p_i \cdot q_2) - (p_i \cdot p_j) 
\,(q_1 \cdot q_2)}{(q_1 \cdot q_2)^2 
\,[p_i\cdot (q_1+q_2)]\, [p_j \cdot (q_1+q_2)]} \,.
\ee
In the case of two final state gluons, the matrix element
factorizes as
\be
\begin{split}
|\cm^{(0)}_{g,g,a_1,\dots}(q_1,q_2,p_1,\dots)|^2 \simeq
\left(4\pi\as\right)^2
\Biggl[ &\f{1}{2}\sum_{ijkl}{\cal S}_{ij}(q_1){\cal S}_{kl}(q_2) \,
\la\cm^{(0)}_{a_1,\dots}(p_1,\dots)|\left\{\cT_i\cdot\cT_j,
\cT_k\cdot\cT_l\right\} |\cm^{(0)}_{a_1,\dots}(p_1,\dots)\ra
\\
&-C_A\sum_{ij}{\cal S}_{ij}(q_1,q_2) \,
\la\cm^{(0)}_{a_1,\dots}(p_1,\dots)|\cT_i\cdot\cT_j
|\cm^{(0)}_{a_1,\dots}(p_1,\dots)\ra
\Biggr]\;.
\end{split}
\ee
The soft function ${\cal S}_{ij}(q_1,q_2)$ can be split into two parts
\be
\label{eq:soft}
{\cal S}_{ij}(q_1,q_2) = {\cal S}^{m=0}_{ij}(q_1,q_2) + \left( m_i^2
\; {\cal S}^{m \neq 0}_{ij}(q_1,q_2) + m_j^2 \; {\cal S}^{m \neq
  0}_{ji}(q_1,q_2) \right) \; ,
\ee
where the first term has been given in \cite{Catani:1999ss}
and reads
\bea
\label{eq:sij0}
{\cal S}^{m=0}_{ij}(q_1,q_2) &=& \f{(1-\ep)}{(q_1 \cdot q_2)^2}
\f{p_i \cdot q_1 \; p_j \cdot q_2 + p_i \cdot q_2 \; p_j \cdot
  q_1}{p_i \cdot (q_1+q_2) \; p_j \cdot (q_1+q_2)} \nn
\\ \nn \\
&& - \f{(p_i \cdot p_j)^2}{2 \; p_i \cdot q_1 \; p_j \cdot q_2 \; p_i
  \cdot q_2 \; p_j \cdot q_1} \left[ 2 - \f{p_i \cdot q_1 \; p_j
    \cdot q_2 + p_i \cdot q_2 \; p_j \cdot q_1}{p_i \cdot (q_1+q_2) \;
    p_j \cdot (q_1+q_2)} \right] \nn \\  \nn \\
&& + \f{p_i \cdot p_j}{2 \; q_1 \cdot q_2} \left[ \f{2}{p_i
    \cdot q_1 \; p_j \cdot q_2} + \f{2}{p_j \cdot q_1 \; p_i \cdot
    q_2} - \f{1}{p_i \cdot (q_1+q_2) \; p_j \cdot (q_1+q_2)}
  \right. \nn \\ \nn \\
&& \times \left. \left( 4 + \f{(p_i \cdot q_1 \; p_j \cdot q_2 +
    p_i \cdot q_2 \; p_j \cdot q_1)^2}{\; p_i \cdot q_1 \; p_j \cdot
    q_2 \; p_i \cdot q_2 \; p_j \cdot q_1} \right) \right] \; .
\eea
The second contribution in Eq.~(\ref{eq:soft}) was derived in
Ref.~\cite{Czakon:2011ve} and represents additional terms generated by
non-vanishing masses. The relevant function is
\bea
\label{eq:sijm}
{\cal S}^{m\neq0}_{ij}(q_1,q_2) &=
&  - \f{1}{4  \; q_1 \cdot q_2 \; p_i \cdot
q_1 \; p_i \cdot q_2} + \f{p_i \cdot p_j \; p_j
\cdot (q_1+q_2)}{2 \; p_i \cdot q_1 \; p_j \cdot q_2 \; p_i \cdot q_2 \;
  p_j \cdot q_1 \; p_i \cdot (q_1+q_2)} \nn \\ \nn \\
&& - \f{1}{2 \; q_1 \cdot q_2 \; p_i \cdot (q_1+q_2) \; p_j \cdot
  (q_1+q_2) } \left( \f{(p_j \cdot q_1)^2}{p_i \cdot q_1 \; p_j
  \cdot q_2} + \f{(p_j \cdot q_2)^2}{p_i \cdot q_2 \; p_j \cdot
  q_1} \right) \; . \nn \\
\eea
%

%%%%%%%%%%%%%%%%%%%%%%%%%%%%%%%%%%%%%%%%%%%%%%%%%%%%%%%%%%%%%%%%%%%%%%%%%%%%%%%%

\section{Infrared limits of one-loop matrix elements}
\label{sec:1Llimits}

\noindent
We consider a one-loop amplitude with $n+1$ partons, which is
integrated over the phase space in the real-virtual
contribution. According to \ref{sec:VirtualIR}, we can separate it
into a divergent part and a finite remainder
\be
\label{RVmatrix}
 2\R\,\la \cm_{n+1}^{(0)} |\cm_{n+1}^{(1)}\ra
 =2\R\,\la \cm_{n+1}^{(0)}
 |\bm{\mathrm{Z}}^{(1)}|\cm_{n+1}^{(0)}\ra +
 2\R\,\la \cm_{n+1}^{(0)} |\cf_{n+1}^{(1)}\ra
 \;.
\ee
In the following we give explicit formulae for the collinear and soft
limits of each of the three contributions. Notice that the expressions
for the finite remainder are only valid at $\ep = 0$.

%%%%%%%%%%%%%%%%%%%%%%%%%%%%%%%%%%%%%%%%%%%%%%%%%%%%%%%%%%%%%%%%%%%%%%%%%%%%%%%%

\subsection{Collinear limit}
\label{sec:collinearM1}

\noindent
The factorization of the one-loop amplitude in the final state
collinear limit, (\ref{clim}), reads \cite{Bern:1994zx, Bern:1998sc,
  Kosower:1999xi, Kosower:1999rx, Bern:1999ry, Somogyi:2006db}
\begin{multline}
\label{coll1LFull}
2\R \, \la\cm_{a_1,a_2,\dots}^{(0)}(p_1,p_2,\dots)|
\cm_{a_1,a_2,\dots}^{(1)}(p_1,p_2,\dots)\ra
\simeq \\ 4 \pi \as \f{2}{s_{12}} \Biggl[ 2\R\;
\la \cm^{(0)}_{a,\dots}(p,\dots) | {\sP}^{(0)}_{a_1 a_2}(z,\kper;\ep) |
\cm^{(1)}_{a,\dots}(p,\dots) \ra
+ \f{\as}{4\pi}
\la \cm^{(0)}_{a,\dots}(p,\dots)
| {\sP}^{(1)}_{a_1 a_2}(z,\kper;\ep) | \cm^{(0)}_{a,\dots}(p,\dots) \ra
\Biggr] \; .
\end{multline}
The one-loop splitting functions, $\sP^{(1)}_{a_1a_2}(z,\kper;\ep)$,
are operators in spin space
\be
\la s | \sP^{(1)}_{a_1a_2}(z,\kper;\ep) | s' \ra =
\Ph^{(1), \,ss'}_{a_1a_2}(z,\kper;\ep)
\; ,
\ee
with
\be
\begin{split}
\Ph_{gg}^{(1), \,\mu \nu}(z,\kper;\ep) &= r^{gg}_{SR}(z)\,\Ph_{gg}^{(0),\mu
  \nu}(z,\kper;\ep) - 4 C_A r^{gg}_{NS}\left[1-2\ep z(1-z)\right]
\f{\kper^{\mu} \kper^{\nu}}{\kper^2} \; , \\
\Ph_{q{\bar q}}^{(1), \,\mu \nu}(z,\kper;\ep) = \Ph_{{\bar q}q}^{(1), \,\mu
  \nu}(z,\kper;\ep) &= r^{{\bar q}q}_{SR}(z)\,\Ph_{q{\bar q}}^{(0),\mu
  \nu}(z,\kper;\ep) \; , \\[.2cm]
\Ph_{qg}^{(1), \,s s'}(z,\kper;\ep) = \Ph_{{\bar q}g}^{(1), \,s
  s'}(z,\kper;\ep) &= r^{qg}_{SR}(z)\,\Ph_{qg}^{(0),s s'}(z,\kper;\ep)
+ C_F r^{qg}_{NS}\left[1-\ep(1-z)\right] \,\delta^{ss'} \;, \\[.2cm]
\Ph_{gq}^{(1), \,s s'}(z,\kper;\ep)=\Ph_{g{\bar q}}^{(1), \,s
  s'}(z,\kper;\ep) &= \Ph_{qg}^{(1), \,s s'}(1-z,\kper;\ep) \; .
\end{split}
\ee
The renormalized singular coefficients $r^{a_1a_2}_{SR}$ are
related to the unrenormalized singular coefficients $r^{a_1a_2}_S$
through
\be
r^{a_1a_2}_{SR}(z) = 2\R \, \left( -\f{\mu_R^2}{s_{12}}
\right)^\ep c_\Gamma \, r^{a_1a_2}_S(z)-\f{\beta_0}{\ep} \; ,
\ee
where
\be
\label{eq:Rlog}
\R \, \left( -\f{\mu_R^2}{s_{12}} \right)^\ep = \left(
\f{\mu_R^2}{s_{12}} \right)^\ep \cos(\pi \ep) \; , \quad c_{\Gamma} =
e^{\ep\gamma_{\mathrm{E}}}
\f{\Gamma^2(1-\ep)\Gamma(1+\ep)}{\Gamma(1-2\ep)} \; ,
\ee
and
\bea
r^{gg}_S(z)&=&-\f{C_A}{\ep^2}\left[\left(\f{z}{1-z}\right)^{\ep}
  \f{\pi \ep}{\sin(\pi \ep)}-\sum_{m=1}^{\infty}{2\ep^{2m-1}\Li_{2m-1}
    \left(-\f{1-z}{z}\right)}\right]\; ,\label{eq:rggs}\\
r^{{\bar q}q}_S(z)&=&\f{1}{\ep^2}\left(C_A-2C_F\right) +
\f{C_A}{\ep^2}\sum_{m=1}^{\infty}{\ep^m\left[\Li_m\left(-\f{z}{1-z}\right)
    +\Li_m\left(-\f{1-z}{z}\right)\right]}\label{eq:rqqs}\nn\\
&&+\f{1}{1-2\ep}\left[\f{1}{\ep}\left(\gamma^q_0-\gamma^g_0\right)+C_A-2C_F
  +\f{C_A+4T_Fn_l}{3(3-2\ep)}\right]\; ,\\
r^{qg}_S(z)&=&-\f{1}{\ep^2}\left[C_A\left(\f{z}{1-z}
  \right)^{\ep}\f{\pi \ep}{\sin(\pi \ep)} +
  \sum_{m=1}^{\infty}{\ep^m\left[\left(1+(-1)^m\right)C_A-2C_F \right]
    \Li_m\left(-\f{1-z}{z}\right)}\right]\; \label{eq:rqgs} .
\eea
The non-singular coefficients read
\be
r^{gg}_{NS} = 2\R \, \left( -\f{\mu_R^2}{s_{12}} \right)^\ep
c_{\Gamma} \, \f{C_A(1-\ep)-2T_Fn_l}{(1-2\ep)(2-2\ep)(3-2\ep)} 
\; , \quad r^{qg}_{NS} = 2\R \, \left( -\f{\mu_R^2}{s_{12}}
\right)^\ep c_{\Gamma} \, \f{C_A-C_F}{1-2\ep}\;.
\ee
The initial state collinear limit can be recovered using crossing
(\ref{crossingrule}), and the following
\be
\R \, \left( -\f{\mu_R^2}{s_{12}} \right)^\ep = \left(
-\f{\mu_R^2}{s_{12}} \right)^\ep \; , \quad \R \, \Bigl( \f{z}{1-z}
\Bigr)^\ep = \Bigl( - \f{z}{1-z} \Bigr)^\ep \cos(\pi \ep) \; .
\ee
Notice that polylogarithms of $-z/(1-z) \in [1, +\infty [$ develop an
imaginary part in this case. The real parts, which are necessary for a
next-to-next-to-leading order calculation are obtained with
\begin{align}
\label{eq:analytic}
&\R\left(\Li_1\left(1/x\right)\right) =
-\R\left(\ln\left(1-1/x\right)\right) =-\ln(1-x)+\ln(x) \; ,
&\R\left(\Li_2\left(1/x\right)\right) =
\f{\pi^2}{3}-\f{\ln^2(x)}{2}-\Li_2(x) \; , \nn \\
&\R\left(\Li_3\left(1/x\right)\right) =
-\f{\pi^2}{3}\ln(x)+\f{\ln^3(x)}{6}+\Li_3(x) \; ,
&\R\left(\Li_4\left(1/x\right)\right) =
\f{\pi^4}{45}+\f{\pi^2}{6}\ln^2(x)-\f{\ln^4(x)}{24}-\Li_4(x) \; ,
\end{align}
with $x = -(1-z)/z \in [0,1]$.

%%%%%%%%%%%%%%%%%%%%%%%%%%%%%%%%%%%%%%%%%%%%%%%%%%%%%%%%%%%%%%%%%%%%%%%%%%%%%%%%

\subsection{Soft limit}
\label{sec:softM1}

\noindent
The factorization of the one-loop amplitude in the soft limit, $q \to 0$,
reads \cite{Catani:2000pi, Bierenbaum:2011gg}
\be
\label{softlimit1L}
\begin{split}
2 \R \,
\la\cm_{g,a_1,\dots}^{(0)}(q,p_1,\dots)|&\cm_{g,a_1,\dots}^{(1)}(q,p_1,\dots)\ra
\simeq \; \\ -4\pi\as &\left\{ \sum_{(i,j)} \Bigl(
  \mathcal{S}_{ij}(q) - \mathcal{S}_{ii}(q) \Bigr) \, 2\R \, \la
  \cm_{a_1,\dots}^{(0)}(p_1,\dots)|\cT_i\cdot\cT_j
  |\cm_{a_1,\dots}^{(1)}(p_1,\dots) \ra  \right. \\ 
&\left. + \f{\as}{4\pi} \left[\sum_{(i,j)} \left({\cal
      S}_{ij}(q)-{\cal S}_{ii}(q)\right) \, R_{ij}\, \la
    \cm_{a_1,\dots}^{(0)}(p_1,\dots)|\cT_i\cdot\cT_j
    |\cm_{a_1,\dots}^{(0)}(p_1,\dots) \ra
    \right. \right.\\ & \hspace{.75cm} \left. \left. -  4\pi\sum_{(i,j,k)} {\cal
      S}_{ik}(q) \, I_{ij} \,\la
    \cm_{a_1,\dots}^{(0)}(p_1,\dots)|f^{abc} T^a_i T^b_j T^c_k
    |\cm_{a_1,\dots}^{(0)}(p_1,\dots)\ra\right]  \right\} \; .
\end{split}
\ee
The coefficients $R_{ij}$ and $I_{ij}$ depend on the kinematics and
the masses of the partons $i$ and $j$. In the case of two massless
partons, they are \cite{Catani:2000pi}
\bea
\label{RImassless}
R_{i_0j_0} &=& 4 C_A \left({\textstyle \f{1}{2}} \mu_R^2
\mathcal{S}_{i_0j_0}(q) \right)^\ep \left( - \f{1}{2\ep^2} s_\Gamma
\cos(\pi \ep) \right) - \f{\beta_0}{\ep}\;,\nn\\ I_{i_0j_0}
&=& 2 \left({\textstyle \f{1}{2}} \mu_R^2 \mathcal{S}_{i_0j_0}(q)
\right)^\ep \left( \f{1}{2\pi\ep^2} s_\Gamma \sin(\pi \ep)
\right) \left(\theta(\sigma_{i_0j_0}) - \theta(\sigma_{i_0q}) -
\theta(\sigma_{j_0q})\right) \; ,
\eea
where
\be
s_{\Gamma}=e^{\ep\gamma_{\mathrm{E}}}\f{\Gamma^3(1-\ep)\Gamma^2(1+\ep)}{\Gamma(1-2\ep)}
\; .
\ee
In the case of at least one massive parton, $R_{ij}$ and $I_{ij}$ are
only known as expansions in $\ep$ \cite{Bierenbaum:2011gg}
\bea
\label{RImassive}
R_{ij}&=& 4C_A\left({\textstyle \f{1}{2}} \mu_R^2 \mathcal{S}_{ij}(q) \right)^\ep
\sum_{n=-2}^{1}\ep^n\,R^{(n)}_{ij}\;-\f{\beta_0}{\ep} \; ,\nn \\
I_{ij}&=& 2 \left({\textstyle \f{1}{2}} \mu_R^2
\mathcal{S}_{ij}(q) \right)^\ep \sum_{n=-1}^{1}\ep^n\,I^{(n)}_{ij}\;.
\label{eq:gij-explicitCase}
\eea
In the case of one massive and one massless parton
\bea
\label{eq:Result-Case-1}
I^{(-1)}_{Ij_0} &=& -{1\over 2} \, , \nn\\
R_S\, I^{(0)}_{Ij_0} &=& 2m_I^2 \pjoq \ln\left({\ai \over 2}\right)\, , \nn\\
R_S\, I^{(1)}_{Ij_0} &=& 4\left[\pipjo \piq - m_I^2 \pjoq\right] \Li_2\left(1- {\ai \over 2} \right)  + m_I^2 \pjoq \ln^2\left({\ai \over 2}\right) 
\nn\\
&& + \pi^2 {-2 \pipjo \piq + m_I^2 \pjoq\over 2} \, , \nn\\
R^{(-2)}_{Ij_0} &=& -{1\over 2} \, , \nn\\
R^{(-1)}_{Ij_0} &=& 0  \, , \nn\\
R_S\, R^{(0)}_{Ij_0} &=&  m_I^2 \pjoq \ln^2\left({\ai \over 2}\right) -\pi^2 {5(2 \pipjo \piq - m_I^2 \pjoq) \over 6}\, , \nn\\
R_S\, R^{(1)}_{Ij_0} &=& 4 \left[\pipjo \piq - m_I^2 \pjoq\right] \Li_3\left({\ai \over 2}\right) - \zeta_3 {4\left[7 \pipjo \piq - 5 m_I^2 \pjoq\right]\over 3} \nn\\
&& + 2 \left[\pipjo \piq - m_I^2 \pjoq\right] \ln\left(1- {\ai \over 2} \right) \ln^2\left({\ai \over 2}\right) \nn\\
&&+ \ln\left({\ai \over 2}\right) \left( \pi^2{-2 \pipjo \piq - 5 m_I^2 \pjoq\over 3 } \right.\\ 
&&\left.+ 4 \left[\pipjo \piq - m_I^2 \pjoq\right] \Li_2\left(1- {\ai \over 2} \right) \right)\, ,\nn
\eea
where
\be
R_S = 4\left[m_I^2 \pjq - 2 \pipjo \piq \right] \; , \quad \alpha_I =
{m_I^2(p_{j_0}\cdot q)\over (p_I\cdot q) (p_I\cdot p_{j_0}) } \; .
\label{eq:def-ai-aj}
\ee
In the case of two massive partons\footnote{The expression for
$R_{IJ}^{(1)}$ published in \cite{Bierenbaum:2011gg} contains three
typos. Twice, there appears a $\ln^4 2$ instead of $4 \ln 2$, and
there is a missing plus sign between two terms. The latter is seen
through the wrong dimension of a term. The expression reproduced here
is free of these misprints. On the other hand, the electronic version
of the result attached to \cite{Bierenbaum:2011gg} is correct.}
\bea
I^{(-1)}_{IJ} &=& -1+{1\over 2v} \, , \nn\\
I^{(0)}_{IJ} &=& 
{\ln(v)\over v} + {\ln(x)\over 2 v} +\left(1 + {1\over 2 v}\right)\ln(1 + x^2)  \nn\\ 
&& + {1\over Q_S}\left[-4 {m_J^2 \piq^2 - m_I^2 \pjq^2\over v} \ln\left({\ai\over \aj}\right) + 16 \pipj \piq \pjq \ln(x) \right]\, , \nn\\
&& \nn\\
I^{(1)}_{IJ} &=& 
\f{1}{v}\left(
\f{1}{16} \ln ^2\left(\f{\ai}{\aj}\right)+\Li_2\left(x^2\right)+\ln (v) \left(\ln \left(x^2+1\right)+\ln (x)\right)+\ln ^2(v)+\f{1}{4} \ln
   ^2\left(x^2+1\right)\right. \nn\\
&&+ \left. \f{1}{2} \ln (x) \ln \left(x^2+1\right)+\f{\ln ^2(x)}{4}-\f{\pi ^2}{8}
\right) \nn\\
&&+
%%%%%%%%%%%%%%%%%%
\f{1}{Q_S}
\Biggl[
\pipj \piq \pjq \left(2 \ln ^2\left(\f{\ai}{\aj}\right)-16 \ln \left(x^2+1\right) 
\ln (x)+8 \ln ^2(x)-\f{8 \pi^2}{3}\right)\nn \\
&&
+\left(m_I^2 \pjq^2+m_J^2 \piq^2\right)\left(8 \ln^2\left(x^2+1\right)-\f{4 \pi ^2}{3}\right)\nn \\
&&
-4 \left(m_J^2 \piq^2-m_I^2 \pjq^2\right) \f{1}{v}
\left(2 \ln (v)+ \ln \left(x^2+1\right)+ \ln (x)\right) \ln \left(\f{\ai}{\aj}\right)\nn \\
&&
+\left(m_J^2 \piq^2+m_I^2 \pjq^2-\pipj \piq \pjq\right)
\Big( \nn \\
&& 
32 \ln (2) \left(- \ln (\ai+v+1)- \ln (\aj+v+1)-2 \ln \left(x^2+1\right)- \ln (x)\right)+64 \ln^2(2) \nn \\
&&
+16 \ln \left(x^2+1\right) ( \ln (\ai+v+1)+ \ln (\aj+v+1)) 
%\nn \\
%&&
+16 \ln (\ai+v+1) \ln (\aj+v+1)\nn \\
&&
+16 \ln (x) \left( \ln (-\aj+v+1)+ \ln (\aj+v-1)+2 \ln \left(x^2+1\right)\right)
% \nn \\
%&&
-16 \ln ^2(x)
 \nn \\
&&
+ 8 \ln \left(\f{\ai}{\aj}\right) ( \ln (\aj+v-1)- \ln (-\aj+v+1))
% \nn \\
%&&
+4 \ln ^2\left(\f{\ai}{\aj}\right) \nn \\
&&
+16 \Li_2\left(\f{-v+\aj+1}{2 \aj}\right) 
%\nn \\
%&&
+16 \Li_2\left(2-\f{2 \aj}{-v+\aj+1}\right) 
%\nn \\
%&&
-16 \Li_2\left(\f{-v+\aj+1}{v+\aj+1}\right) \nn \\
&&
+16 \Li_2\left(\f{v+\aj+1}{2 v+2}\right) 
%\nn \\
%&&
-16 \Li_2\left(-\f{(v-1) (v+\aj+1)}{(v+1) (-v+\aj+1)}\right) 
%\nn \\
%&&
+16 \Li_2\left(\f{2 \aj}{v+\aj+1}\right) 
\Biggr)
\Biggr]\, , \nn \\
%%%%%%%%%%%%%%%%%%%%%
R^{(-2)}_{IJ} &=& -{1\over 2} \, , \nn\\
R^{(-1)}_{IJ} &=& {1\over 2} \left(-1 + {1\over v}\right) \ln(x) +{1\over 2} \ln(1 + x^2)  \, , \nn\\
R^{(0)}_{IJ} &=&  {1\over 2 v} \Li_2(x^2) + \pi^2 \left({19\over 24} - {7\over 12 v}\right) + {1\over v} \ln(v) \ln(x)   +{1\over 2}\left(1 +{1\over v}\right) \ln(x) \ln(1 + x^2) \nn\\
&&   - {1\over 4} \ln^2(1 + x^2)   + {1\over Q_S}\left[ \left(m_J^2 \piq^2 + m_I^2 \pjq^2\right) \ln^2\left({\ai\over \aj}\right) \right.\nn\\
&&\left. + 4\left(m_J^2 \piq^2 + m_I^2 \pjq^2\right) \ln^2(x) - 4{m_J^2 \piq^2 - m_I^2 \pjq^2\over v} \ln\left({\ai\over \aj}\right) \ln(x) \right] \, , \nn
\nn\\
R^{(1)}_{IJ} &=& 
\f{1}{v}
\left(
-\ln (v) \left(\ln (x) \ln \left(x^2+1\right)+\pi ^2\right)
+\f{\ln ^3(x)}{12}+\f{\zeta (3)}{2}
\right.
\nn\\
&& \left.
+\ln (x) 
\left(
\f{1}{16} \ln ^2\left(\f{\ai}{\aj}\right)
+\f{\Li_2\left(x^2\right)}{2}
-\f{3}{4} \ln ^2\left(x^2+1\right)
-\f{5 \pi^2}{24}
\right)\right.
\nn\\
&&\left. 
-\left(\f{\Li_2\left(x^2\right)}{2}+\f{5 \pi ^2}{12}\right) \ln \left(x^2+1\right)
-\f{1}{2} \left(2 \Li_3\left(1-x^2\right)+\Li_3\left(x^2\right)\right) 
\right)
\nn\\
%%%%%%%%%%%%%%%%%%%%%%
&&
+\f{1}{Q_S}
\Big[
\pipj\piq\pjq
\Big(
\f{32 \ln ^3(x)}{3}-\f{280 \zeta (3)}{3}-32 \ln \left(x^2+1\right) \ln ^2(x) \nn\\
&&
\nn\\
&&
+\ln \left(x^2+1\right) \left(36 \pi ^2-4 \ln ^2\left(\f{\ai}{\aj}\right)\right) 
+\left(48 \ln^2\left(x^2+1\right)-\f{40 \pi ^2}{3}\right) \ln (x)
\nn\\
&&
-\f{88}{3} \ln ^3\left(x^2+1\right)
\Big)
\nn\\
&&
%%%%%%%%%%%%%%%%%%%%%%
+(m_J^2 \piq^2 + m_I^2 \pjq^2)
\Big(
\left(3 \ln \left(x^2+1\right)+\ln (x)\right) \ln ^2\left(\f{\ai}{\aj}\right)
\nn\\
&&
+28 \ln ^3\left(x^2+1\right)
-44 \ln^2\left(x^2+1\right) \ln (x)
-\f{70}{3} \pi ^2 \ln \left(x^2+1\right)
+28 \ln \left(x^2+1\right) \ln ^2(x)
\nn\\
&&
-\f{28}{3} \ln ^3(x)
+\f{2}{3} \pi ^2 \ln (x)
+\f{224 \zeta (3)}{3}
\Big)
\nn\\
&&
%%%%%%%%%%%%%%%%%%%%%%
-\f{(m_J^2 \piq^2 - m_I^2 \pjq^2)}{v}
\ln \left(
\f{\ai}{\aj}\right) 
\left(
4 \Li_2\left(x^2\right)
+4 \ln \left(x^2+1\right) \ln (x)
+8 \ln (v) \ln (x)
-\f{14 \pi ^2}{3}
\right)
\nn\\
&&
%%%%%%%%%%%%%%%%%%%%%%
+(m_I^2 \pjq^2 + m_J^2 \piq^2 - \pipj \piq \pjq)\Big(
\ln ^3\left(\f{\ai}{\aj}\right)
- \ln ^2\left(\f{\ai}{\aj}\right)2 \ln (v)
\nn\\
&&
+ \ln ^2\left(\f{\ai}{\aj}\right)2 (\ln (\ai+v+1)+\ln (-\aj+v+1)+\ln (\aj+v-1)-3 \ln (2))
\nn\\
&&
\nn\\
&&
+\ln \left(\f{\ai}{\aj}\right)
\Big(
2 \ln ^2(v) 
-12 \ln ^2(\ai+v+1)
+6 \ln^2(\aj+v+1)
-4 \ln ^2(x)
-12 \ln ^2\left(x^2+1\right)
\nn\\
&&
+28 \ln (2) \ln (\ai+v+1)
+4 \ln (v) (\ln (\ai+v+1)-2 \ln (\aj+v+1)+\ln (2))
-10 \ln ^2(2)
\nn\\
&&
-4 \ln (\ai+v+1) \ln (\aj+v+1)
-8 \ln (2) \ln (\aj+v+1)
\nn\\
&&
+8 (2 \ln (\ai+v+1)-\ln (-\aj+v+1)+\ln (\aj+v-1)+\ln (\aj+v+1)-3 \ln (2)) \ln (x)
\nn\\
&&
+24 (\ln (2)-\ln (\ai+v+1)) \ln \left(x^2+1\right)
+24 \ln (x) \ln \left(x^2+1\right)
-\f{2}{3} \pi ^2
\Big)
\nn \\
&&
+\f{32}{3} \ln ^3(\ai+v+1)
+12 \ln ^3(\aj+v+1)
-32 \ln (2) \ln ^2(\ai+v+1)
-8 \ln (v) \ln ^2(\aj+v+1)
\nn\\
&&
+4 \ln (\ai+v+1) \ln ^2(\aj+v+1)
-40 \ln (2) \ln ^2(\aj+v+1)
-8 \ln (v) \ln ^2(x)
-\f{4}{3} \pi ^2 \ln (v)
\nn\\
&&
+8 (3 \ln (\ai+v+1)+\ln (-\aj+v+1)+\ln (\aj+v-1)-5 \ln (2)) \ln ^2(x)
\nn\\
&&
+40 (\ln (\ai+v+1)+\ln (\aj+v+1)-2 \ln (2)) \ln ^2\left(x^2+1\right)
+36 \ln ^2(2) \ln (\ai+v+1)
\nn\\
&&
+8 \ln (v) \ln (\ai+v+1) \ln \left(\f{1}{2} (\aj+v+1)\right)
-4 \ln ^2(v) (\ln (\ai+v+1)-\ln (\aj+v+1))
\nn\\
&&
+\f{4}{3} \pi ^2 (-3 \ln (\ai+v+1)-4 \ln (\aj+v+1)+7 \ln (2))
+8 \ln (2) \ln (v) \ln (\aj+v+1)
\nn\\
&&
-8 \ln (2) \ln (\ai+v+1) \ln (\aj+v+1)
+44 \ln ^2(2) \ln (\aj+v+1)
+4 \ln ^2(v) \ln (x)
\nn\\
&&
-24 \ln ^2(\ai+v+1) \ln (x)
-4 \ln ^2(\aj+v+1) \ln (x)
+8 \ln (v) (\ln (2)-\ln (\ai+v+1)) \ln (x)
\nn\\
&&
+56 \ln (2) \ln (\ai+v+1) \ln (x)
-8 \ln (\ai+v+1) \ln (\aj+v+1) \ln (x)
-36 \ln ^2(2) \ln (x)
\nn\\
&&
+16 \ln (2) \ln (\aj+v+1) \ln (x)
+32 \ln ^2(\ai+v+1) \ln \left(x^2+1\right)
+80 \ln ^2(2) \ln \left(x^2+1\right)
\nn\\
&&
+32 \ln ^2(\aj+v+1) \ln \left(x^2+1\right)
-80 \ln (2) \ln (\ai+v+1) \ln \left(x^2+1\right)
\nn\\
&&
+16 \ln (\ai+v+1) \ln (\aj+v+1) \ln \left(x^2+1\right)
-80 \ln (2) \ln (\aj+v+1) \ln \left(x^2+1\right)
\nn\\
&&
+16 (-4 \ln (\ai+v+1)-\ln (\aj+v+1)+5 \ln (2)) \ln (x) \ln \left(x^2+1\right)
-\f{80 \ln ^3(2)}{3}
\nn\\
&&
+\left(8 \ln \left(\f{\ai}{\aj}\right)+16 \ln (x)\right) \Li_2\left(\f{1-v}{\aj}\right)
+\left(16 \ln (x)-8 \ln \left(\f{\ai}{\aj}\right)\right) \Li_2\left(\f{\aj}{v+1}\right)
\nn\\
&&
+\left(4 \ln \left(\f{\ai}{\aj}\right)-8 \ln (\ai+v+1)-8 \ln (\aj+v+1)+24 \ln (x)-16 \ln \left(x^2+1\right)+16 \ln  (2)\right) 
\nn\\
&&
\quad \times \left( 
\Li_2\left(\f{v-1}{\aj}\right)-\Li_2\left(\f{\aj}{\aj-v+1}\right)
\right)
\nn\\
&&
+\left(4 \ln \left(\f{\ai}{\aj}\right)-8 \ln (\ai+v+1)-8 \ln (\aj+v+1)-8 \ln (x)-16 \ln \left(x^2+1\right)+16 \ln  (2)\right) 
\nn\\
&&
\quad \times \left( 
\Li_2\left(-\f{v+1}{\aj}\right)- \Li_2\left(\f{\aj}{\aj+v+1}\right)
\right)
\nn\\
&&
+8( \ln (v)- \ln (\aj+v+1)+ \ln (2)) \Li_2\left(-\f{(v-1) (\aj+v+1)}{(\aj-v+1) (v+1)}\right)
-16 \ln (x) \Li_2\left(x^2\right)
\nn\\
&&
-16 \Li_3\left(\f{1-v}{\aj}\right)
-16 \Li_3\left(\f{\aj}{\aj-v+1}\right)
+8 \Li_3\left(\f{\aj}{v-1}\right)
-16 \Li_3\left(\f{v-1}{\aj}\right)
+8 \Li_3\left(-\f{\aj}{v+1}\right)
\nn\\
&&
-8 \Li_3\left(-\f{2 v}{\aj-v+1}\right)
-16 \Li_3\left(\f{\aj}{v+1}\right)
+8 \Li_3\left(x^2\right)
-16 \Li_3\left(-\f{v+1}{\aj}\right)
\nn\\
&&
-16 \Li_3\left(\f{v-1}{-\aj+v-1}\right)
-16 \Li_3\left(\f{\aj}{\aj+v+1}\right)
-8 \Li_3\left(\f{2 v}{\aj+v+1}\right)
\nn\\
&&
-8 \Li_3\left(\f{2 \aj v}{(v-1) (\aj+v+1)}\right)
-16 \Li_3\left(\f{v+1}{\aj+v+1}\right)
+8 F_c \left(\f{\aj}{\aj-v+1},\f{\aj}{\aj+v+1}\right)
%%%%%%%%%%%%%%%%%%%%%%
\Biggr)
\Biggr]\, , 
\eea
where 
\be
\begin{split}
Q_S &=  16\left( m_J^2 \piq^2 - 2 \pipj \piq \pjq + m_I^2 \pjq^2
\right) \; , \\ 
\alpha_I &= {m_I^2(p_J\cdot q)\over (p_I\cdot q) (p_I\cdot p_J) } \;
, \quad
\alpha_J = {m_J^2(p_I\cdot q)\over (p_J\cdot q) (p_I\cdot p_J) } \;
, \quad
x = \sqrt{\f{1-v}{1+v}} \; , \quad v = v_{IJ}\, . 
\label{eq:variables}
\end{split}
\ee
Furthermore
\be
F_c(x_1,x_2) = \int_0^1 dt  {\ln(1 - t) \ln\left(1 - t{x_2\over
    x_1}\right) \over {1\over x_2} - t} \; .
\ee
%

%%%%%%%%%%%%%%%%%%%%%%%%%%%%%%%%%%%%%%%%%%%%%%%%%%%%%%%%%%%%%%%%%%%%%%%%%%%%%%%%

\subsection{Limits of matrix elements of $\bm{\mathrm{Z}}^{(1)}$}
\label{sec:limitsZ1}

\noindent
The matrix element of the $\bm{\mathrm{Z}}^{(1)}$ operator can be
obtained from Eq.~(\ref{result})
\be
\begin{split} \label{Z1explicit}
2\R \, \la \cm_{n+1}^{(0)} | \bm{\mathrm{Z}}^{(1)} | \cm_{n+1}^{(0)}
\ra  & = \\ \f{\as}{4\pi} \, \f{1}{\ep} \, \Biggl[ & \biggl( -\f{2}{\ep}
  \sum_{i_0}C_{i_0} + \sum_{i} \gamma_0^i \biggr) \,
  |\cm_{n+1}^{(0)}|^2 + 2 \sum_{(i_0,j_0)} \ln\left|
  \f{\mu_R^2}{s_{i_0j_0}} \right| \, \la \cm_{n+1}^{(0)} | \cT_{i_0}
  \cdot \cT_{j_0} | \cm_{n+1}^{(0)} \ra \\ &- \sum_{(I,J)}
  \f{1}{v_{IJ}} \ln\left( \f{1+v_{IJ}}{1-v_{IJ}} \right) \, \la
  \cm_{n+1}^{(0)} |\cT_I \cdot \cT_J | \cm_{n+1}^{(0)} \ra  + 4
  \sum_{I,j_0} \ln\left| \f{m_I\mu_R}{s_{Ij_0}} \right| \, \la
  \cm_{n+1}^{(0)} | \cT_I \cdot \cT_{j_0} | \cm_{n+1}^{(0)} \ra
  \Biggr] \; .
\end{split}
\ee
The factorization of Eq.~(\ref{Z1explicit}) in the collinear limit,
(\ref{clim}), reads
\be
\begin{split}
2\R \, \la \cm_{a_1,a_2,\dots}^{(0)}(p _1,p_2,\dots) |
\bm{\mathrm{Z}}^{(1)} | \cm_{a_1,a_2,\dots}^{(0)}&(p_1,p_2,\dots) \ra
\simeq \\ 4 \pi \as \f{2}{s_{12}} \, \Biggl\{ & 2\R \, \la
\cm_{a,\dots}^{(0)}(p,\dots) | \sP^{(0)}_{a_1a_2}(z,\kper;\ep) \,
\bm{\mathrm{Z}}^{(1)} | \cm_{a,\dots}^{(0)}(p,\dots) \ra \\ & +
\f{\as}{4\pi} \, \f{1}{\ep} \, \Biggl[ 2\left( C_a - C_{a_1}  -
  C_{a_2} \right) \Biggl( \f{1}{\ep} + \ln\biggl| \f{\mu_R^2}{s_{12}}
  \biggr| \Biggr) - \left( \gamma_0^{a} - \gamma_0^{a_1} -
  \gamma_0^{a_2} \right) \\ & \hspace{1.25cm} + 2
  C_a\ln\bigl|z(1-z)\bigr| + 2 \left( C_{a_1} - C_{a_2} \right)
  \ln\left|\f{z}{1-z}\right|  \Biggl] \\ & \hspace{1.25cm} \times \la
\cm_{a,\dots}^{(0)}(p,\dots) | \sP^{(0)}_{a_1a_2}(z,\kper;\ep) |
\cm_{a,\dots}^{(0)}(p,\dots) \ra \, \Biggr\} \; .
\end{split}
\ee
This expression is valid for both final- and initial state collinear
limits, if the crossing relation, (\ref{crossingrule}), is taken into
account in the initial state case.

\noindent
The factorization of Eq.~(\ref{Z1explicit}) in the soft limit, $q \to
0$, reads
\be
\begin{split}
2\R \, \la \cm_{g,a_1,\dots}^{(0)}&(q,p_1,\dots) |
\bm{\mathrm{Z}}^{(1)} | \cm_{g,a_1,\dots}^{(0)}(q,p_1,\dots) \ra
\simeq \\ - 4 \pi \as \Biggl\{ & \sum_{(i,j)} \biggl(
\mathcal{S}_{ij}(q) - \mathcal{S}_{ii}(q) \biggr) \, 2\R \, \la
\cm_{a_1,\dots}^{(0)}(p_1,\dots) | \cT_i \cdot \cT_j \,
\bm{\mathrm{Z}}^{(1)} | \cm_{a_1,\dots}^{(0)}(p_1,\dots) \ra \\ & +
\f{\as}{4\pi} \, \f{1}{\ep} \, \Biggl[ \sum_{(i,j)} \biggl(
  \mathcal{S}_{ij}(q) - \mathcal{S}_{ii}(q) \biggr) \left(  -2C_A
  \biggl( \f{1}{\ep} + \ln\bigl( {\textstyle \f{1}{2}} \mu_R^2
  \mathcal{S}_{ij}(q) \bigr) \biggr) +\gamma_0^g \right) \, \la
  \cm_{a_1,\dots}^{(0)}(p_1,\dots) | \cT_i \cdot \cT_j |
  \cm_{a_1,\dots}^{(0)}(p_1,\dots) \ra \\ & \hspace{1.25cm} - C_A
  \sum_{(I,J)} \biggl( \mathcal{S}_{IJ}(q) - \mathcal{S}_{II}(q)
  \biggr) \left( \f{1}{v_{IJ}} \ln \biggl( \f{1+v_{IJ}}{1-v_{IJ}}
  \biggr) + 2 \ln\biggl( \f{m_I m_J}{s_{IJ}} \biggr) \right) \, \la
  \cm_{a_1,\dots}^{(0)}(p_1,\dots) | \cT_I \cdot \cT_J |
  \cm_{a_1,\dots}^{(0)}(p_1,\dots) \ra \\ & \hspace{1.25cm} - 4 \pi
  \sum_{(i,j,k)} \mathcal{S}_{ik}(q) \left( \f{1}{v_{ij}}
  \theta(\sigma_{ij}) -\theta(\sigma_{iq})-\theta(\sigma_{jq}) \right)
  \, \la \cm_{a_1,\dots}^{(0)}(p_1,\dots) | f^{abc} T^a_i T^b_j T^c_k
  | \cm_{a_1,\dots}^{(0)}(p_1,\dots) \ra \Biggr] \Biggr\} \; .
\end{split}
\ee
%

%%%%%%%%%%%%%%%%%%%%%%%%%%%%%%%%%%%%%%%%%%%%%%%%%%%%%%%%%%%%%%%%%%%%%%%%%%%%%%%%

\subsection{Limits of the finite remainder}
\label{sec:limitsF1}

\noindent
The factorization of the finite remainder in the collinear limit,
(\ref{clim}), reads
\be
\begin{split}
2\R \, &\la\cm_{a_1,a_2,\dots}^{(0)}(p_1,p_2,\dots)|
\cf_{a_1,a_2,\dots}^{(1)}(p_1,p_2,\dots)\ra \simeq \\ &4 \pi \as
\f{2}{s_{12}} \Biggl[ 2\R\; \la \cm^{(0)}_{a,\dots}(p,\dots) |
  {\sP}^{(0)}_{a_1 a_2}(z,\kper;\ep = 0) |
  \cf^{(1)}_{a,\dots}(p,\dots) \ra + \f{\as}{4\pi} \la
  \cm^{(0)}_{a,\dots}(p,\dots) | {\sP}^{(1)}_{Fa_1 a_2}(z,\kper)
  | \cm^{(0)}_{a,\dots}(p,\dots) \ra \Biggr] \; .
\end{split}
\ee
The finite one-loop splitting functions, $\sP^{(1)}_{Fa_1a_2}(z,\kper)$,
are operators in spin space
\be
\la s | \sP^{(1)}_{Fa_1a_2}(z,\kper) | s' \ra =
\Ph^{(1), \,ss'}_{Fa_1a_2}(z,\kper)
\; ,
\ee
with
\be
\begin{split}
\Ph_{Fgg}^{(1), \,\mu \nu}(z,\kper) &= r^{gg}_{SF}(z)\,\Ph_{gg}^{(0),\mu
  \nu}(z,\kper;\ep = 0) - \f{4}{3} C_A \big(C_A - 2T_Fn_l\big)
\f{\kper^{\mu}\kper^{\nu}}{\kper^2} \; , \\ \Ph_{Fq{\bar q}}^{(1), \,\mu
  \nu}(z,\kper) = \Ph_{F{\bar q}q}^{(1), \,\mu \nu}(z,\kper) &=  r^{{\bar
    q}q}_{SF}(z)\,\Ph_{q{\bar q}}^{(0),\mu \nu}(z,\kper;\ep = 0) \; ,
\\[.2cm] \Ph_{Fqg}^{(1), \,s s'}(z,\kper) = \Ph_{F{\bar q}g}^{(1), \,s
  s'}(z,\kper) &= r^{qg}_{SF}(z)\,\Ph_{qg}^{(0),s s'}(z,\kper;\ep = 0) +
2C_F\big(C_A-C_F\big) \,\delta^{ss'} \; , \\[.2cm] \Ph_{Fgq}^{(1), \,s
  s'}(z,\kper)=\Ph_{Fg{\bar q}}^{(1), \,s s'}(z,\kper) &=
\Ph_{Fqg}^{(1), \,s s'}(1-z,\kper) \; .
\end{split}
\ee
The finite coefficients $r_{SF}^{a_1a_2}(z)$ are given by
\bea
r^{gg}_{SF}(z) &=& C_A \left( \f{5\pi^2}{6} - \ln^2 \biggl| \f{z}{1-z}
\biggr| + 2 \ln\big| z(1-z) \big| \ln\biggl| \f{\mu_R^2}{s_{12}}
\biggr| - \ln^2\biggl| \f{\mu_R^2}{s_{12}} \biggr| \right) \; , \\[0.2cm]
r^{{\bar q}q}_{SF}(z) &=& C_A \, \biggl( \f{152}{9} - \f{3\pi^2}{2}
\biggr) + C_F \, \biggl( \f{7\pi^2}{3}-16 \biggr) - \f{40}{9} T_F n_l
- C_A \ln^2\biggl| \f{z}{1-z} \biggr| \nn\\ && + 2 \, \Bigl( \beta_0 - 3
C_F + C_A \ln\bigl| z(1-z) \bigr| \Bigr) \ln\biggl|
\f{\mu_R^2}{s_{12}} \biggr| + \bigl( C_A - 2 C_F \bigr) \ln^2\biggl|
\f{\mu_R^2}{s_{12}} \biggr| + 2 \bigl( C_A - C_F \bigr) \, \pi^2
\, \theta(-s_{12}) \; , \\[0.2cm]
r^{qg}_{SF}(z) &=& \f{5\pi^2}{6} C_A + 4 C_F \ln |z| \ln\biggl|
\f{\mu_R^2}{s_{12}} \biggr| - C_A \, \Biggl( \ln\biggl| \f{z}{1-z}
\biggr| + \ln\biggl| \f{\mu_R^2}{s_{12}} \biggr| \Biggr)^2 + 4
\bigl( C_F - C_A \bigr) \, \R \, \Li_2\biggl( -\f{1-z}{z} \biggr) \; .
\eea
These expressions are valid for both final- and initial state collinear
limits, if the crossing relation, (\ref{crossingrule}), is taken into
account in the initial state case. Furthermore, the real part of the
dilogarithm in the case, when $z > 1$ can be obtained using
Eq.~(\ref{eq:analytic}).

\noindent
The factorization of the finite remainder in the soft limit, $q \to
0$, reads
\be
\begin{split}
2 \R \,
\la\cm_{g,a_1,\dots}^{(0)}(q,p_1,\dots)|&\cf_{g,a_1,\dots}^{(1)}(q,p_1,\dots)\ra
\simeq \; \\ -4\pi\as &\left\{ \sum_{(i,j)} \Bigl(
  \mathcal{S}_{ij}(q) - \mathcal{S}_{ii}(q) \Bigr) \, 2\R \, \la
  \cm_{a_1,\dots}^{(0)}(p_1,\dots)|\cT_i\cdot\cT_j
  |\cf_{a_1,\dots}^{(1)}(p_1,\dots) \ra  \right. \\ 
&\left. + \f{\as}{4\pi} \left[\sum_{(i,j)} \left({\cal
      S}_{ij}(q)-{\cal S}_{ii}(q)\right) \, R^F_{ij}\, \la
    \cm_{a_1,\dots}^{(0)}(p_1,\dots)|\cT_i\cdot\cT_j
    |\cm_{a_1,\dots}^{(0)}(p_1,\dots) \ra
    \right. \right.\\ & \hspace{.75cm} \left. \left. -  4\pi\sum_{(i,j,k)} {\cal
      S}_{ik}(q) \, I^F_{ij} \,\la
    \cm_{a_1,\dots}^{(0)}(p_1,\dots)|f^{abc} T^a_i T^b_j T^c_k
    |\cm_{a_1,\dots}^{(0)}(p_1,\dots)\ra\right]  \right\} \; ,
\end{split}
\ee
where the functions $R^F_{ij}$ and $I^F_{ij}$ are the
$\mathcal{O}(\ep^0)$ coefficients of (\ref{RImassless}) and
(\ref{RImassive}) after expansion in $\ep$
\bea
R^F_{ij} &=& 4C_A\left( R^{(0)}_{ij} + R^{(-1)}_{ij} \ln \Big(
{\textstyle \f{1}{2}} \mu_R^2 \mathcal{S}_{ij}(q) \Big) + {\textstyle
  \f{1}{2}} R^{(-2)}_{ij} \ln^2 \Big( {\textstyle \f{1}{2}} \mu_R^2
\mathcal{S}_{ij}(q) \Big) \right) \; ,\nn \\
I^F_{ij} &=& 2\left( I^{(0)}_{ij} + I^{(-1)}_{ij} \ln \Big(
{\textstyle \f{1}{2}} \mu_R^2 \mathcal{S}_{ij}(q) \Big) \right) \;.
\eea
%

%%%%%%%%%%%%%%%%%%%%%%%%%%%%%%%%%%%%%%%%%%%%%%%%%%%%%%%%%%%%%%%%%%%%%%%%%%%%%%%%

\section{Splitting functions}
\label{sec:AltarelliParisi}

\noindent
For the collinear factorization contribution, we need the splitting
functions up to ${\cal O}(\as)$ \cite{Ellis}
\begin{eqnarray}
P_{q_iq_j}(x,\as)&=&\delta_{ij}P^{(0)}_{qq}(x)+\f{\as}{2 \pi} P_{q_iq_j}^{(1)}(x)+\dots \; ,\\
P_{qg}(x,\as)&=&P^{(0)}_{qg}(x)+\f{\as}{2 \pi} P_{qg}^{(1)}(x)+\dots \; ,\\
P_{gq}(x,\as)&=&P^{(0)}_{gq}(x)+\f{\as}{2 \pi} P_{gq}^{(1)}(x)+\dots \; ,\\
P_{gg}(x,\as)&=&P^{(0)}_{gg}(x)+\f{\as}{2 \pi} P_{gg}^{(1)}(x)+\dots \; .
\end{eqnarray}
The leading order contributions are
\begin{eqnarray}
P^{(0)}_{qq}(x)&=&C_F\left[\f{1+x^2}{(1-x)_+}+\f{3}{2}\delta(1-x)\right] \; ,\\
P^{(0)}_{qg}(x)&=&T_F\left[x^2+(1-x)^2\right] \; ,\\
P^{(0)}_{gq}(x)&=&C_F\left[\f{1+(1-x)^2}{x}\right] \; ,\\
P^{(0)}_{gg}(x)&=&2C_A\left[\f{x}{(1-x)_+}+\f{1-x}{x}+x(1-x)\right] +\delta(1-x)\f{11C_A-4T_Fn_l}{6}\;.
\end{eqnarray}
Beyond leading order one writes the splitting function $P_{q_iq_j}$ in
terms of a flavor singlet (S) and non-singlet (V) contribution
\begin{eqnarray}
P_{q_iq_j}(x,\as)&=&\delta_{ij}P^{\mathrm{V}}_{qq}(x,\as)+P^{\mathrm{S}}_{qq}(x,\as)
\; ,\\
P_{q_i\bar{q}_j}(x,\as)&=&\delta_{ij}P^{\mathrm{V}}_{q\bar{q}}(x,\as)+P^{\mathrm{S}}_{q\bar{q}}(x,\as)
\; .
\end{eqnarray}
The next-to-leading order contribution to the splitting functions are
\begin{eqnarray}
P^{\mathrm{V}(1)}_{qq}(x)&=&C_F^2\left\{ -\left[2\ln x \ln(1-x)
  +\f{3}{2}\ln x\right]p_{qq}(x)-\left(\f{3}{2}+\f{7}{2}x\right)\ln x
-\f{1}{2}(1+x)\ln^2x -5(1-x)\right\}\nonumber \\
&&+C_FC_A\left\{\left[\f{1}{2}\ln^2x+\f{11}{6}\ln x +\f{67}{18}-\f{\pi^2}{6}\right]p_{qq}(x) +(1+x)\ln x + \f{20}{3}(1-x)\right\}\nonumber \\
&&+C_FT_Fn_l\left\{-\left[\f{2}{3}\ln x +\f{10}{9}\right]p_{qq}(x)-\f{4}{3}(1-x)\right\}+\delta P^{(1)}_{qq}(x)\;,
\\
\nonumber \\[.6cm]
P^{\mathrm{V}(1)}_{q\bar{q}}(x)&=&C_F\left(C_F-\f{C_A}{2}\right)\left\{2p_{qq}(-x)S_2(x)
+2 (1+x)\ln x +4(1-x)\right\}\; ,
\\
\nonumber \\[.6cm]
P^{\mathrm{S}(1)}_{qq}(x) &=& P^{\mathrm{S}(1)}_{q\bar{q}}(x)\,\, =\,\, C_F T_F
\left[-2+\f{20}{9x}+6x-\f{56}{9}x^2 +
  \left(1+5x+\f{8}{3}x^2\right)\ln x -(1+x)\,\ln^2x\right] \; ,
\\
%P^{(1)}_{qq}(x)&=&C_F^2\left\{-1+x+\left(\f{1}{2}-\f{3}{2}x\right)\ln x-\f{1}{2}(1+x)\ln^2x  -\left[\f{3}{2}\ln x + 2\ln x \ln(1-x) \right]p_{qq}(x) + 2p_{qq}(-x)S_2(x)\right\}\nonumber \\
%&&+ C_FC_A\left\{\f{14}{3}(1-x)+\left[\f{11}{6}\ln x + \f{1}{2}\ln^2x+\f{67}{18}-\f{\pi^2}{6} \right]p_{qq}(x)  - p_{qq}(-x)S_2(x)\right\}\nonumber \\
%&&+C_FT_Fn_l\left\{-\f{16}{3}+\f{40}{3}x+\left(10x+\f{16}{3}x^2+2\right)\ln x \right.\nonumber \\
%&&\left. -\f{112}{9}x^2+\f{40}{9x}-2(1+x)\ln^2x-\left[\f{10}{9}+\f{2}{3}\ln x\right]p_{qq}(x)\right\}+\delta P^{(1)}_{qq}(x)\;,
%\\
%
%
\nonumber \\[.6cm]
P^{(1)}_{qg}(x)&=&\f{C_FT_F}{2}\left\{4-9x-(1-4x)\ln x
-(1-2x)\ln^2x+4\ln(1-x)\right.\nonumber \\
&&\left. +\left[2\ln^2\left(\f{1-x}{x}\right)-4\ln\left(\f{1-x}{x}\right)-\f{2}{3}\pi^2+10\right]p_{qg}(x) \right\} \nonumber \\
&& +\f{C_AT_F}{2}\left\{\f{182}{9}+\f{14}{9}x+\f{40}{9x}+\left(\f{136}{3}x-\f{38}{3}\right)\ln x -4\ln(1-x) -(2+8x)\ln^2x+2p_{qg}(-x)S_2(x)\right.\nonumber \\
&&\left. +\left[-\ln^2x+\f{44}{3}\ln x -2\ln^2(1-x) +4 \ln(1-x) + \f{\pi^2}{3}-\f{218}{9}\right]p_{qg}(x)\right\}\;,
\\
\nonumber \\[.6cm]
P^{(1)}_{gq}(x)&=&C_F^2\left\{-\f{5}{2}-\f{7x}{2}+\left(2+\f{7}{2}x\right)\ln x-\left(1-\f{1}{2}x\right)\ln^2x -2x\ln(1-x)-\left[3 \ln(1-x)+\ln^2(1-x)\right]p_{gq}(x)\right\}\nonumber \\
&&+C_FC_A\left\{\f{28}{9}+\f{65}{18}x+\f{44}{9}x^2-\left(12+5x+\f{8}{3}x^2\right)\ln x +(4+x) \ln^2 x +2x\ln(1-x)+S_2(x)p_{gq}(-x)\right.\nonumber \\
&&\left.+\left[\f{1}{2}-2\ln x \ln(1-x)+\f{1}{2}\ln^2x+\f{11}{3}\ln(1-x)+\ln^2(1-x)-\f{\pi^2}{6}\right]p_{gq}(x)\right\}\nonumber \\
&&+C_FT_Fn_l\left\{-\f{4}{3}x-\left[\f{20}{9}+\f{4}{3}\ln(1-x)\right]p_{gq}(x)\right\}\;,
\\
\nonumber \\[.6cm]
P^{(1)}_{gg}(x)&=&C_FT_Fn_l\left\{-16+8x+\f{20}{3}x^2+\f{4}{3x}-(6+10x)\ln x -(2+2x)\ln^2x\right\}\nonumber \\
&&+C_AT_Fn_l\left\{2-2x+\f{26}{9}\left(x^2-\f{1}{x}\right)-\f{4}{3}(1+x)\ln x-\f{20}{9}p_{gg}(x)\right\}\nonumber \\
&&+C_A^2\left\{\f{27}{2}(1-x)+\f{67}{9}\left(x^2-\f{1}{x}\right)-\left(\f{25}{3}-\f{11}{3}x+\f{44}{3}x^2\right)\ln x \right.\nonumber \\
&&\left. +4(1+x)\ln^2x+2p_{gg}(-x)S_2(x)+\left[\f{67}{9}-4\ln x \ln(1-x)+\ln^2x-\f{\pi^2}{3}\right]p_{gg}(x)\right\}+\delta P_{gg}^{(1)}(x)\;,
\end{eqnarray}
where
\begin{equation}
S_2(x)=-2\Li_2(-x)+\f{1}{2}\ln^2x-2\ln x\ln(1+x)-\f{\pi^2}{6}\;.
\end{equation}
The functions $p_{qq}$, $p_{qg}$, $p_{gq}$ and $p_{gg}$ read
\begin{align}
p_{qg}(x)&=x^2+(1-x)^2\;, & p_{gq}(x)\,\,\,&=\f{1+(1-x^2)}{x}\;,\\
p_{qq}(x)&=\f{2}{(1-x)_+}-1-x\;, & p_{qq}(-x)&=\f{2}{1+x}-1+x\;,\\
p_{gg}(x)&=\f{1}{(1-x)_+}+\f{1}{x}-2+x(1-x)\;, & p_{gg}(-x)&=\f{1}{(1+x)}-\f{1}{x}-2-x(1+x)\;.
\end{align}
The terms proportional to the $\delta$-functions are
\begin{eqnarray}
\delta
P^{(1)}_{qq}(x)&=&\left[C_F^2\left\{\f{3}{8}-\f{\pi^2}{2}+6\zeta_3\right\}+C_FC_A\left\{\f{17}{24}+\f{11\pi^2}{18}-3\zeta_3\right\}-C_FT_Fn_l\left\{\f{1}{6}+\f{2\pi^2}{9}\right\}
  \right]\delta(1-x) \; 
\end{eqnarray}
and
\begin{eqnarray}
\delta
P_{gg}^{(1)}(x)&=&\left[C_A^2\left\{\f{8}{3}+3\zeta_3\right\}-C_FT_Fn_l-\f{4}{3}C_AT_Fn_l\right]\delta(1-x)
\; .
\end{eqnarray}
%

%%%%%%%%%%%%%%%%%%%%%%%%%%%%%%%%%%%%%%%%%%%%%%%%%%%%%%%%%%%%%%%%%%%%%%%%%%%%%%%%

%\section*{References}

\end{document}